\documentclass[3p]{elsarticle} %style used for arXiv
\usepackage[active]{srcltx} %enables inverse search
\usepackage{hyperref}
\usepackage{graphics}
\usepackage{amsmath}
\usepackage{amsfonts}
\usepackage{amssymb}

\newcommand{\be}{\begin{eqnarray}}
\newcommand{\ee}{\end{eqnarray}}

\newcommand{\non}{\nonumber\\}
\newcommand{\ave}[1]{\left\langle #1 \right\rangle}

\newcommand{\GeV}{\hbox{\,GeV}}
\newcommand{\MeV}{\hbox{\,MeV}}

\newcommand{\mh}{\hat{\mu}}

\newcommand{\beq}{\begin{equation}}
\newcommand{\eeq}{\end{equation}}
\newcommand{\bea}{\begin{eqnarray}}
\newcommand{\eea}{\end{eqnarray}}
\newcommand {\bc}{\begin{center}}
\newcommand {\ec}{\end{center}}

\def\lsim{\mathrel{\rlap{\lower4pt\hbox{\hskip1pt$\sim$}}
    \raise1pt\hbox{$<$}}}               
\def\gsim{\mathrel{\rlap{\lower4pt\hbox{\hskip1pt$\sim$}}
    \raise1pt\hbox{$>$}}}      
\title{Properties of hot and dense matter from relativistic
heavy ion collisions}

\author[GSI,TUD,FIAS]{Peter Braun-Munzinger}
\ead{p.braun-munzinger@gsi.de}
\author[LBL]{Volker Koch}
\ead{vkoch@lbl.gov}
\author[NCS]{Thomas Sch\"afer}
\ead{tmschaef@ncsu.edu}
\author[HD]{Johanna Stachel}
\ead{stachel@physi.uni-heidelberg.de}

\address[GSI]{EMMI, GSI Helmholtzzentrum fuer Schwerionenforschung, 64291 Darmstadt, Germany}
\address[TUD]{Technical University, 64287 Darmstadt, Germany}
\address[FIAS]{FIAS, 60438 Frankfurt, Germany}
\address[LBL]{Nuclear Science Division,
Lawrence Berkeley National Laboratory,
Berkeley, CA 94720, USA}
\address[HD]{Physikalisches Institut, Universit\"at Heidelberg, 69120 Heidelberg, Germany}
\address[NCS]{Department of Physics, 
North Carolina State University,
Raleigh, NC 27695, USA}

\begin{document}

\begin{abstract}
We review the progress achieved in extracting the properties of hot
and dense matter from relativistic heavy ion collisions at the
relativistic heavy ion collider (RHIC) at Brookhaven National
Laboratory and the large hadron collider (LHC) at CERN. We 
focus on bulk properties of the medium, in
particular the evidence for thermalization, aspects of the equation of
state, transport properties, as well as fluctuations and
correlations. We also discuss the in-medium properties of hadrons with
light and heavy quarks, and measurements of dileptons and
quarkonia. This review is dedicated to the memory of Gerald E.~Brown.
\end{abstract}

\maketitle
\tableofcontents

\graphicspath{{plots/}}

%%%%%%%%%%%%%%%%%%%%%%%%%%%%%%%%%%%%%%%%%%%%%%%%%%%%%%%%%%%%%%%%%%%%%%%%%%%%%
\section{Introduction}
\label{sec:intro}
%%%%%%%%%%%%%%%%%%%%%%%%%%%%%%%%%%%%%%%%%%%%%%%%%%%%%%%%%%%%%%%%%%%%%%%%%%%%%

Soon after the discovery of QCD \cite{Fritzsch:1973pi}, and following
the realization that QCD exhibits asymptotic freedom
\cite{Gross:1973id,Politzer:1973fx}, is was recognized that QCD
implies the existence of a new high temperature phase of weakly interacting 
quarks and gluons, termed the quark-gluon plasma 
\cite{Collins:1974ky,Cabibbo:1975ig,Shuryak:1977ut}.
The idea of a limiting temperature for hadronic matter predates 
the discovery of QCD, and a quantitative prediction $T\simeq 170 \MeV$ 
was obtained in the statistical bootstrap model of Hagedorn 
\cite{Hagedorn:1965st}. The existence of a new phase was confirmed 
in the first calculations using the lattice formulation of QCD, initially 
for pure $SU(2)$ gauge theory \cite{Creutz:1980zw,Creutz:1979dw,McLerran:1980pk,Kuti:1980gh}. 

 These results inspired the community to explore the possibility to
create and study the quark-gluon plasma  by colliding heavy
nuclei at high energy, see for example \cite{Chin:1978gj}. Early
ideas of creating thermodynamically equilibrated matter in high
energy hadronic collisions go back to Fermi \cite{Fermi:1950jd},
Landau \cite{Landau:1953gs}, and Hagedorn \cite{Hagedorn:1965st}.
The idea of colliding $U+U$ at the CERN ISR was considered, 
but not pursued, in the late 1960s. The subject received 
``subtle stimulation'' \cite{Baym:2001in} from a workshop on ``GeV/
nucleon collisions of heavy ions'' at Bear Mountain, New York
\cite{BearMountain}. 
A meeting on ultra-relativistic heavy ion physics was convened 
in Berkeley in 1979 \cite{:1900mid}, which spawned a series of 
Quark Matter conferences that continue to this day.

 An experimental relativistic heavy ion program began at the Bevalac
 facility at Lawrence Berkeley National Laboratory in the mid
 nineteen-seventies, initially motivated by the study of compressed
 nuclear matter and the search for ``abnormal'' states of matter, such
 as pion condensed matter or Lee-Wick matter
 \cite{Stock:1977uy,Nagamiya:1984vk}.  These experiments discovered a
 number of collective phenomena \cite{Gustafsson:1984ka}, such as
 hydrodynamic flow, that are still being studied today. Exploratory
 experiments in the highly relativistic regime, initially carried out
 with rather small nuclei, began at the Brookhaven AGS and the CERN
 SPS accelerator in 1986. These experiments confirmed that a
 significant amount of energy is being deposited at mid-rapidity. It
 was also found that the observed particle yields are well described
 be the Hagedorn inspired hadron resonance model
 \cite{BraunMunzinger:1994xr,BraunMunzinger:1995bp}. There were
 already some surprises, such as an unexpected enhancement of low mass
 lepton pairs \cite{Agakishiev:1995xb}.
 
 The availability of $Pb$ beams at the SPS, and the beginning of the 
collider era at the dedicated Relativistic Heavy Ion Collider (RHIC) 
at Brookhaven, mark the beginning of the current era in relativistic
heavy ion physics. A wealth of phenomena were discovered, many of 
them surprising. At CERN this includes the observation of anomalous
$J/\psi$ suppression \cite{Abreu:1999qw}, the enhanced, compared to pp
collisions, production of strange hadrons \cite{Andersen:1999ym}, as
well as a low-mass enhancement coupled with the disappearance of the
rho-peak in dilepton measurements \cite{Agakishiev:1997au}.  

The central discoveries at RHIC are the observation of a large
azimuthal asymmetry, known as elliptic flow $v_2$, in the particle
yields \cite{Ackermann:2000tr}, as well as a strong suppression of
high energy jets and heavy quarks \cite{Adcox:2001jp}.  The observed
elliptic flow was consistent with predictions from ideal
hydrodynamics, which was puzzling, since one expected to find a weakly
interacting quark-gluon plasma, which should not exhibit fluid dynamic
behavior. Further analysis of this effect, together with the large
opacity of the QGP implied by the jet quenching data, forced a
paradigm shift.  In particular, it was argued that, instead of the
originally anticipated weakly coupled system of quarks and gluons, the
experiments had discovered a strongly interacting quark-gluon plasma
(sQGP) \cite{Back:2004je,Arsene:2004fa,Adcox:2004mh,Adams:2005dq}.
 
These experimental advances were accompanied by important theoretical
developments and breakthroughs. For example it was realized that by
using methods developed in string theory, the holographic duality
between gravitational theories in warped higher dimensional space-time 
and gauge theories in flat space on its boundary, one could study certain 
strongly interacting theories \cite{Maldacena:1997re}. Using these 
techniques it was shown that theories that can be realized using holographic
dualities saturate a lower bound on the shear viscosity over entropy
density ratio \cite{Policastro:2001yc}. In addition, inspired by the
RHIC data, the theory community revisited the long-standing problem of
relativistic viscous fluid dynamics and turned it into a practical
tool \cite{Dusling:2007gi,Romatschke:2007mq}. And, last but not least,
improved actions and algorithms together with increased computing
power allowed lattice simulations of real QCD with realistic quark
masses. These calculations for example showed that the transition at
vanishing baryon density is a cross-over \cite{Aoki:2006we} with a
pseudo-critical temperature of $T\simeq 150 \MeV$
\cite{Aoki:2006br,Aoki:2009sc,Bazavov:2011nk,Bazavov:2014pvz}.

 Experiments at the CERN LHC at about ten times the energy of RHIC
 confirmed several of the main results obtained at RHIC, such as
 elliptic flow and energy loss, while providing improved statistics
 and a much larger kinematic range for many key observables. An
 example is the detailed measurement of higher flow harmonics, first
 carried out at the LHC, which opened up a window for testing the
 fluctuating initial conditions with the possibility to probe the
 structure of nuclei at the partonic scale. LHC experiments also made
 a number of surprising discoveries. For example, features 
 resembling collective flow have been observed in small
  systems, such as $p+Pb$ and possibly even $p+p$
  \cite{Abelev:2012ola,Khachatryan:2015waa}. Furthermore, the J/$\psi$ 
 suppression in Pb-Pb collisions at the LHC \cite{Abelev:2013ila} is  
 much reduced compared to measurements at RHIC.
 Although this has been predicted many years before the start
  of the LHC \cite{BraunMunzinger:2000px,Thews:2000rj}, 
  and thus should not have been a surprise, it took the
  actual measurement for these ideas to be taken seriously.

 While the experiments at the highest energies focus on detailed
 measurements of the properties of a QGP, a new program at RHIC has
 started with the goal to explore the QCD phase diagram at finite net
 baryon density. To achieve this a systematic beam energy scan down to
 the lowest energies available at RHIC has been carried out
 \cite{Odyniec:2013aaa}. The first set of measurements found
 intriguing non-monotonic dependence on the beam energy of some of the 
 key observables, such as proton number fluctuation and system size at
 freeze out as determined by Hanbury-Brown Twiss type pion
 correlation, as one would expect from a phase change at finite
 density. At the same time flow observables are remarkably insensitive
 to the collision energy, adding to the puzzle raised by the
 unexpected flow in small systems.

In this review we wish to summarize some of the recent development
discussed above, and provide an introduction to recent observations
and ideas. We wish to dedicate this review to the memory of Gerry
Brown, our teacher, mentor, and friend. After many years of working
in nuclear structure and nuclear astrophysics, Gerry developed an
interest in relativistic heavy ion collisions as a way of pinning
down the nuclear equation of state at densities above nuclear
saturation density, which is of interest for type II supernova
explosions \cite{Ainsworth:1987hc,Brown:1990pp}. Gerry was well 
aware of the theoretical and experimental developments in the field. 
He had been a speaker at the Bear Mountain workshop, describing 
his work on pion condensation in nuclear matter \cite{BearMountain}. 
He applied his expertise in the theory of collective modes to the
problem of hadrons in hot and dense matter \cite{Bertsch:1988xu}, and,
in collaboration with M.~Rho, developed the idea of Brown-Rho scaling
\cite{Brown:1991kk}, which drove much of the interest in dilepton
experiments for many years. In general, his interests focused on bulk
phenomena, like the equation of state \cite{Brown:1993dm}, and the
properties and spectra of hadrons \cite{Brown:1991en}.  Gerry followed
the early RHIC results, as well as improved results from CERN, with
great interest, but he fell seriously ill before the start of the
LHC. He certainly would have been excited to see the first results. We
spent many hours discussing the physics of relativistic heavy ions
with him in the office, at lunch in the nuclear theory common room, or
at dinner in his Setauket home.

 This review focuses on issues that were closest to his interests, the
 bulk properties of hot and dense matter, the spectra of produced
 particles and the evidence for thermalization, as well as the
 in-medium properties of hadrons. The review is organized as
 follows. In Section~\ref{sec_phase} we provide an overview of the
 phase diagram of QCD, and discuss the equation of state. In 
 Section~\ref{sec:PJ_thermal} we summarize experimental results on hadron
 spectra, as well as the evidence that thermalization is achieved in
 relativistic heavy ion collisions.  The theory of locally
 equilibrated matter, relativistic fluid dynamics, is discussed in
 Section~\ref{sec_hydro}, together with ongoing efforts to determine
 transport properties of the QGP, and discover the limits of
 applicability of fluid dynamics. In Section~\ref{sec:fluct} we
 consider fluctuations and correlations, as well as their role is the
 extraction of freeze-out properties and the search for the critical
 point. We close with Sections devoted to the in-medium properties of
 hadrons, Section~\ref{sec_hadrons}, the production of dileptons,
 Section~\ref{sec:dilepton}, and the spectra of hadrons containing
 heavy quarks, Section~\ref{sect:quarkonium}. Finally, we provide a
 brief outlook in Section~\ref{sec:out}.

%%%%%%%%%%%%%%%%%%%%%%%%%%%%%%%%%%%%%%%%%%%%%%%%%%%%%%%%%%%%%%%%%%%%
\section{The phase structure of QCD}
\label{sec_phase}
\subsection{The phases of QCD}
\subsubsection{The QCD Vacuum}
%%%%%%%%%%%%%%%%%%%%%%%%%%%%%%%%%%%%%%%%%%%%%%%%%%%%%%%%%%%%%%%%%%%%

 Strongly interacting matter has a rich phase structure, which includes 
a nuclear liquid phase, a hadronic gas, and the quark-gluon plasma. 
All these states of matter are described by quantum chromodynamics
(QCD), which is the theory of quarks and gluons and their interactions. 
The complicated phenomenology of the strong interaction is encoded in 
a deceptively simple Lagrangian. The Lagrangian is formulated in terms 
of quark fields $q_{\alpha\, f}^c$ and gluon fields $A_\mu^a$. Here, $\alpha=1,
\ldots,4$ is a Dirac spinor index, $c=1,\ldots,N_c$ with $N_c=3$ is a 
color index, and $f={\it up}, {\it down},{\it strange}, {\it charm},
{\it bottom}, {\it top}$ is a flavor index. 

 The dynamics of the theory is governed by the color degrees of freedom. 
The gluon field $A_\mu^a$ is a vector field labeled by an adjoint color 
index $a=1,\ldots,8$. The octet of gluon fields can be used to construct 
a matrix valued field $A_\mu=A_\mu^a \frac{\lambda^a}{2}$, where $\lambda^a$ 
is a set of traceless, Hermitian, $3\times 3$ matrices. The QCD Lagrangian is
\be
\label{l_qcd}
 {\cal L } =  - \frac{1}{4} G_{\mu\nu}^a G_{\mu\nu}^a
  + \sum_f^{N_f} \bar{q}_f ( i\gamma^\mu D_\mu - m_f) q_f\, ,
\ee
where $G^a_{\mu\nu}$ is the QCD field strength tensor defined by 
\be
 G_{\mu\nu}^a = \partial_\mu A_\nu^a - \partial_\nu A_\mu^a
  + gf^{abc} A_\mu^b A_\nu^c\, ,
\ee
and $f^{abc}=4i\,{\rm Tr}([\lambda^a,\lambda^b]\lambda^c)$ is a
set of numbers called the $SU(3)$ structure constants. The
covariant derivative acting on the quark fields is
\be
 i D_\mu q =  \left(
 i\partial_\mu + g A_\mu^a \frac{\lambda^a}{2}\right) q\, ,
\ee
and $m_f$ is the mass of the quarks. The terms in Eq.~(\ref{l_qcd}) 
describe the interaction between quarks and gluons, as well as nonlinear 
three and four-gluon interactions. We observe that, except for the number 
of flavors and their masses, the structure of the QCD Lagrangian is
completely fixed by the local $SU(3)$ color symmetry.

 For the purpose of understanding hadronic matter and the quark-gluon 
plasma we can consider the light flavors (up, down, and strange) 
to be approximately massless, and the heavy flavors (charm, bottom, top) 
to be infinitely massive. In this limit the QCD Lagrangian contains a 
single dimensionless parameter, the coupling constant $g$. If quantum 
effects are taken into account the coupling becomes scale dependent
\cite{Gross:1973id,Politzer:1973fx}. At leading order the running 
coupling constant is
\be
\label{g_1l}
 g^2(q^2) = \frac{16\pi^2}
  {b_0\log(q^2/\Lambda_{\it QCD}^2)}\, , \hspace{1cm}
 b_0=\frac{11}{3}N_c-\frac{2}{3}N_f\, ,
\ee
where $q$ is a characteristic momentum and $N_f$ is the number of active 
flavors ($N_f=3$ in the approximation considered here). The running 
coupling implies that, as a quantum theory, QCD is not characterized 
by a dimensionless coupling but by a dimensionful scale, the QCD scale 
parameter $\Lambda_{\it QCD}$. This effect is known as dimensional 
transmutation~\cite{Coleman:1973jx}. We also observe that the coupling 
decreases with increasing momentum. This is the phenomenon of asymptotic 
freedom~\cite{Gross:1973id,Politzer:1973fx}. The flip side of asymptotic 
freedom is anti-screening, or confinement: The effective interaction 
between quarks increases with distance.

 In massless QCD the scale parameter is an arbitrary parameter (a QCD
``standard kilogram''), 
and all observables are dimensionless ratios
like $m_p/\Lambda_{\it QCD}$, where $m_p$ is the mass of the proton. If
QCD is embedded into the electroweak sector of the standard model, and 
quarks acquire masses by electroweak symmetry breaking, then the QCD 
scale is fixed by the choice of units in the standard model. A number
that is commonly quoted is the value of the QCD fine structure constant
$\alpha_s=g^2/(4\pi)$ at the $Z$ boson pole, $\alpha_s(m_z)= 0.1184\pm
0.0007$ \cite{Nakamura:2010zzi}. The numerical value of $\Lambda_{QCD}$ 
depends on the renormalization scheme used to derive Eq.~(\ref{g_1l}). 
Physical masses, as well as the value of $b_0$, are independent of this 
choice. In the modified minimal subtraction ($\overline{MS}$) scheme
one finds $\Lambda_{QCD}\simeq 200$ MeV \cite{Nakamura:2010zzi}.

%%%%%%%%%%%%%%%%%%%%%%%%%%%%%%%%%%%%%%%%%%%%%%%%%%%%%%%%%%%%%%%%%%%%%%%%%
\begin{figure}[t]
\bc\includegraphics[width=0.7 \textwidth]{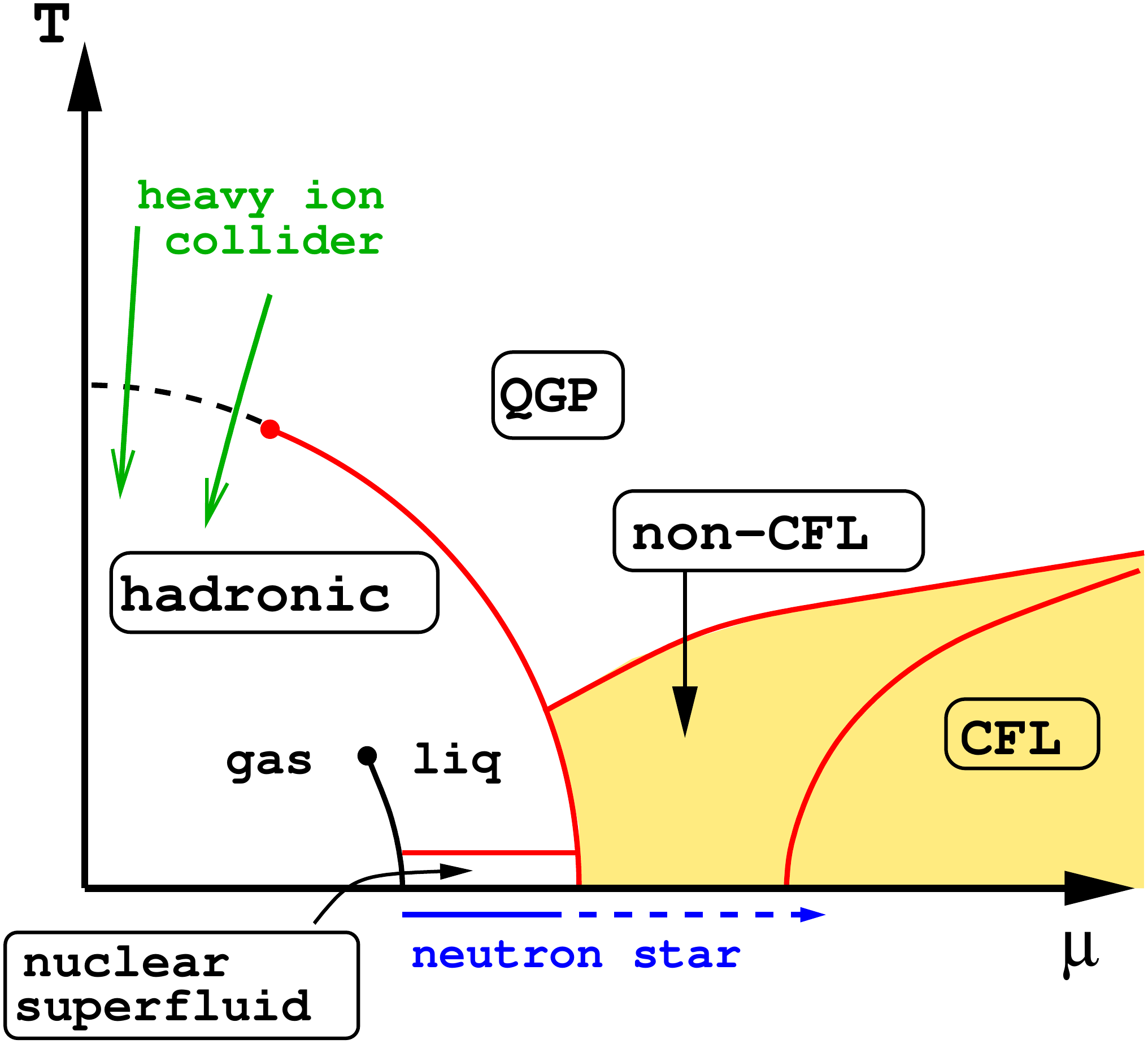}\ec
\caption{\label{fig_qcd_phase}
Schematic phase diagram of QCD as a function of temperature $T$ and
baryon chemical potential $\mu$. QGP refers to the quark-gluon plasma. 
The CFL (color-flavor locked) phase is the color superconducting phase 
that occurs at asymptotically large chemical potential. The red and 
black points denote the critical endpoints of the chiral and nuclear 
liquid-gas phase transitions, respectively. The dashed line is the 
chiral pseudo-critical line associated with the crossover transition
at low temperature. The green arrows denote the regions of the phase 
diagram that are being explored by the experimental heavy ion programs 
at the LHC and RHIC.}
\end{figure}
%%%%%%%%%%%%%%%%%%%%%%%%%%%%%%%%%%%%%%%%%%%%%%%%%%%%%%%%%%%%%%%%%%%%%%%%%

 Asymptotic freedom and the symmetries of QCD determine the basic
phases of strongly interacting matter that appear in the QCD phase
diagram shown in Fig.~\ref{fig_qcd_phase}. In this figure we show the
phases of QCD as a function of the temperature $T$ and the baryon
chemical potential $\mu$. The chemical potential $\mu$ controls the
baryon density $\rho$, defined as 1/3 times the number density of
quarks minus the number density of anti-quarks.

 At zero temperature and chemical potential the interaction between
quarks is dominated by large distances and the effective coupling is
large. As a consequence, quarks and gluons are permanently confined in
color singlet hadrons, with masses of order $\Lambda_{QCD}$. For example, 
the proton has a mass of $m_p=935$ MeV. If we view the proton as composed 
of three constituent quarks this implies that quarks have effective masses 
$m_Q\simeq m_P/3\simeq \Lambda_{QCD}$. This should be compared to the bare 
up and down quark masses which are of the order 10~MeV.

 Strong interactions between quarks, anti-quarks, and gluons lead
to the formation of vacuum condensates of color-singlet bosonic states. 
In particular, the QCD ground state supports a condensate of $\bar{q}q$ 
pairs, $\langle\bar{q}q\rangle\simeq
-\Lambda^3_{QCD}$~\cite{GellMann:1968rz,Coleman:1980mx,tHooft:1979bh}.
The quark condensate couples left and right handed fermions, $\bar{q}q
=\bar{q}_Lq_R+\bar{q}_Rq_R$, and it is diagonal in flavor space, 
$\langle \bar{q}_fq_g\rangle = \delta_{fg}\langle\bar{q}q\rangle$.  
Quark-anti-quark condensation spontaneously breaks the approximate chiral
$SU(3)_L\times SU(3)_R$ flavor symmetry of the QCD Lagrangian down to
its vectorial subgroup, the flavor symmetry $SU(3)_V$. Chiral symmetry 
breaking implies the existence of Goldstone bosons, massless modes 
with the quantum numbers of the generators of the broken axial
symmetry $SU(3)_A$. These particles are pions, kaons, and etas.
The $SU(3)_L\times SU(3)_R$ is explicitly broken by quark masses, 
and the mass of the charged pion is $m_\pi=139$ MeV, which is not 
much smaller than $\Lambda_{QCD}$. The lightest non-Goldstone particle 
is the rho meson, which has a mass $m_\rho=770$ MeV.

%%%%%%%%%%%%%%%%%%%%%%%%%%%%%%%%%%%%%%%%%%%%%%%%%%%%%%%%%%%%%%%%%%%%%%%%
\subsubsection{High temperature QCD}
%%%%%%%%%%%%%%%%%%%%%%%%%%%%%%%%%%%%%%%%%%%%%%%%%%%%%%%%%%%%%%%%%%%%%%%%

At very high temperature quarks and gluons have thermal momenta $p\sim 
T\gg\Lambda_{QCD}$. Asymptotic freedom implies that these particles are 
weakly interacting, and that they form a plasma of mobile color charges, 
the quark-gluon plasma~\cite{Shuryak:1977ut,Shuryak:1978ij}. We note 
that the argument that the QGP at asymptotically high temperature is 
weakly coupled is somewhat more subtle than it might appear at first 
sight. If two particles in the plasma interact via large angle scattering 
then the momentum transfer is large, and the effective coupling is weak
because of asymptotic freedom. However, the color Coulomb interaction
is dominated by small angle scattering, and it is not immediately clear
why the effective interaction that governs small angle scattering is
weak. The important point is that in a high temperature plasma there
is a large thermal population ($n\sim T^3$) of mobile charges that
screen the interaction at distances beyond the Debye length $r_D\sim
1/(gT)$. We also note that even in the limit $T\gg\Lambda_{QCD}$ the
QGP contains a non-perturbative sector of static magnetic color
fields~\cite{Linde:1980ts}. This sector is strongly coupled, but
it does not contribute to thermodynamic or transport properties
of the plasma in the limit $T\to\infty$, see the discussion in 
Section~\ref{sec_qp_qgp}.

 The plasma phase exhibits neither color confinement nor chiral
symmetry breaking. This means that the high temperature QGP phase must
be separated from the low temperature hadronic phase by a phase transition.
The nature of this transition is very sensitive to the values of the quark
masses. In QCD with massless $u,d$ and infinitely massive $s,c,b,t$ quarks
the transition is second order~\cite{Pisarski:1983ms}. In the case of
massless (or sufficiently light) $u,d,s$ quarks the transition is first
order. Lattice simulations show that for realistic quark masses, $m_u
\simeq m_d\simeq 10$ MeV and $m_s\simeq 120$ MeV, the phase transition
is a rapid crossover~\cite{Aoki:2006we,Bazavov:2011nk}. A pseudo-critical
transition temperature can be defined by locating the maximum of the chiral 
susceptibility, that means by identifying the point at which fluctuations
of the chiral order parameter are largest. The result is $T_c\simeq 
151\pm 3 \pm 3 \MeV$ \cite{Aoki:2006br,Aoki:2009sc}, consistent with
the determination $154 \pm 9 \MeV$ published in
\cite{Bazavov:2011nk,Bazavov:2014pvz}.

 The transition is believed to strengthen as a function of chemical
potential, so that there is a critical $\mu$ at which the crossover
turns into a first order phase transition~\cite{Stephanov:2004wx}.
This point is the critical endpoint of the chiral phase transition.
Due to the fermion sign problem it is very difficult to locate the
critical endpoint using simulations on the lattice. Model calculations 
typically predict the existence of a critical point, but do not 
constrain its location. A number of exploratory lattice calculations 
have been performed \cite{Fodor:2001pe,Allton:2002zi,Karsch:2003va,Fodor:2004nz,Gavai:2008zr,Datta:2012pj},
but at this point it is not even clear whether the idea that the
transition strengthens with increasing baryon chemical potential 
is correct~\cite{deForcrand:2010he}. The critical endpoint is
interesting because it is the only thermodynamically stable point 
on the phase transition line at which the correlation length diverges 
(there is a similar endpoint on the nuclear liquid-gas transition line). 
This means that the critical point may manifest itself in heavy ion 
collisions in terms of enhanced fluctuations~\cite{Stephanov:1998dy}, see
Section~\ref{sec:fluct}.
 
 The $T=\mu=0$ point in the phase diagram corresponds to the vacuum 
state of QCD. If the chemical potential is increased at $T=0$ then 
initially there is no change, because at zero temperature the chemical
potential $\mu$ is the energy required to add a baryon to the system, 
and QCD has a large mass gap for baryonic states. The first non-vacuum 
state one encounters along the $\mu$ axis of the phase diagram is nuclear 
matter, a strongly correlated superfluid composed of approximately 
non-relativistic neutrons and protons. Nuclear matter is self-bound, 
and the baryon density changes discontinuously at the onset transition,
from $\rho_B=0$ to nuclear matter saturation density $\rho_B=\rho_0 \simeq 
0.15\,{\rm fm}^{-3}$. The discontinuity decreases as nuclear matter is 
heated, and the nuclear-liquid gas phase transition ends in a critical 
point at $T\simeq 18$ MeV \cite{mosel_liquid,Pochodzalla:1995xy,Elliott:2013pna}. 
Hot hadronic or nuclear matter can be described quite accurately as a 
weakly interacting gas of hadronic resonances, see Section~\ref{sec:PJ_thermal}. 
Empirically, the density of states for both mesons and baryons grows 
exponentially. This is reminiscent of the old string picture of hadronic 
resonances, and suggests that hadronic matter below $T_c$ can be viewed 
a Hagedorn gas. 

 We will show more detailed comparisons between lattice results and the 
hadronic resonance gas model in Section~\ref{sec:fluct}. One can try to 
make the resonance gas model more precise by considering the limit $N_c
\to \infty$. Witten and 't Hooft argued that in this limit hadronic 
resonances become narrow and weakly interacting, and that the $1/N_c$ 
expansion in gauge theory can be mapped onto the perturbative expansion
of a weakly coupled string theory \cite{tHooft:1973jz,Witten:1979kh}.
Non-interacting relativistic strings are known to have an exponential
density of states, and a limiting temperature $T_H$. However, $T_H$ 
cannot be precisely equal to the critical temperature of large $N_c$
QCD. Large $N_c$ QCD has a first order phase transition between a 
hadronic phase at low $T$ with pressure $O(N_c^0)$, and a deconfined
phase at large $T$ with pressure $O(N_c^2)$. The string gas, on the other 
hand, has a pressure that diverges as $T\to T_H$ \cite{Cohen:2006qd}.
Indeed, lattice calculations in large $N_c$ QCD suggest that 
$T_c<T_H$, and that $T_H$ corresponds to the endpoint of a 
meta-stable hadronic phase above $T_c$ \cite{Lucini:2005vg}. 

%%%%%%%%%%%%%%%%%%%%%%%%%%%%%%%%%%%%%%%%%%%%%%%%%%%%%%%%%%%%%%%%%%%%%%%
\subsubsection{High baryon density QCD}
%%%%%%%%%%%%%%%%%%%%%%%%%%%%%%%%%%%%%%%%%%%%%%%%%%%%%%%%%%%%%%%%%%%%%%%
 
 At very large chemical potential we can use arguments similar
to those in the high temperature limit to establish that quarks and 
gluons are weakly coupled. The main difference between cold quark 
matter and the hot QGP is that because of the large density of states 
near the quark Fermi surface even weak interactions can cause qualitative 
changes in the ground state of dense matter. In particular, attractive
interactions between quark pairs lead to color superconductivity and the 
formation of a $\langle qq\rangle$ condensate. Since quarks carry color, 
flavor, and spin labels, many superconducting phases are possible. The 
most symmetric of these, known as the color-flavor locked (CFL) phase, is 
predicted to exist at very high density~\cite{Alford:1998mk,Schafer:1999fe}. 
In the CFL phase the diquark order parameter is $\langle q^A_{\alpha f} 
q^B_{\beta g}\rangle \sim \epsilon_{\alpha\beta} \epsilon^{ABC}\epsilon_{fgC}$.
This order parameter has a number of interesting properties. It
breaks the $U(1)$ symmetry associated with baryon number, leading
to superfluidity, and it breaks the chiral $SU(3)_L \times SU(3)_R$
symmetry. Except for Goldstone modes the spectrum is fully gapped;
fermions acquire a BCS-pairing gap, and gauge fields are screened
by the Meissner effect. This implies that the CFL phase, even
though it arises from a superdense liquid of quarks, shares many
properties of superfluid nuclear matter.

 The CFL phase involves equal pair-condensates $\langle ud\rangle
=\langle us\rangle = \langle ds\rangle$  of all three light quark
flavors. As the density is lowered effects of the non-zero strange quark
mass become more important, and less symmetric phases are likely to
appear~\cite{Alford:2007xm}. Possible phases include Bose condensates
of pions and kaons, hyperon matter, states with inhomogeneous 
quark-anti-quark or diquark condensates, and less symmetric color 
superconducting phases. The intermediate $\mu$ regime in the phase
diagram shown in Fig.~\ref{fig_qcd_phase} is therefore largely 
conjecture. We know that at low $\mu$ there is a nuclear matter 
phase with broken chiral symmetry and zero strangeness, and that 
at high $\mu$ we find the CFL phase with broken chiral symmetry
but non-zero strangeness. In principle the two phases could 
be separated only by a continuous onset transition for strangeness
\cite{Schafer:1998ef,Hatsuda:2006ps}, but model calculation suggest a 
more complicated picture in which one or more first order transitions
intervene, as shown in Fig.~\ref{fig_qcd_phase}.

%%%%%%%%%%%%%%%%%%%%%%%%%%%%%%%%%%%%%%%%%%%%%%%%%%%%%%%%%%%%%%%%%%%%
\subsection{The equation of state}
\label{sec_EOS}
%%%%%%%%%%%%%%%%%%%%%%%%%%%%%%%%%%%%%%%%%%%%%%%%%%%%%%%%%%%%%%%%%%%%

 The most basic property of a phase of QCD, and the observable
that enters most directly in the theoretical description of an 
expanding quark-gluon plasma, is its equation of state (EOS). The
EOS governs the dependence of the pressure of the system on the 
energy and baryon density, $P=P({\cal E},n_B)$, or equivalently, 
on the temperature and chemical potential, $P=P(T,\mu)$. Here
we have used the fact that in thermodynamic equilibrium the 
total electric charge must be zero, and strangeness is not 
conserved. In a heavy ion collision the system has a net charge
and strangeness is approximately conserved. However, at mid-rapidity 
both net isospin and strangeness are approximately 
zero, and these are the conditions we will consider in the 
following.  

 A fundamental quantity that determines the expansion of hot dense 
matter is the speed of sound, 
\be
c_s^2= \left. \frac{\partial P}{\partial {\cal E}}\right|_{s/n_B}\, , 
\ee
where the derivative is taken at constant entropy per baryon. Note 
that at $n_B=0$ the speed of sound is simply a function of temperature.
The EOS determines how gradients in the energy density profile are 
translated into pressure gradients. In hydrodynamics, pressure 
gradients lead to acceleration, and generate collective expansion. 

There are several regimes in which we can analytically control the 
calculation of $c_s$. One is the regime of very high temperature, 
$T\gg T_c$. In this regime the running of the coupling is slow, and
QCD is approximately scale invariant. This implies that the equation 
of state is ${\cal E}=3P$ and $c_s^2=1/3$. Perturbative corrections
to this result are computable and start at $O(\alpha_s^2)$ 
\cite{Kapusta:1979fh,Arnold:2006fz}. At very low temperature and $n_B=0$ 
the pressure is dominated by weakly interacting pions. If pions are 
massless we also find $c_s^2=1/3$. In practice, pions are non-relativistic 
for $T\lsim 100$ MeV, and the speed of sound approaches that of a 
classical gas, $c_s^2\simeq T/m_\pi$.

 At zero baryon density we then expect the following behavior of the 
speed of sound: At low temperature the speed of sound is rising 
towards $c_s^2\sim 1/3$. Near the crossover temperature matter is 
very compressible and the speed of sound has a minimum. As a function
of baryon density, the minimum speed of sound tends to zero as we
approach the critical point. At high temperature $c_s^2$ increases
towards the perturbative value $1/3$. This behavior implies that 
a system produced with an initial energy density far above the 
critical density will initially accelerate quite rapidly, and 
then coast through the phase transition regime. More importantly, 
a systems produced near the critical energy density will tend to 
spend some amount of time in the critical regime. 

 The behavior of the speed of the sound in cold dense matter is
quite different. Cold nuclear matter can be understood as a Fermi
liquid of protons and neutrons. If we ignore interactions, then
the velocity of sound is given by $c_s^2=k_F^2/(3m)$, where the 
Fermi momentum is defined in terms of the baryon density by 
$\rho_B=2k_F^3/(3\pi^2)$ and we have assumed that the system is 
isospin symmetric. Near nuclear matter saturation density the
ideal Fermi gas speed of sound is $c_s\simeq 0.15$. Interactions
between nucleons can be described using effective masses and 
Landau parameters, and lead to modest corrections to the free Fermi
gas result. The equation of state at supra-nuclear densities is 
constrained by neutron star masses and radii. The existence of 
neutron stars with masses close to two solar masses indicates
that the high density EOS is quite stiff, and that the speed
of sound at several times nuclear matter density is most likely
close to the speed of light \cite{Bedaque:2014sqa}. We note that
the speed of sound in asymptotically dense quark matter approaches 
the scale invariant value $c_s^2=1/3$. This implies that whereas
the EOS is very soft, and $c_s^2$ has a minimum, in hadronic
matter below the finite temperature transition, the EOS is very
stiff, and $c_s^2$ has a maximum, in hadronic matter below the
finite baryon density transition.

%%%%%%%%%%%%%%%%%%%%%%%%%%%%%%%%%%%%%%%%%%%%%%%%%%%%%%%%%%%%%%%%%%%%
\subsection{Lattice QCD}
\label{sec_lQCD}
%%%%%%%%%%%%%%%%%%%%%%%%%%%%%%%%%%%%%%%%%%%%%%%%%%%%%%%%%%%%%%%%%%%%

 In the high temperature regime corrections to the equation of state 
of an ideal quark-gluon plasma can be calculated in perturbation theory. 
The perturbative expansion is based on the separation of scales $m_M\ll 
m_D \ll T$, where $m_M\sim g^2T$ and $m_D\sim gT$ are the effective masses 
of magnetic and electric modes in the plasma. Strict perturbation theory 
in $g$ works only for very small values of the coupling constant, $g\lsim 
1$ \cite{Kajantie:2002wa}. However, quasi-particle models that rely on 
the separation of scales, but not on strict perturbation theory, describe 
the thermodynamics of the plasma quite well, even for temperatures close 
to the phase transition to a hadronic gas \cite{Blaizot:2003tw}. 
Quasi-particle models are quite useful, in particular in connecting 
equilibrium to non-equilibrium properties of the plasma, but reliable 
results for the equation of state in the vicinity of $T_c$ can only 
be obtained using numerical calculations on the lattice.

%%%%%%%%%%%%%%%%%%%%%%%%%%%%%%%%%%%%%%%%%%%%%%%%%%%%%%%%%%%%%%%%%%%%%%%%%
\begin{figure}[t]
\bc\includegraphics[width=0.7 \textwidth]{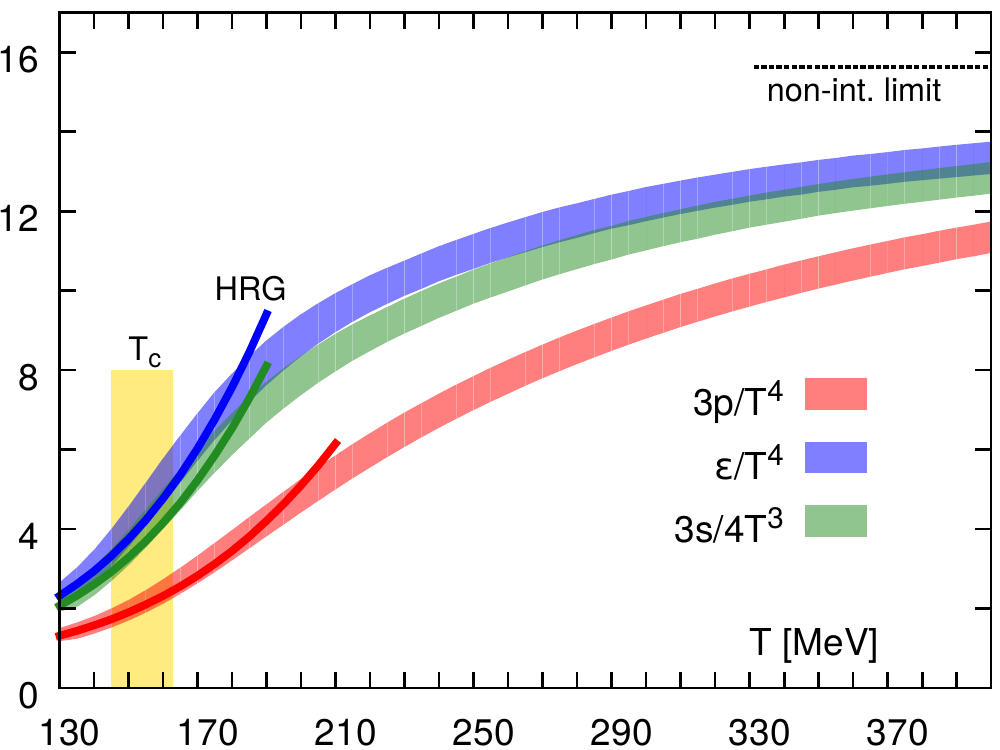}\ec
\caption{\label{fig_eos}
Equation of state of QCD with 2+1 flavors, from \cite{Bazavov:2014pvz}. The 
shows the normalized pressure, energy density, and entropy density as a 
function of the temperature. The bands indicate systematic and statistical
errors. The lines show the prediction of the hadron resonance gas model. 
The horizontal band at $T_c=(154\pm 9)$ MeV indicates critical
regime. These results are in very good agreement with previous
calculations based on a different lattice discretization  scheme \cite{Borsanyi:2010cj}.}
\end{figure}
%%%%%%%%%%%%%%%%%%%%%%%%%%%%%%%%%%%%%%%%%%%%%%%%%%%%%%%%%%%%%%%%%%%%%%%%%

 Lattice QCD is based on the euclidean path integral representation
of the partition function, see \cite{Fodor:2009ax,Ding:2015ona}
for recent reviews. We have
\be
\label{Z_latt}
 Z(T,\mu,V) = \int {\cal D}A_\mu\, {\cal D}q_f\, {\cal D}\bar{q}_f
 \; \exp(-S_E) 
\ee
where $S_E$ is the euclidean action 
\be 
 S_E = -\int_0^\beta d\tau \int_V d^3x\; {\cal L}^E\, , 
\ee
$\beta=T^{-1}$ is the inverse temperature and ${\cal L}^E$ is the 
euclidean Lagrangian, obtained by analytically continuing Eq.~(\ref{l_qcd})
to imaginary time $\tau=it$. We observe that the temperature enters
through the periodicity of the euclidean path integral in the imaginary
time direction. Gauge fields and fermions obey periodic and anti-periodic
boundary conditions, respectively. The chemical potential couples to the 
conserved baryon density in the Lagrangian, 
\be 
 {\cal L}^E(\mu) =  {\cal L}^E(0) + \mu \bar{q}_f\gamma_0 q_f\, . 
\ee
Following Wilson's original suggestion, the Lagrangian is discretized
on an $N_\tau\times N_\sigma^3$ space-time lattice with lattice spacings
$a_\tau$ and $a_\sigma$. In many calculations $a_\sigma=a_\tau=a$, but this
condition is not necessary. In finite temperature calculations we 
choose $\beta<L$ with $\beta=N_\tau a_\tau$ and $L=N_\sigma a_\sigma$, 
where $V=L^3$ is the volume. Eq.~(\ref{Z_latt}) provides a lattice 
definition of the partition function $Z=\exp(-\beta(H-\mu N))$. 
Thermodynamic quantities are determined by taking suitable derivatives, 
for example
\begin{eqnarray}
{\cal E} &=& -\frac{1}{V} \left.\frac{\partial\log Z}{\partial\beta}
  \right|_{\beta\mu}\, , \\
 n_B  &=& \;\frac{1}{\beta V} \left.\frac{\partial\log Z}{\partial\mu}
  \right|_{\beta}\, .
\end{eqnarray}
The gauge fields are discretized on links and the fermion fields reside 
on sites. This allows the gauge invariance of QCD to be maintained exactly, 
even on a finite lattice, but Lorentz invariance is only restored in the 
continuum limit. We note that because of classical scale invariance the 
massless QCD action is independent of $a$. The continuum limit is taken 
by adjusting the bare coupling at the scale of the lattice spacing 
according to asymptotic freedom, see Eq.~(\ref{g_1l}). In practical
calculations the lattice spacing is not quite small enough to ensure
the accuracy of this method, and more sophisticated scale setting 
procedures are used \cite{Fodor:2009ax,Ding:2015ona}.

 Formally, the integration over the fermion fields can be performed
exactly, resulting in the determinant of the Dirac operator $\det(M
(A_\mu,\mu))$. Several methods exist for discretizing the Dirac operator 
$M$, and for sampling the determinant. Different discretization
schemes differ in the degree to which chiral symmetry is maintained
on a finite lattice. The original formulation due to Wilson 
\cite{Wilson:1974sk} preserves no chiral symmetry, the staggered Fermion 
scheme \cite{Kogut:1974ag} maintains a subset of the full chiral symmetry, 
while the domain wall \cite{Kaplan:1992bt} and overlap methods
\cite{Neuberger:1997fp} aim to preserve the full chiral symmetry on 
a discrete lattice. 

%%%%%%%%%%%%%%%%%%%%%%%%%%%%%%%%%%%%%%%%%%%%%%%%%%%%%%%%%%%%%%%%%%%%%%%%%
\begin{figure}[t]
\bc\includegraphics[width=0.8\textwidth]{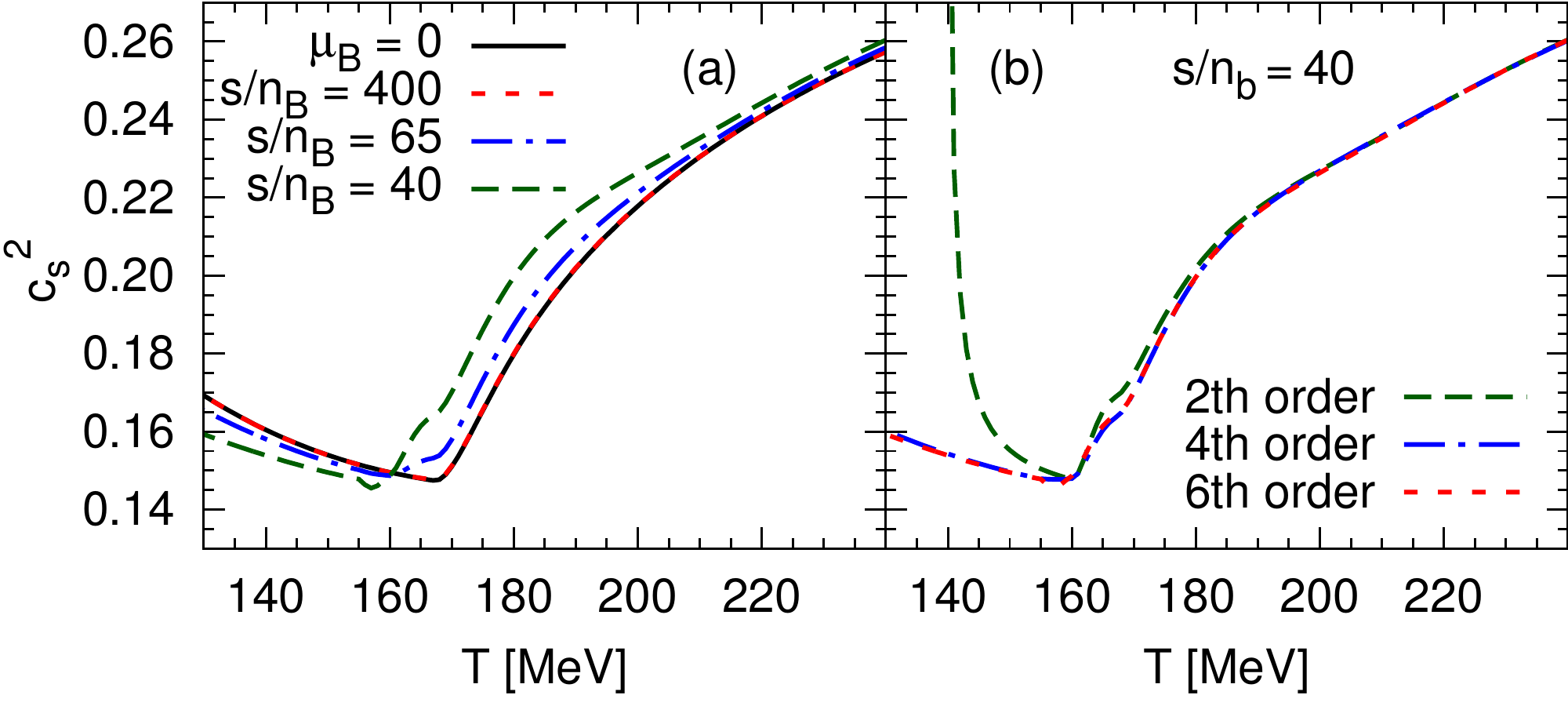}\ec
\caption{\label{fig_cs}
The square of the speed of sound, $c_s^2$, as a function of
temperature on various isentropic curves with constant entropy per
baryon. Fig. (a) shows a lattice calculations using a Taylor expansion
to explore the regime of finite baryon density finite baryon density 
\cite{Huovinen:2014woa}. Fig. (b) illustrates the convergence of the 
method in the case $s/n_B = 40$.}
\end{figure}
%%%%%%%%%%%%%%%%%%%%%%%%%%%%%%%%%%%%%%%%%%%%%%%%%%%%%%%%%%%%%%%%%%%%%%%%%

 A second issue with fermions is that for $\mu\neq 0$ the fermion 
determinant is no longer real, so that standard importance sampling methods
fail. This is the ``sign'' problem already mentioned in Section~\ref{sec_EOS}.
There are many attempts to find direct solutions to the sign problem, 
but at this time the only regime in which controlled calculations
are feasible is the regime of small $\mu$ and high $T$. In this region
the partition function can be expanded in a Taylor series in $\mu/T$. 
The corresponding expansion coefficients are generalized susceptibilities
that can be determined from lattice simulations at zero chemical
potential. The susceptibilities not only determine the equation of state
at finite baryon density, but also control fluctuations of conserved
charges as explained in Section~\ref{sec:fluct}. 

 As examples of lattice results that are central to the analysis
of heavy ion collisions we show a calculation of equation of state
at $\mu=0$ in Fig.~(\ref{fig_eos} \cite{Bazavov:2014pvz}, and 
the speed of sound for several values of $\mu$  in Fig.~(\ref{fig_cs})
\cite{Huovinen:2014woa,Borsanyi:2010cj,Bazavov:2014pvz}. We observe
that the energy density, pressure, and entropy density, normalized
to suitable powers of $T$, vary rapidly in the critical regime defined
by fluctuations of the chiral order parameter. We also note that the
agreement with the hadron resonance gas model is very good up to 
temperature $T\lsim 180$ MeV. Finally, we observe that the rise
of ${\cal E},P$ and $s$ towards the perturbative limit is quite 
slow. In particular, even at temperatures as large as $T=400$ MeV,
the pressure remains about 25\% below the Stefan-Boltzmann limit. 

 The speed of sound at $\mu=0$ was determined from direct lattice 
simulations of $P(T)$. The results at non-zero baryon density were 
obtained using the Taylor expansion method. We observe that $c_s^2$ 
indeed shows the expected behavior, a soft point near the phase 
transitions where $c_s^2\simeq 0.15$, followed by a gradual rise 
towards $c_s^2=1/3$. In the regime accessible with Taylor expansions 
the dependence of $c_s^2$ on $n_B$ is not very pronounced. The main 
effect is a slight reduction in the temperature of the softest point, 
corresponding to the curvature of the phase transition line in the 
$T-\mu$ plane. 

%%%%%%%%%%%%%%%%%%%%%%%%%%%%%%%%%%%%%%%%%%%%%%%%%%%%%%%%%%%%%%%%%%%%
\subsection{Lattice QCD: Frontiers and challenges}
\label{sec_lQCD_front}
%%%%%%%%%%%%%%%%%%%%%%%%%%%%%%%%%%%%%%%%%%%%%%%%%%%%%%%%%%%%%%%%%%%%

  The equation of state in three flavor QCD with physical quark 
masses and vanishing baryon density is now fairly well established,
but many new challenges for lattice QCD have emerged. One challenge
is clearly to extend calculations of the EOS into the regime of finite 
baryon density and to locate or exclude the presence of a critical 
point. In addition to methods that are restricted to the regime $\mu
\lsim \pi T$, a number of proposals to explore QCD at high baryon 
density are being pursued. This includes new approaches, like integration 
over Lefshetz thimbles \cite{Cristoforetti:2012su,Aarts:2014nxa},
as well as novel variants of old approaches, like the complex Langevin
method \cite{Aarts:2009uq,Sexty:2013ica}, or the use of dual variables 
\cite{Kloiber:2013rba}. The ultimate promise of these methods is still 
unclear, but the central importance of the sign problem to computational 
physics continues to attract new ideas. 

 Some progress has also been achieved in a different area, the 
calculation of near-equilibrium real time properties of the 
plasma using response functions \cite{Meyer:2011gj}. The prototypical
example is the calculation of the shear viscosity using the retarded 
correlation function of the stress tensor $T_{xy}$, 
\be
\label{G_ret}
G^{xy,xy}_R(\omega,\vec{k}) = -i\int dt\int d^3x\, 
  e^{i(\omega t-\vec{k}\cdot\vec{x})} \Theta(t)
  \langle \left[ T^{xy}(\vec{x},t),T^{xy}(0,0)\right]\rangle\, , 
\ee
The associated spectral function is defined by $\rho(\omega,\vec{k})
=-\,{\rm Im}\,G_R(\omega,\vec{k})$. As we will explain in more 
detail in Section~\ref{sec_npfl} the imaginary part of the retarded 
correlator is a measure of dissipation. Matching the correlation 
function from linear response theory to the hydrodynamic correlator 
gives the Kubo relation
\be 
\label{eta_kubo}
\eta = \lim_{\omega\to 0} \lim_{k\to 0} 
   \frac{\rho^{xy,xy}(\omega,\vec{k})}{\omega}\, . 
\ee
The formula for the bulk viscosity involves the trace of the 
energy momentum tensor
\be 
\label{zeta_kubo}
\zeta = \frac{1}{9}\lim_{\omega\to 0} \lim_{k\to 0} 
   \frac{\rho^{ii,jj}(\omega,\vec{k})}{\omega}\, ,
\ee
and analogous results can be derived for the thermal conductivity 
and diffusion constants. 

 The spectral function contains information about the physical
excitations that carry the response. Lattice calculations are 
based on the relation between the spectral function and the Matsubara 
(imaginary energy) correlation function 
\be 
\label{G_E_w}
G_E(i\omega_n)= \int \frac{d\omega}{2\pi} \frac{\rho(\omega)}
 {\omega-i\omega_n}\, , 
\ee
where $\omega_n=2\pi nT$ is the Matsubara frequency. The imaginary
time correlation function is given by 
\be 
\label{G_E_tau}
G_E(\tau)= \int \frac{d\omega}{2\pi} K(\omega,\tau) \rho(\omega) \, , 
\ee
where the kernel $K(\omega,\tau)$ is defined by
\be 
\label{Ker}
K(\omega,\tau) = \frac{\cosh[\omega(\tau-1/(2T))]}{\sinh[\omega/(2T)]}
 =  \left[1+n_B(\omega)\right] e^{-\omega\tau}
      + n_B(\omega)e^{\omega\tau}\, ,
\ee
and $n_B(\omega)$ is the Bose distribution function. The imaginary time 
correlation function (\ref{G_E_tau}) was determined in a number of lattice 
studies \cite{Karsch:1986cq,Meyer:2007ic,Meyer:2007dy,Sakai:2007cm}. The 
basic approach for extracting transport properties is to compute $G_E(
\tau)$ numerically, invert the integral transform in Eq.~(\ref{G_E_tau}) 
to obtain $\rho(\omega)$, and finally obtain the transport coefficient 
from the limit $\omega\to 0$ of the spectral function. The difficulty 
is that $G_E(\tau)$ is typically only known on a small number of lattice
sites, and that the imaginary time correlator is not very sensitive to 
the slope of the spectral function at low energy. Many recent calculations 
make use of the maximum entropy method to obtain numerically stable spectral 
functions and reliable error estimates \cite{Aarts:2007wj,Aarts:2007va}. 
It was also observed that one can minimize the contribution from continuum 
states to the imaginary time Green function by studying the correlators 
of conserved charges, energy and momentum density, at non-zero spatial 
momentum \cite{Aarts:2006wt,Meyer:2008gt}. In physical terms this means 
that one can extract the viscosity from the sound pole rather than the 
shear pole in the retarded correlator. 

 Pioneering calculations of the shear viscosity were performed by 
Karsch and Wyld \cite{Karsch:1986cq}. More recently, the problem
of determining shear and and bulk viscosity near $T_c$ was revisited
by Meyer \cite{Meyer:2007ic,Meyer:2008gt}. He finds $\eta/s=0.102(56)$ 
and $\zeta/s=0.065(17)$ at $T=1.24T_c$. The shear viscosity is only 
weakly temperature dependent, but bulk viscosity grows strongly near 
$T_c$, and becomes very small at large temperature. The value of $\eta/s$ 
is consistent with the experimental determinations discussed in
Section~\ref{sec_flow}, 
and the proposed holographic bound $\eta/s=1/(4\pi)$
\cite{Kovtun:2004de}. An interesting aspect of these calculations is 
that it is easier to numerically determine a small shear viscosity as
compared to a large one. In weak coupling $\eta/s$ is large (see
Eq.~\ref{eta_s_w}), but this result is encoded in a very narrow 
peak in the spectral function, which is hard to resolve numerically. A
small shear viscosity, on the other hand, corresponds to a very
smooth spectral function, which is much easier to reconstruct. This
implies that the reported lattice determinations of $\eta/s$ near
$T_c$ may well be reliable, but that it is also difficult to demonstrate
the accuracy of the method by studying the weak coupling limit 
$T\gg T_c$. Recent calculations of spectral functions have focused on 
other observables, in particular the heavy quark diffusion constant, 
the dilepton spectral function, and the spectrum of charmonia with 
different quantum numbers, see \cite{Ding:2015ona} for an overview.

%%%%%%%%%%%%%%%%%%%%%%%%%%%%%%%%%%%%%%%%%%%%%%%%%%%%%%%%%%%%%%%%%%%%%%%%%%%%%
\begin{figure}[t]
\centering\includegraphics[width=.6\textwidth]{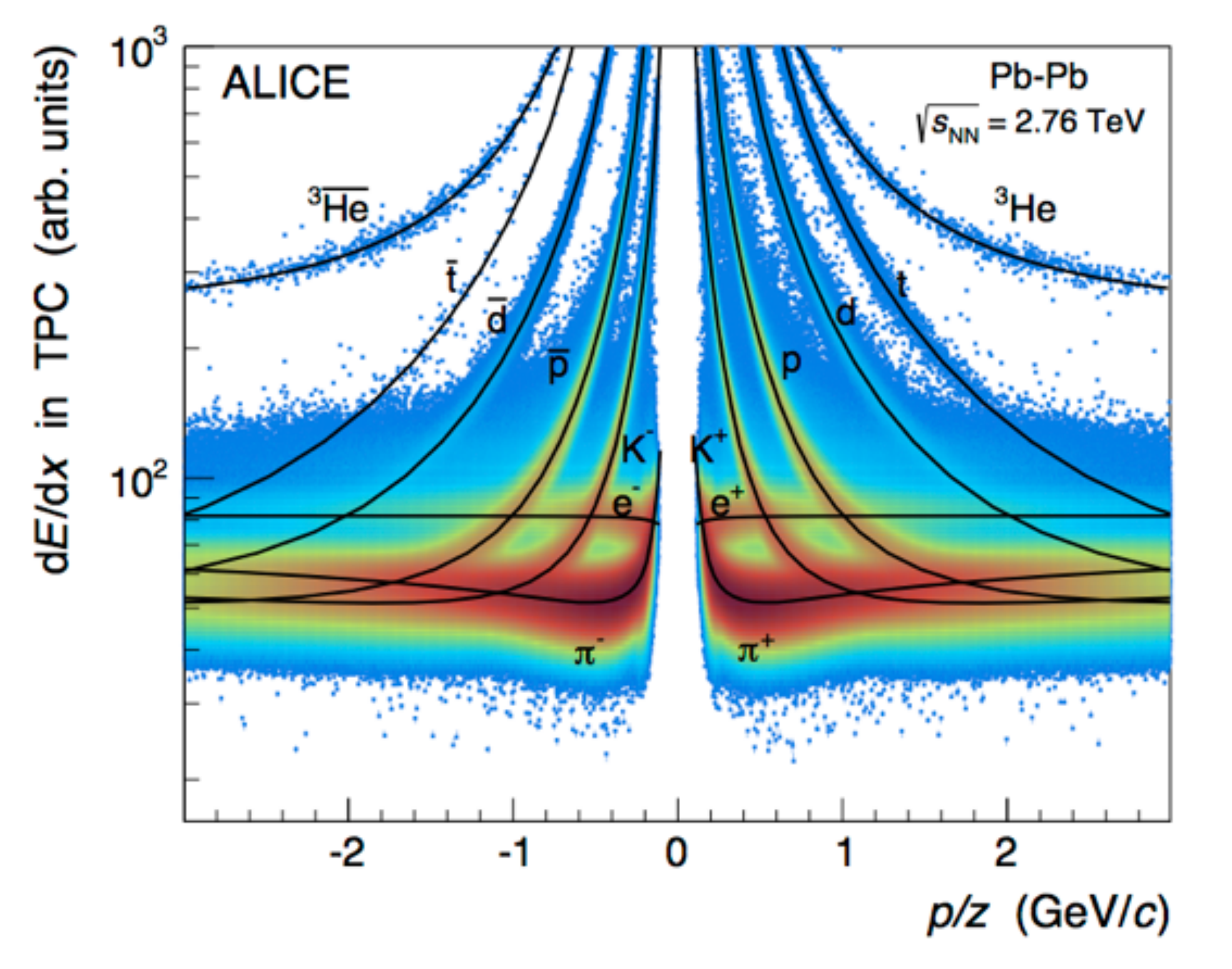}
\caption{Particle identification in the ALICE experiment via the
  specific energy loss and momentum measurement in the ALICE TPC and
  inner tracking system (figure taken from \cite{Adam:2015vda}).}
\label{fig:TPCpid}
\end{figure}
%%%%%%%%%%%%%%%%%%%%%%%%%%%%%%%%%%%%%%%%%%%%%%%%%%%%%%%%%%%%%%%%%%%%%%%%%%%%%

%%%%%%%%%%%%%%%%%%%%%%%%%%%%%%%%%%%%%%%%%%%%%%%%%%%%%%%%%%%%%%%%%%%%%%%%%%%%%
\section{Hadrons with (u,d,s) quarks}
\label{sec:PJ_thermal}
%%%%%%%%%%%%%%%%%%%%%%%%%%%%%%%%%%%%%%%%%%%%%%%%%%%%%%%%%%%%%%%%%%%%%%%%%%%%%

\begin{figure}[t]
\centering\includegraphics[width=.8\textwidth]{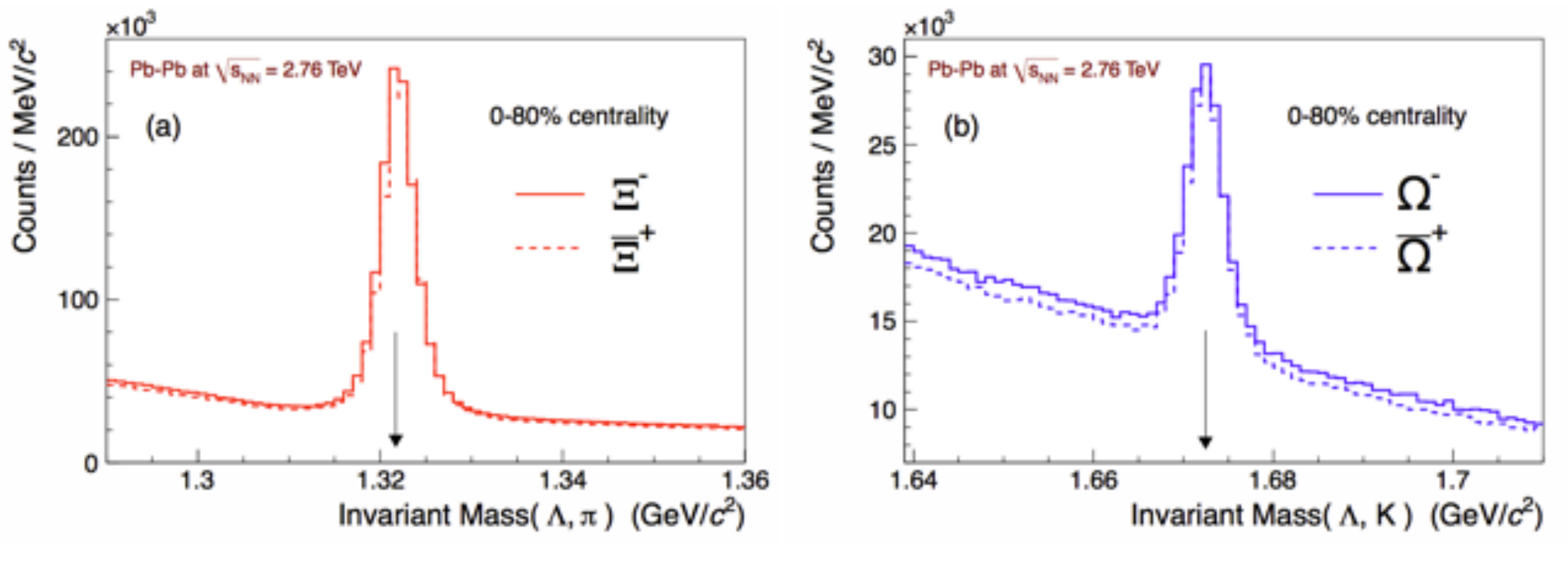}
\caption{Reconstruction of multi-strange baryons produced in central
  Pb--Pb collisions with ALICE at the LHC via the invariant mass of
  their weak decay products (figure taken from \cite{ABELEV:2013zaa}).}
\label{fig:masses}
\end{figure}

Hadron production in ultra-relativistic nucleus-nucleus collisions has
been studied now for nearly 30 years. The first successful description
of a comprehensive set of data within the framework of statistical
hadronization was achieved in 1994 for data from the Brookhaven AGS
and Si+Au(Pb) collisions \cite{BraunMunzinger:1994xr}. The same approach 
was applied to more data soon thereafter. In this context we also point 
to an early article by Gerry Brown and co-workers \cite{Brown:1991en}, 
where a thermal approach was successfully used to describe the
influence of resonance decays on pion spectra.  Excellent surveys of
the data from the pre-LHC era, together with an analysis and interpretation 
in the framework of the statistical hadronization model, can be found in 
\cite{Andronic:2005yp,Becattini:2005xt,Andronic:2008gu}.  The
first three years of data taking at the CERN Large Hadron Collider
(LHC) have brought a wealth of new, high precision data on hadron
production in ultra-relativistic nuclear collisions at the TeV energy
scale. As an example of the data quality we show, in Figs.~\ref{fig:TPCpid} 
and \ref{fig:masses}, the particle identification with the ALICE TPC and 
the reconstructed mass distributions for strange baryons in Pb--Pb 
collisions with ALICE at the LHC.

%%%%%%%%%%%%%%%%%%%%%%%%%%%%%%%%%%%%%%%%%%%%%%%%%%%%%%%%%%%%%%%%%%%%%%%%%%%%%
%%%%%%%%%%%%%%%%%%%%%%%%%%%%%%%%%%%%%%%%%%%%%%%%%%%%%%%%%%%%%%%%%%%%%%%%%%%%%

\begin{figure}[t]
\centering\includegraphics[width=.7\textwidth]{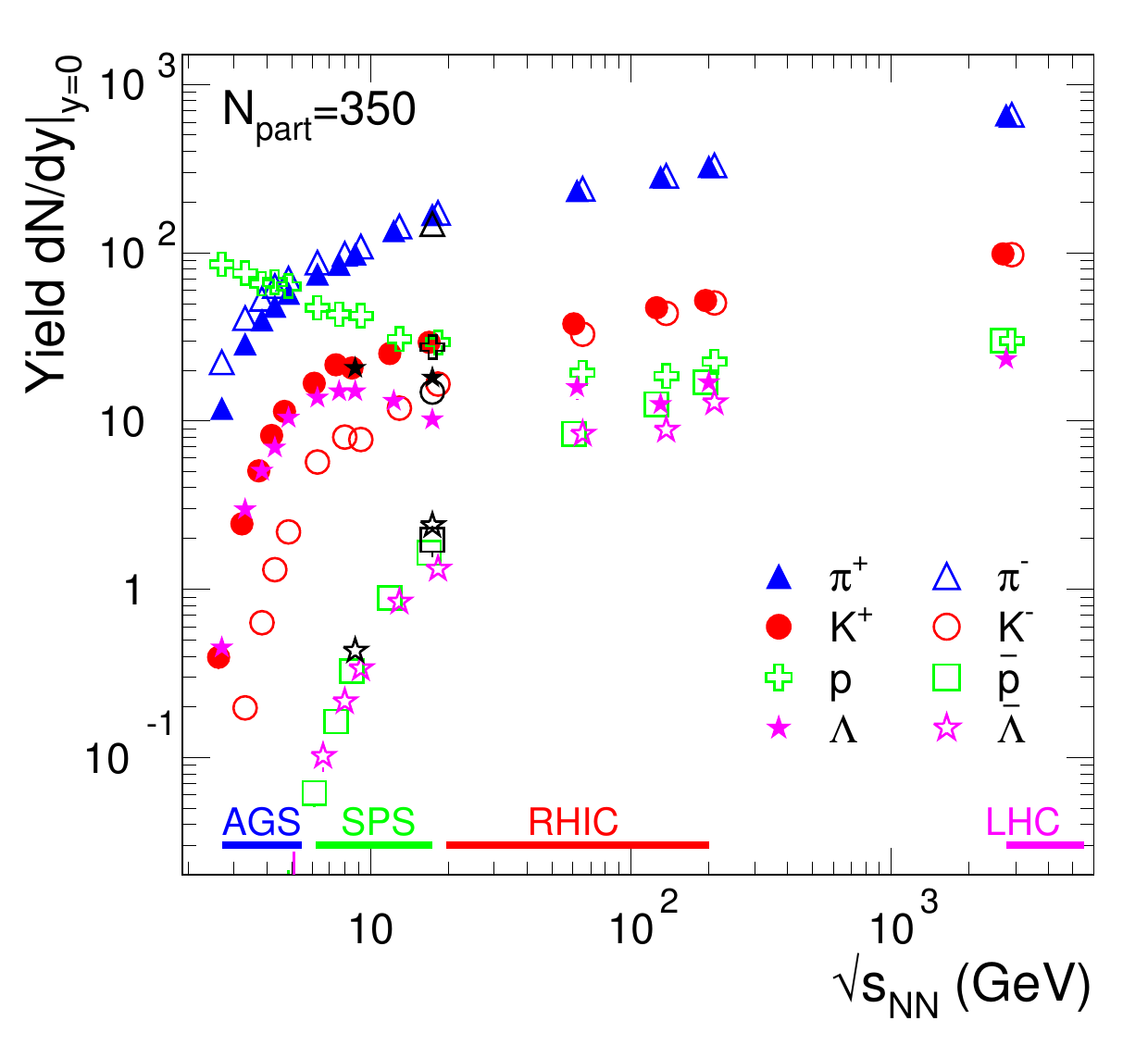}
\caption{Energy dependence of the rapidity density for identified
  hadrons produced in central nucleus-nucleus collisions. Figure taken
  from \cite{Andronic:2005yp,Andronic:2014zha}. The colliding systems
  are either Pb--Pb or Au--Au and central collisions are selected by
  the requirement of at least 350 participating nucleons in each
  collision.}
\label{fig:rap_den}
\end{figure}

A compilation of the available data on hadron production at
mid-rapidity in central nucleus-nucleus collisions is presented in
Fig.~\ref{fig:rap_den}. This comprehensive plot contains the work of
essentially all large collaborations in the field of
ultra-relativistic heavy ion physics, performed over a period of
approximately 30 years. Some regularities are obvious: At lower
energies protons from the colliding nuclei dominate the yield at
mid-rapidity while all produced particles are strongly
suppressed. With increasing energy baryon pair production becomes more
and more dominant, and the charge and baryon number in the colliding
nuclei (fragmentation regions) becomes irrelevant at mid-rapidity.  As
a consequence, at LHC energies, the central fireball formed in the
collisions contains equal amounts of matter and anti-matter: Big Bang
matter produced in the laboratory\footnote{Big Bang matter contains,
  in addition to hadrons, also leptons (including neutrinos) and
  photons and, at very high temperatures, electro-weak bosons in
  equilibrium.}. It is this rich data sample on which all further
interpretations are based. Two of us (pbm and js) remember discussing
versions of this plot with Gerry Brown during the early years of RHIC
operation while trying to convince him (successfully, we believe) to
focus his mind again on the beautiful phenomenology of particle
production in ultra-relativistic nuclear collisions.

%%%%%%%%%%%%%%%%%%%%%%%%%%%%%%%%%%%%%%%%%%%%%%%%%%%%%%%%%%%%%%%%%%%%%%%%%%%%%
%%%%%%%%%%%%%%%%%%%%%%%%%%%%%%%%%%%%%%%%%%%%%%%%%%%%%%%%%%%%%%%%%%%%%%%%%%%%%

The data, of which some systematics is shown in
Fig.~\ref{fig:rap_den}, can be economically and concisely described
over the whole energy range in the framework of the statistical
hadronization model.  This thermal or statistical hadronization model
\cite{BraunMunzinger:2003zd} (we will use the term synonymously here)
describes a snapshot of the collision, namely the chemical freeze-out,
which is assumed to be driven by rapid changes in energy and entropy
density near the phase boundary \cite{BraunMunzinger:2003zz}. The
fireball formed in the collision is assumed to be in chemical
equilibrium when the dramatic changes in density near the phase
boundary lead to (nearly) simultaneous freeze-out of all hadrons at
the chemical freeze-out temperature $T$ and baryo-chemical potential
$\mu_b$. The energy dependence of $T$ and $\mu_b$ and of the rapidity
density of charged pions determine the thermal parameters $T$, $\mu_b$
and $V$ and, hence, the rapidity density of all hadron species. In
general, the precision of this description is on the order of
10\%. Due to the data sets available, the energy dependence of the
thermal parameters is measured at discrete energies and interpolated
in between, see below.

This approach provides a phenomenological link between the data and
the QCD phase diagram shown in Fig.~\ref{fig_qcd_phase}, a link surmised 
a long time ago \cite{Cabibbo:1975ig,Hagedorn:1984hz} but explored and 
discussed in quantitative detail only more recently
\cite{BraunMunzinger:1998cg,Stock:1999hm,BraunMunzinger:2001mh,
BraunMunzinger:2003zz,Andronic:2008gu,Andronic:2009gj,Floerchinger:2012xd}.
In this review we use the most recent data and the latest update of
the model as described in \cite{Stachel:2013zma}.

We note that, for the first time, the data obtained by the ALICE 
collaboration at the LHC are corrected in hardware for feed-down from
weakly decaying resonances via the use of the excellent ALICE inner tracking 
detector, see \cite{Abelev:2014ffa}. Consequently, for a description of ALICE 
data no feed-down correction is applied to the thermal model calculations. 
For analysis of the data from the RHIC, SPS and AGS accelerators, feeding
from weak decays needs to be taken into account. For details of this procedure 
see, e.g., \cite{Andronic:2005yp,Andronic:2008gu}. The uncertainties resulting 
from this correction lead to significantly increased uncertainties in the 
data from RHIC and the lower energy accelerators compared to those from 
the LHC.

%%%%%%%%%%%%%%%%%%%%%%%%%%%%%%%%%%%%%%%%%%%%%%%%%%%%%%%%%%%%%%%%%%%%%%%%%%%%%
\begin{figure}[t]
\centering\includegraphics[width=.7\textwidth]{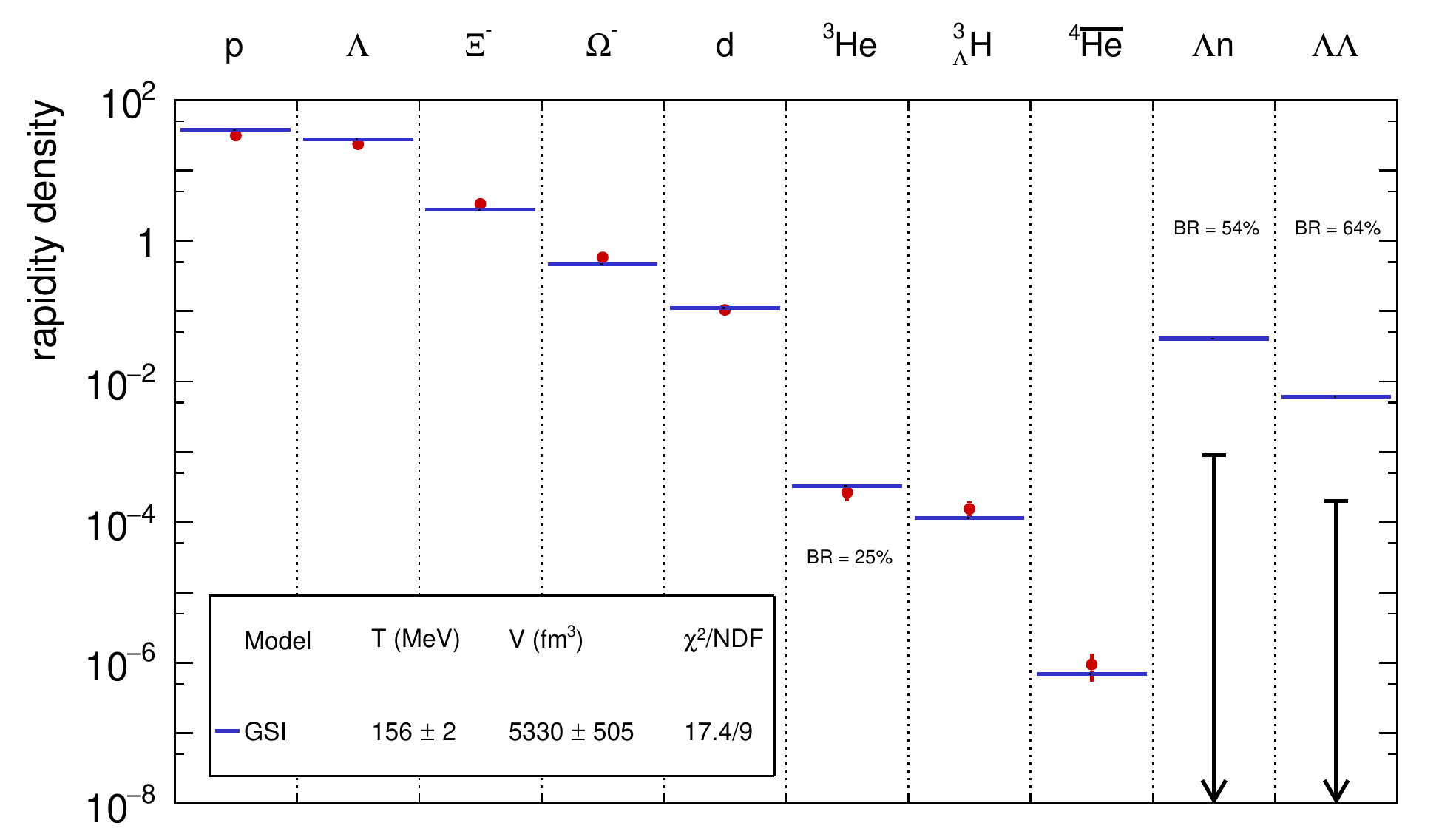}
\caption{Measured hadron abundances in comparison with thermal model
  calculations for the best fit to ALICE data \cite{Floris:2014pta}
  for central Pb--Pb collisions at the LHC.  Plotted are the ``total"
  thermal model yields, including all contributions from strong decays
  of high-mass resonances (for the $\Lambda$ hyperon, the contribution
  from the electromagnetic decay $\Sigma^0\rightarrow\Lambda\gamma$,
  which cannot be resolved experimentally, is also included) (figure
  taken from \cite{BraunMunzinger:2015zz}).}
\label{fig:fit}
\end{figure}
%%%%%%%%%%%%%%%%%%%%%%%%%%%%%%%%%%%%%%%%%%%%%%%%%%%%%%%%%%%%%%%%%%%%%%%%%%%%%

Good fits of the measurements are achieved with the thermal model
\cite{BraunMunzinger:2003zd} with 3 parameters: Temperature $T$,
baryochemical potential $\mu_B$, and volume $V$, as shown in
Fig.~\ref{fig:fit} for the fit of data at the LHC
\cite{Stachel:2013zma,Floris:2014pta}.  Remarkably, multiply-strange
hyperons and light nuclei and (hyper)nuclei are well described by the
model. At LHC energy, the baryochemical potential turns out be zero
within uncertainties, implying \cite{Andronic:2011yq} equal production
of matter and antimatter at the LHC \cite{Abelev:2012wca}.  Note that
also loosely bound systems such as the deuteron (with binding energy
$E_b$ = 2.23 MeV) and hypertriton (binding energy $E_b$ = 2.35 MeV,
$\Lambda$ separation energy $S_{\Lambda}$ = 0.13 MeV) are well
reproduced for a $T$ value of 156 MeV, i.e. $T\gg E_b \gg S_{\Lambda}$.
The rates for such loosely bound systems are indeed fixed near the
phase boundary, implying that the expansion after chemical freeze-out
is isentropic. It is then immaterial for the calculation whether 
the (hyper) nuclei are droplets of quark matter \cite{Chapline:1978kg} 
or are formed via nucleon (and hyperon) coalescence.

An interesting and to-date not fully appreciated (or understood)
outcome of these thermal model analyses is that the best fits are
obtained if all hadron masses are kept at their vacuum values
irrespective of the values of $T$ and $\mu_b$. As discussed below, the
chemical freeze-out temperature $T$ approaches, for center-of-mass
energies $\sqrt{s_{nn}} > 10$ GeV closely the values predicted from
lattice QCD calculations for the (pseudo-)critical temperature of the
quark-hadron transition, yet no deviation from the vacuum mass
scenario is seen.

%%%%%%%%%%%%%%%%%%%%%%%%%%%%%%%%%%%%%%%%%%%%%%%%%%%%%%%%%%%%%%%%%%%%%%%%%%%%%
\begin{figure}[t]
\centering\includegraphics[width=.8\textwidth]{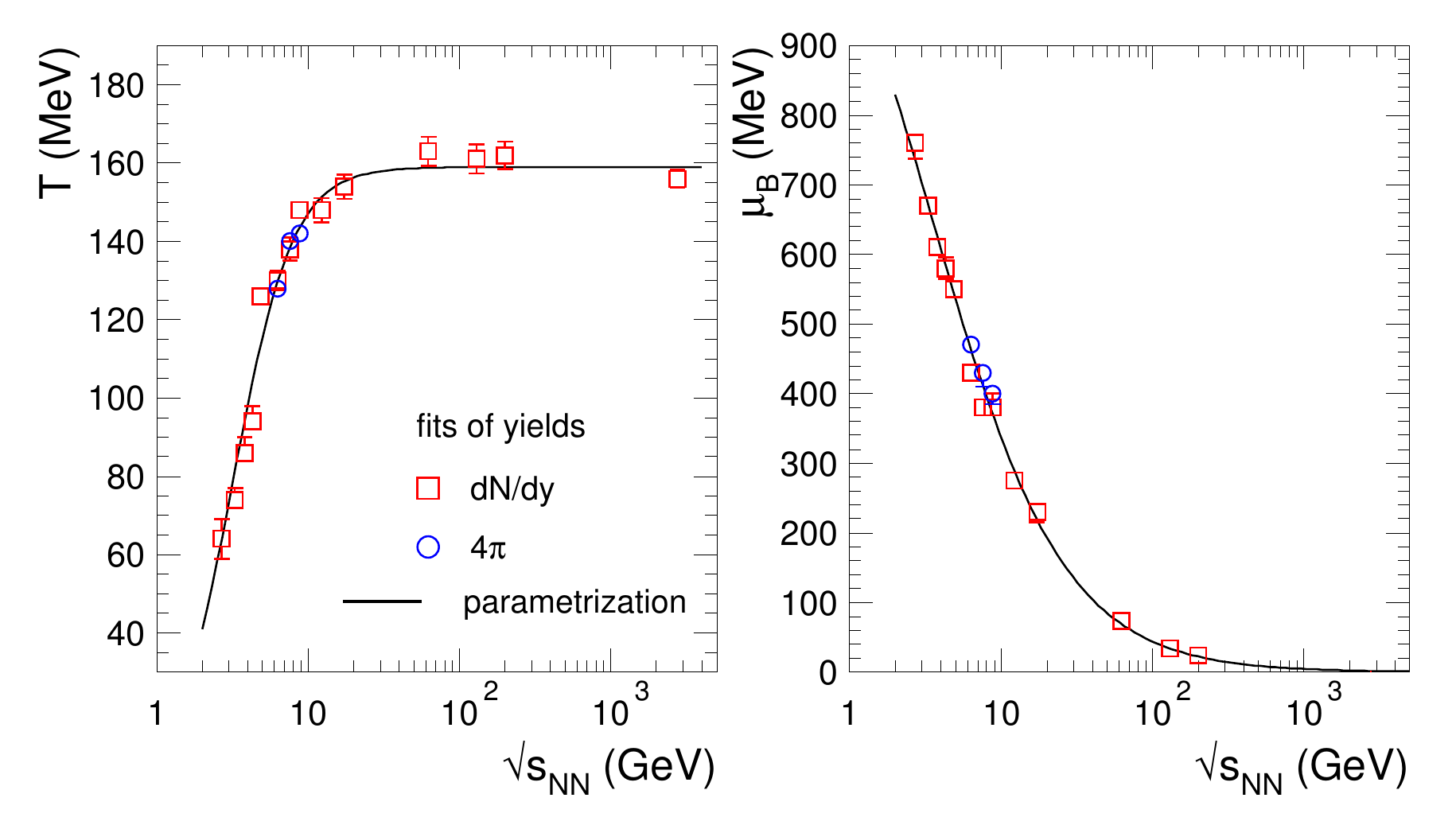}
\caption{Energy dependence of the thermal model parameters $T$ and
$\mu_b$ \cite{Andronic:2005yp} updated to include the most recent LHC 
results. For more details see text. Figure taken from
\cite{Andronic:2014zha}.}
\label{fig:energydep}
\end{figure}
%%%%%%%%%%%%%%%%%%%%%%%%%%%%%%%%%%%%%%%%%%%%%%%%%%%%%%%%%%%%%%%%%%%%%%%%%%%%%

\begin{figure}[t]
\centering\includegraphics[width=.65\textwidth,height=.62\textwidth]{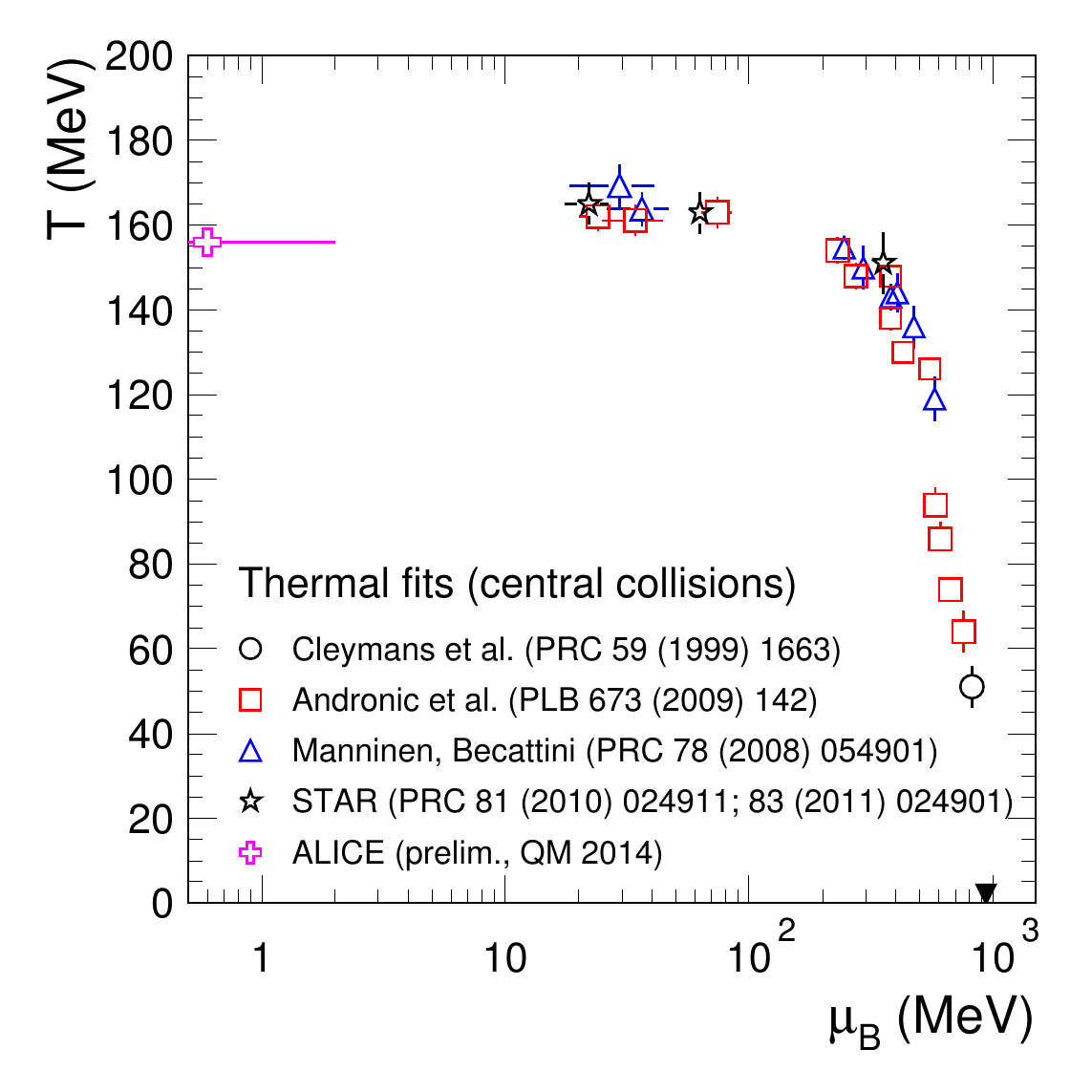}
\caption{The phase diagram of strongly interacting matter with the
  points representing the thermal fits of hadron yields at various
  collision energies
  \cite{Cleymans:1998yb,Andronic:2008gu,Manninen:2008mg,Abelev:2009bw,
    Aggarwal:2010pj,Floris:2014pta}.  For the LHC, $\mu_B$=0 is the
  value resulting from the fit. In the plot, a value of 0.6 MeV is
  used here for adaptation to the logarithmic scale.  The
  down-pointing triangle indicates the baryo-chemical potential of
  bound nuclei, i.e. $\mu_b = 931.4 MeV$.  }
\label{fig:t-mu}
\end{figure}

A remarkable outcome of these fits is that $T$ increases with
increasing energy and decreasing $\mu_B$ from about 50 MeV to about
160 MeV, where it exhibits, see Fig.~\ref{fig:energydep}, a saturation
for $\sqrt{s_{nn}} > 10$ GeV and $\mu_B\lesssim 300$ MeV. The
freeze-out points can be put together for an experimental version of
the phase diagram, see Fig.~\ref{fig:t-mu}. An
interpretation of the saturation of the freeze-out temperature was
put forward in \cite{Andronic:2009gj}, based on the conjecture that
the chemical freeze-out temperature is the hadronization temperature
\cite{Andronic:2008gu}, and therefore probes the QCD phase
boundary. The proposal is that the two regimes in the phase diagram,
see Fig.~\ref{fig_qcd_phase} and
Fig.~\ref{fig:t-mu}, that of approximately constant $T$ for small
$\mu_B$ values, and that of the strong increase in $T$ at low energy
and large $\mu_B$, may reflect the existence of a triple point in the
QCD phase diagram \cite{Andronic:2009gj}, see Section~\ref{sec_phase}.
Various criteria for the chemical freeze-out were proposed
\cite{Tawfik:2004ss,Cleymans:2005xv}. In our understanding it is
linked to the rapid drop in energy, entropy, and particle densities
near the pseudocritical temperature \cite{BraunMunzinger:2003zz},
leading first to equilibrium hadron population at or just below $T_C$
and then to rapid fall-out of equilibrium (i.e. freeze-out).

%%%%%%%%%%%%%%%%%%%%%%%%%%%%%%%%%%%%%%%%%%%%%%%%%%%%%%%%%%%%%%%%%%%%%%%%%%%%%
%%%%%%%%%%%%%%%%%%%%%%%%%%%%%%%%%%%%%%%%%%%%%%%%%%%%%%%%%%%%%%%%%%%%%%%%%%%%%

\begin{figure}[t]
\begin{tabular}{lr} \begin{minipage}{.49\textwidth}\hspace{-0.5cm}
\includegraphics[width=1.05\textwidth,height=1.1\textwidth]{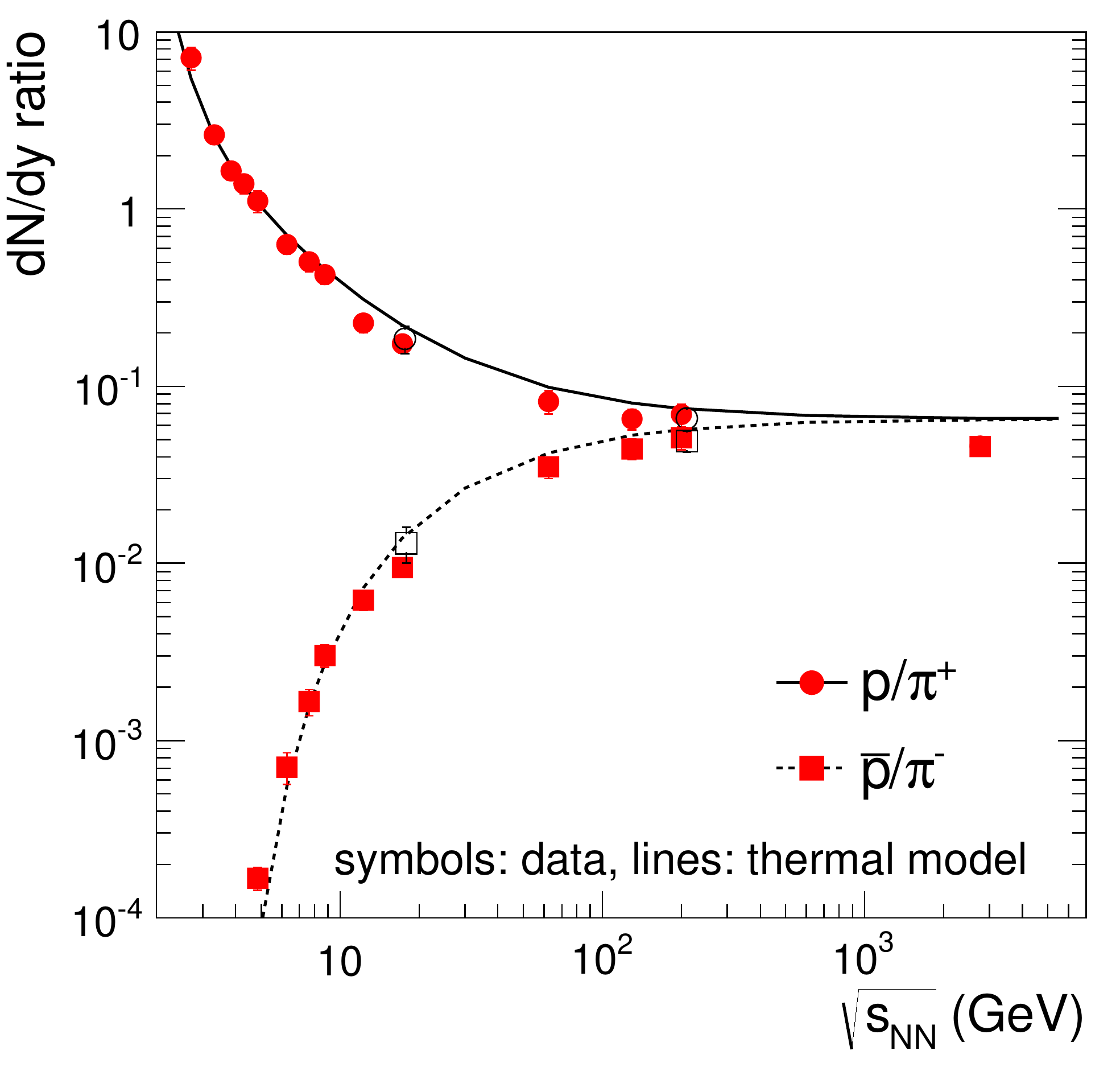}
\end{minipage} & 
\begin{minipage}{.49\textwidth}\hspace{-0.5cm}
\includegraphics[width=1.05\textwidth,height=1.1\textwidth]{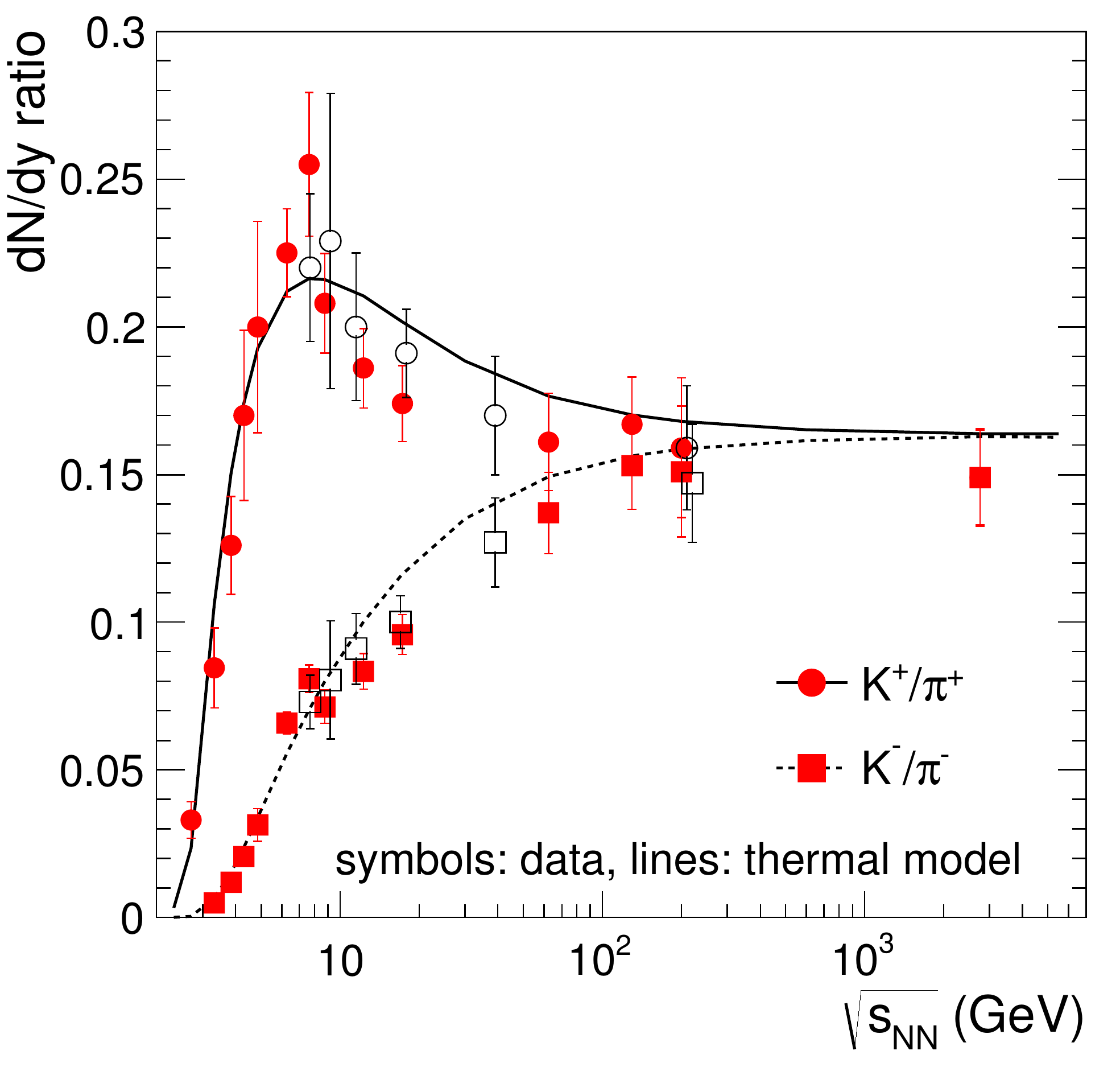}
\end{minipage}\end{tabular}
\caption{Collision energy dependence of ratios of yields of protons
  and antiprotons (left panel) and of kaons (right panel) to yields of
  pions (figure taken from \cite{Andronic:2014zha}. The symbols are
  data, the lines are thermal model calculations for energy-dependent
  parametrizations of $T$ and $\mu_B$ (as in
  Ref. \cite{Andronic:2008gu}).  }
\label{fig:ratios}
\end{figure}

In Fig.~\ref{fig:ratios} we demonstrate that the thermal model
approach can be successfully used to reproduce, over the full energy
range where data have been measured, the ratios of production yields
for various hadron species. The calculations are being performed with
parametrizations for the energy dependence of $T$ and $\mu_B$ as
obtained in Ref. \cite{Andronic:2008gu} and updated for
\cite{Stachel:2013zma}. The striking energy dependence of $T$ vs
$\sqrt{s}$ is shown in Fig.~\ref{fig:energydep} along with the very
smooth decrease with energy of $\mu_b$.   

%%%%%%%%%%%%%%%%%%%%%%%%%%%%%%%%%%%%%%%%%%%%%%%%%%%%%%%%%%%%%%%%%%%%%%%%%%%%%
%%%%%%%%%%%%%%%%%%%%%%%%%%%%%%%%%%%%%%%%%%%%%%%%%%%%%%%%%%%%%%%%%%%%%%%%%%%%%

The non-monotonic dependence on energy of the $K^+/\pi^+$ yield ratio
was originally proposed as a signature \cite{Gazdzicki:1998vd}, and
the measurement by the NA49 collaboration taken as evidence
\cite{Alt:2007aa} for the onset of deconfinement. However, the results
including the rather pronounced maximum near $\sqrt{s_{nn}}$= 8 GeV
are well understood within the above described framework of the
thermal model \cite{Andronic:2008gu}, as shown in
Fig.~\ref{fig:ratios} (right panel).  Based on this success, the
thermal model predictions provide a reliable guidance for experimental
searches for other exotic nuclei \cite{Andronic:2010qu}.

The phenomenological phase diagram obtained from these hadron yield
analyses within the statistical hadronization model is shown in
Fig.~\ref{fig:t-mu}. Each point corresponds to a fit of hadron yields
in central Au--Au or Pb--Pb collisions at a given collision energy. 
The agreement between the results from several independent analyses 
\cite{Andronic:2008gu,Manninen:2008mg,Abelev:2009bw,Aggarwal:2010pj} 
is remarkable.  In some cases 
\cite{Manninen:2008mg,Abelev:2009bw,Aggarwal:2010pj} an additional
ad-hoc fit parameter, the 'strangeness suppression' factor $\gamma_s$,
is used to search for a departure from equilibrium of hadrons
containing strange quarks.  Values of $\gamma_s$ (slightly) below
unity are found although no statistically significant improvement in
the fit is obtained, especially when the fit is restricted to data at
mid-rapidity.  An approach with more non-equilibrium parameters
\cite{Letessier:2005qe,Petran:2013qla} also does not lead to
significant improvements but results in a non-monotonic energy
dependence and generally decreased values for $T$. Fits considering a
spread in $T$ and $\mu_B$ were also performed \cite{Dumitru:2005hr}
and are currently again under consideration because of the
cross-over nature of the transition.

\begin{figure}[t]
\centering\includegraphics[width=.55\textwidth]{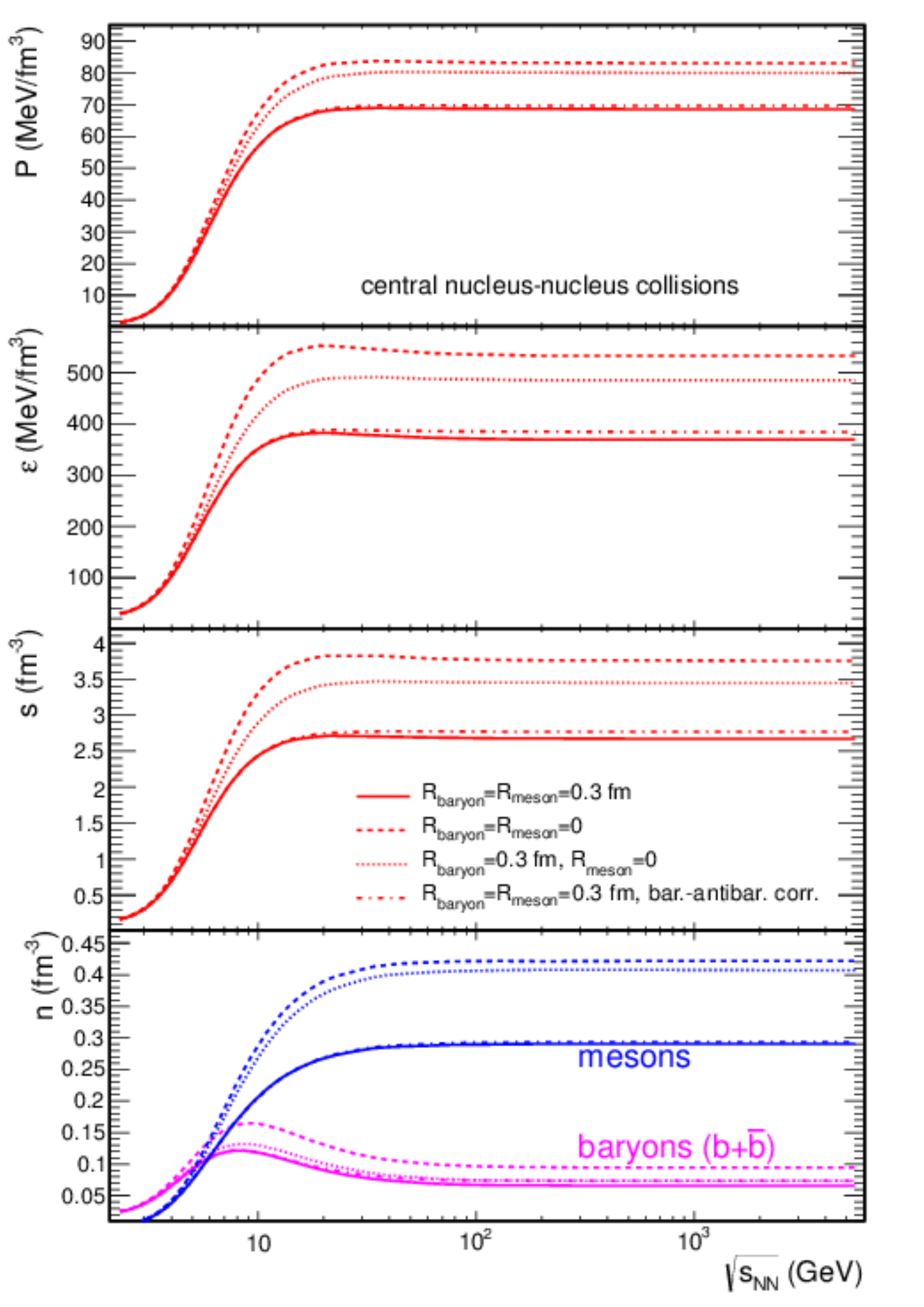}
%%% VK: changed width to .55 from .65. Otherwise this and all
%%% subsequent figs will wind up at the end of the document
\caption{Energy dependence of thermodynamic variables resulting from
  the thermal model analysis. The various curves correspond to
  different treatment of the excluded volume correction, which is used
  to model repulsive interactions in the hadron resonance gas. Figure
  taken from \cite{Andronic:2012ut}.}
\label{fig:therm_parameters}
\end{figure}

The thermal model inherently provides information on the underlying
thermodynamic quantities characterizing the state of the fireball
under consideration. While the absolute yields are insensitive to the
question whether or not repulsive interactions are considered (to
first order they are renormalized via the volume) quantities like
energy density, pressure, entropy density and particle density
significantly depend on the choice of interaction. Here we follow the
approach taken in \cite{Andronic:2012ut}. There, the repulsive part of
the interactions is taken into account via excluded volumes. Each
meson or baryon 'excludes' a spherical volume with radius $R_{meson}$
or $R_{baryon}$. The corresponding interaction correction is then
obtained following the procedure of \cite{Rischke:1991ke}. The 
derived thermodynamic quantities are shown as a function of energy in
Fig.~\ref{fig:therm_parameters} for various values of the excluded
volume radii. At LHC energy, corresponding to the ``limiting
temperature" $T_{lim}=159$ MeV, the following values are obtained for
$R_{meson} = R_{baryon} = 0.3$ fm: pressure $P\simeq 60$ MeV/fm$^3$,
energy density $\varepsilon\simeq 330$ MeV/fm$^3$, entropy density
$s=2.4$ fm$^{-3}$, density of mesons $n_m=0.26$ fm$^{-3}$, total
baryon density (particles plus antiparticles) $n_{B+\bar{B}}=0.06$ fm$^{-3}$ .

%%%%%%%%%%%%%%%%%%%%%%%%%%%%%%%%%%%%%%%%%%%%%%%%%%%%%%%%%%%%%%%%%%%%%%%%%%%%%
%%%%%%%%%%%%%%%%%%%%%%%%%%%%%%%%%%%%%%%%%%%%%%%%%%%%%%%%%%%%%%%%%%%%%%%%%%%%%

While the fit quality reported in Fig.~\ref{fig:fit} is impressive
indeed, we would like to point out that there is currently a 2.7
$\sigma$ excess, not visible on the log plot in the figure, of the
calculated thermal model yields over the data for protons and
antiprotons. Given the overall excellent agreement at LHC energy, this
discrepancy led to a number of theoretical investigations, mostly
stressing the possible importance of the role of interactions after
chemical freeze-out \cite{Becattini:2014hla} (in the hybrid model of
Ref. \cite{Becattini:2014hla} higher $T$ values are obtained for the
LHC case); also the effect of a possible extension of the hadronic
mass spectrum beyond the currently established hadron states
\cite{Stachel:2013zma,Bazavov:2014xya,Bazavov:2014yba} was
discussed. To date, there is no consistent explanation of the apparent
'LHC proton puzzle', especially also considering the excellent
description of light nuclei and antinuclei as well as the situation in
the multi-strange baryon sector, where final state annihilation,
should it exist, would also be visible. The connection to fit results
for data from e$^+$e$^-$ (see Ref. \cite{Andronic:2008ev} and
references therein) and in elementary hadronic collisions
\cite{Becattini:2010sk} remains also to be better understood.

The results reported up to this point only deal with the mean number of
hadrons produced for a given collision centrality, i.e. with the first
moment of the thermal model distributions. It is clearly interesting
to extend these studies into the measurement of higher moments and to
investigate whether these can also be described within the framework
of the grand-canonical hadron resonance gas, i.e. the thermal model
described here. This will be discussed in detail in Section~\ref{fluct:sec:freeze}.

%%%%%%%%%%%%%%%%%%%%%%%%%%%%%%%%%%%%%%%%%%%%%%%%%%%%%%%%%%%%%%%%%%%%
\section{Fluid dynamics and collective flow}
\label{sec_hydro}
%%%%%%%%%%%%%%%%%%%%%%%%%%%%%%%%%%%%%%%%%%%%%%%%%%%%%%%%%%%%%%%%%%%%

%%%%%%%%%%%%%%%%%%%%%%%%%%%%%%%%%%%%%%%%%%%%%%%%%%%%%%%%%%%%%%%%%%%%
\subsection{Introduction}
\label{sec_fl_intro}
%%%%%%%%%%%%%%%%%%%%%%%%%%%%%%%%%%%%%%%%%%%%%%%%%%%%%%%%%%%%%%%%%%%%

 The non-equilibrium evolution of a heavy ion collision, beginning
from the initial production of partons or fields, the possible 
formation of a quark-gluon plasma and its subsequent expansion, to
a hadronization stage and the final decoupling, is clearly a very
complicated process. However, the situation simplifies greatly if
the interaction in the initial state is strong, and rapid local 
thermalization occurs. In this case microscopic details of the state 
of the system are not relevant, and the dynamics is completely 
determined by the distribution of energy, momentum, and baryon density. 
The time evolution of these quantities is governed by the corresponding 
conservation laws, as encoded in the equations of fluid dynamics. The 
reason that conserved charges are special is that generic observables 
can relax locally, on a microscopic time scale, while conserved 
quantities can only relax by collective motion or diffusion, which takes 
place on a macroscopic time scale.  Fluid dynamics depends on a small 
number of equilibrium and near-equilibrium parameters, most notably 
the equation of state and the transport coefficients of the fluid, see 
\cite{Heinz:2009xj,Teaney:2009qa,Romatschke:2009im,Schafer:2009dj,Schaefer:2014awa}
for recent reviews. The collective expansion of the hot and dense matter 
created in the collision therefore constrains these parameters. 

 In the late stages of the collision matter hadronizes, interactions
become weak, and the fluid dynamic description must break down. 
However, because the conversion of a fluid to weakly interacting
particles is a local process, the collective flow pattern established
in the fluid dynamic stage is robust. Furthermore, if the breakdown
of fluid dynamics occurs in a regime in which kinetic theory based
on hadronic quasi-particles is valid, then a more microscopic 
description is possible and the combination of fluid dynamics 
and kinetic theory can be used to predict particle spectra, and 
the flow of identified particles.

%%%%%%%%%%%%%%%%%%%%%%%%%%%%%%%%%%%%%%%%%%%%%%%%%%%%%%%%%%%%%%%%%%%%
\subsection{Relativistic fluid dynamics}
\label{sec_ns}
%%%%%%%%%%%%%%%%%%%%%%%%%%%%%%%%%%%%%%%%%%%%%%%%%%%%%%%%%%%%%%%%%%%%

 In the ultra-relativistic regime we can neglect the baryon density
and the conserved densities are the energy density ${\cal E}$ and
the momentum density $\vec{\pi}$. These two hydrodynamic variables
can be embedded in the energy momentum tensor $T_{\mu\nu}$, where 
$T^{00}={\cal E}$ and $T^{0i}=\pi^i$. Conservation of energy 
and momentum is expressed by the equation
\be 
\label{rel_euler}
 \partial^\mu T_{\mu\nu}=0\, . 
\ee
In order to solve this equation we have to determine the remaining
components of $T_{\mu\nu}$. The basic idea is to make use of the 
fact that hydrodynamics is a macroscopic, coarse-grained, description
of the underlying microscopic dynamics. This implies that we can
systematically expand the currents in gradients of the fluid dynamic 
variables. The leading order expression contains no gradients, and 
is completely fixed by symmetries. The basic observation is that in 
the fluid rest frame, $T_{\mu\nu}={\rm diag}({\cal E},P,P,P)$. In 
a general frame
\be 
\label{T_id}
T^{(0)}_{\mu\nu} = ({\cal E}+P)u_\mu u_\nu + P g_{\mu\nu}\, ,
\ee
where we have introduced the fluid velocity $u^\mu$, and we use the 
convention $u^2=-1$. The fluid dynamic equations $\partial^\mu T_{\mu\nu}
=0$ close once we provide an equation of state, $P=P({\cal E})$. 

 Equation (\ref{rel_euler}) is the relativistic Euler equation. We can 
split this equation into longitudinal and transverse parts using the 
projectors
\be
\label{Delta_uu}
\Delta_{\mu\nu}^{||}=-u_\mu u_\nu\, ,\hspace{0.5cm}
\Delta_{\mu\nu} = g_{\mu\nu}+u_\mu u_\nu\, . 
\ee
The longitudinal and transverse projections of $\partial^\mu T_{\mu\nu}=0$
can be viewed as the equation of energy (or entropy) conservation, and
the relativistic Euler equation. We get $\partial^\mu\left(su_\mu\right)
=0$ and 
\be
\label{rel_euler_2}
Du_\mu = -\frac{1}{{\cal E}+P}\,\partial_\mu^\perp P\, , 
\ee
where $D=u^\mu\partial_\mu$ and $\partial_\mu^\perp=\Delta_{\mu\nu}\partial^\nu$.
Eq.~(\ref{rel_euler_2}) has the same structure as the non-relativistic
Euler equation, where $D$ plays the role of the comoving derivative, and
${\cal E}+P$ is the inertia of the fluid. 

 Gradient corrections to Eq.~(\ref{T_id}) are important for a number 
of reasons. The most important is that the Euler equation is exactly 
time reversal invariant, and no entropy is produced (there is an 
exception to this statement if shocks are present). Gradient terms,
on the other hand, violate time reversal invariance and lead to the
production of entropy. This implies, in particular, that even if 
gradient terms are small at any given time, their effects can 
exponentiate and the late time flow is qualitatively different. 

 In order to identify gradient corrections to $T_{\mu\nu}$ we have 
to define the fluid velocity more carefully. In the ultra-relativistic
domain we can define $u^\mu$ through the condition $u^\mu T_{\mu\nu}=
{\cal E}u_\nu$. This relation, called the Landau frame condition, 
implies that the energy current in the rest frame does not receive 
any dissipative corrections, $T_{0i}|_{\it rf}=0$. In the non-relativistic
domain it is more natural to define the fluid velocity in terms
of the baryon current $\jmath_B^\mu = u^\mu n_B$. This condition, 
called the Eckart frame, implies that there are no dissipative 
corrections to the baryon current but allows gradient corrections
to the energy current in the rest frame of the fluid. Physically, 
the two frames are of course equivalent. For example, heat conduction
appears as a correction to the energy current in the Eckart frame, and
as a correction to the baryon current in the Landau frame. 

 After these preliminaries, we can state the possible first order 
gradient terms in the stress tensor. We have
\be 
 \delta T^{(1)}_{\mu\nu}= -\eta \sigma_{\mu\nu} 
     - \zeta g_{\mu\nu} (\partial\cdot u)\, , 
\ee
where $\eta$ is the shear viscosity, $\zeta$ is the bulk viscosity, 
and we have defined
\be 
\label{sig_vis}
\sigma^{\mu\nu} = 
  \Delta^{\mu\alpha}\Delta^{\nu\beta} 
  \left(\partial_\alpha u_\beta+\partial_\beta u_\alpha 
  - \frac{2}{3}\eta_{\alpha\beta}\partial\cdot u \right) \,  ,
\ee
where $\Delta_{\mu\nu}$ is the projector defined in Eq.~(\ref{Delta_uu}).
This definition ensures that in the local rest frame there is no 
dissipative correction to the energy current, and only the momentum
density current is modified. This modification describes friction 
and viscous heating in the fluid. We also note that in a scale invariant 
fluid the bulk viscosity vanishes. Kinetic theory suggests that the 
shear viscosity scales as 
\be 
 \eta \sim sT\tau_{\it coll}\, , 
\ee
where $\tau_{\it coll}$ is a typical collision time scale. In perturbative
QCD $\tau_{\it coll}\sim 1/(g^4T)$, see Section~\ref{sec_npfl}. Bulk 
viscosity is controlled by the same collision rate, but is also 
sensitive to deviations of the equation of state from exact scale
invariance, ${\cal E}\neq 3P$. A simple estimate is $\zeta \sim 
({\cal E}-3P)^2 \eta$ \cite{Arnold:2006fz,Dusling:2011fd}, but 
deviations from this relation may appear in the strong coupling
limit. The factor ${\cal E}-3P$ occurs squared because deviations
from scale invariance are needed for collisions to build up a
scalar contribution to non-equilibrium distribution functions, and
they are also required for the non-equilibrium distribution 
function to feed into a non-equilibrium pressure term.

 The formalism presented in this Section is well established, and 
has been known for many years. However, before a flurry of activity 
triggered by the first RHIC data led to a re-evaluation of the theory, 
it was widely believed that the relativistic Navier-Stokes equation 
is unstable and probably ill-defined, and that there is no hope for 
the gradient expansion to convergence in a relativistic heavy ion 
collision (see \cite{Rischke:1998fq,Muronga:2001zk} for some notable 
exceptions). In the following we will describe how our understanding 
of relativistic fluid dynamics has evolved in response to both the 
data, and to progress in the theory of relativistic fluid dynamics. 

%%%%%%%%%%%%%%%%%%%%%%%%%%%%%%%%%%%%%%%%%%%%%%%%%%%%%%%%%%%%%%%%%%%%
\subsubsection{Consistency of the Navier-Stokes equation}
\label{sec_fl_grad}
%%%%%%%%%%%%%%%%%%%%%%%%%%%%%%%%%%%%%%%%%%%%%%%%%%%%%%%%%%%%%%%%%%%%
 
 Once the equations of fluid dynamics are determined it is natural
to study the motion of small fluctuations on a fixed background flow
\cite{Landau_Hydro}. One obvious excitation is sound. In a fluid at 
rest longitudinal sound waves are propagating with the speed of sound 
$c_s$. A second, transverse, excitation is a diffusive shear wave. 
The dispersion relation of a shear wave is $\omega = i\nu k^2$, where 
$\nu =\eta/(sT)$. Consequently the ``speed'' $(\partial|\omega|)/
(\partial k)$ of a diffusive wave is \cite{Romatschke:2009im}
\be 
 v_D \simeq 2\nu k  \, , 
\ee
and a very sharp diffusive front can move at arbitrarily high velocity. 
This implies that the relativistic Navier-Stokes equation is acausal 
for $k\gsim sT/(2\eta)$. It was also found that in this regime the 
Navier-Stokes equation has unstable modes \cite{Hiscock:1985zz}.

 We emphasize that there is nothing fundamentally wrong here: 
Fluid dynamics is an effective low energy, low momentum description, 
and we will see that these modes occur outside the regime of validity
of the theory. However, as a practical method for studying the 
evolution of a heavy ion collision we are interested in a causal, 
stable set of differential equations that can be solved on a 
computer. 

 The problem with causality is related to the fact that in the 
Navier-Stokes equation the dissipative stresses are instantaneously
equal to spatial gradients of the fluid velocity. In any microscopic
treatment this is not the case -- stresses are generated by strains
in the flow velocity over some characteristic time. This effect
automatically appears at second order in the gradient expansion, as 
we will see below. In a scale invariant fluid the most general form 
of the stress tensor is \cite{Baier:2007ix}
\begin{eqnarray}
\label{del_Pi_2}
 \delta T^{\mu\nu} &=& -\eta\sigma^{\mu\nu}
 +\eta\tau_{R} \left[ ^\langle D\sigma^{\mu\nu\rangle}
 +\frac{1}{3}\,\sigma^{\mu\nu} (\partial\cdot u) \right] \\
& & \hspace{0.2cm}\mbox{}
 +\lambda_1\sigma^{\langle\mu}_{\;\;\;\lambda} \sigma^{\nu\rangle\lambda}
 +\lambda_2\sigma^{\langle\mu}_{\;\;\;\lambda} \Omega^{\nu\rangle\lambda}
 +\lambda_3\Omega^{\langle\mu}_{\;\;\;\lambda} \Omega^{\nu\rangle\lambda}
 \, ,
\nonumber 
\end{eqnarray}
where ${\cal O}^{\langle\mu\nu\rangle}= \frac{1}{2}\Delta^{\mu\alpha}
\Delta^{\nu\beta} ( {\cal O}_{\alpha\beta}+{\cal O}_{\beta\alpha} 
- \frac{2}{3}\Delta^{\mu\nu}\Delta^{\alpha\beta}{\cal O}_{\alpha\beta})$ 
denotes the transverse traceless part of ${\cal O}^{\alpha\beta}$. The 
relativistic vorticity tensor is 
\be 
\Omega^{\mu\nu} = \frac{1}{2}
\Delta^{\mu\alpha}\Delta^{\nu\beta} (\partial_\alpha u_\beta
 - \partial_\beta u_\alpha). 
\ee
The quantities $\tau_R$ and $\lambda_i$ are second order 
transport coefficients. We note that Eq.~\ref{del_Pi_2} can also
be obtained in kinetic theory \cite{Israel:1979wp,Denicol:2012cn}, 
but in this case certain terms allowed by the symmetries, like the 
$\lambda_3$-term in Eq.~\ref{del_Pi_2}, are absent.

 The coefficient $\tau_R$ describes the relaxation of the fluid
stresses to the Navier-Stokes form. This can be seen more explicitly
by writing the equations of fluid dynamics as
\be 
\label{hydro_visc}
 \partial^\mu \left( T^{(0)}_{\mu\nu}+\pi_{\mu\nu}\right) = 0 \, , 
\ee
where the viscous stresses $\pi_{\mu\nu}$ satisfy the dynamical equation
\begin{eqnarray}
\label{del_Pi_3}
 \tau_{R}  ^\langle D\pi^{\mu\nu\rangle}
  &=& - \Big( \pi^{\mu\nu}+\eta\sigma^{\mu\nu} \Big)
      -  \frac{4}{3}\, \tau_R\,\pi^{\mu\nu} (\partial\cdot u) 
 \\
& & \hspace{0.2cm}\mbox{}
 +\frac{\lambda_1}{\eta^2}
        \pi^{\langle\mu}_{\;\;\;\lambda} \pi^{\nu\rangle\lambda}
 -\frac{\lambda_2}{\eta}
        \pi^{\langle\mu}_{\;\;\;\lambda} \Omega^{\nu\rangle\lambda}
 +\lambda_3
        \Omega^{\langle\mu}_{\;\;\;\lambda} \Omega^{\nu\rangle\lambda}
  \, .
\nonumber 
\end{eqnarray}
This equation is equivalent to Eq.~\ref{del_Pi_2} at second order in 
gradients of the fluid velocity. It has the form of a relaxation
equation: The viscous stress $\pi^{\mu\nu}$ relaxes to the Navier-Stokes
result $-\eta\sigma^{\mu\nu}$ on a time scale controlled by $\tau_R$.

The set of equations given by (\ref{hydro_visc}-\ref{del_Pi_3})
provides a causal and stable set of fluid dynamic equations. In
particular, the dispersion relation of the shear mode is given by
\be 
\omega = \frac{i\nu k^2}{1+i\omega\tau_R}\, , 
\ee
and the limiting speed is $v_D^{\it max}=[\eta/(s\tau_R T)]^{1/2}$. 
We will see below that for reasonable values of $\tau_R$ we 
get $v_D^{\it max}< c$, and the fluid dynamic equations are causal.
Formally, the equations have second order accuracy in gradients, 
and the sensitivity to poorly constrained higher order transport
coefficients can be checked by varying $\tau_R$ and $\lambda_i$. 

%%%%%%%%%%%%%%%%%%%%%%%%%%%%%%%%%%%%%%%%%%%%%%%%%%%%%%%%%%%%%%%%%%%%
\subsubsection{Nearly perfect fluidity and the convergence of the 
gradient expansion}
\label{sec_npfl}
%%%%%%%%%%%%%%%%%%%%%%%%%%%%%%%%%%%%%%%%%%%%%%%%%%%%%%%%%%%%%%%%%%%%

 Another important issue is whether the gradient expansion converges
in the case of relativistic heavy ion collisions. This is far from
obvious, because the systems are small, initial energy density gradients
are large, and the evolution is very rapid. We will provide more 
detailed estimates for real heavy ion collisions in Section~\ref{sec_flow}.
As a warm-up, we consider an equilibrium response function in a static
medium. The shear stress response function
is given by Eq.~(\ref{G_ret}). In linear response theory this function 
controls the stress induced by an external strain. In fluid dynamics 
$T_{xy}\simeq ({\cal E}+P) u_xu_y$ and we can compute the correlation 
function from linearized hydrodynamics and fluctuation relations. We 
find \cite{Schaefer:2014awa,Kubo:1957,Son:2007vk}
\be 
\label{Kubo}
 G_R^{xyxy}(\omega,k) = P -i\eta \omega +\tau_R\eta \omega^2
  + O(\omega^3,k^2)\, .
\ee
This result, called the Kubo formula, can be used to relate the shear 
viscosity and other transport coefficients to an equilibrium (but 
time-dependent) correlation function in the microscopic theory, see 
Eq.~(\ref{eta_kubo}). We can now read off the criterion that the
gradient expansion converges. From the first two terms in 
Eq.~(\ref{Kubo}) we get
\be
\label{grad_exp}
 \omega \lsim \frac{sT}{\eta} \, ,
\ee
where we have used $P\sim sT$. Analogously, from the second and
third term we get $\omega\lsim \tau_R^{-1}$. This constraint is 
consistent with Eq.~(\ref{grad_exp}) provided $\tau_R\sim sT/\eta$,
which is indeed what one finds both in the weak coupling, kinetic
theory, and in the strong-coupling, AdS/CFT-like, regime
\cite{York:2008rr,Baier:2007ix}. We note that this estimate of
$\tau_R$ also implies that the limiting speed of a diffusive wave
is the speed of light, consistent with causality. 

 In the early stages of a relativistic heavy ion collision the 
characteristic expansion time of the plasma, $\tau \sim (\partial
\cdot u)^{-1}$, is significantly less than 1 fm. In order for the 
expansion parameter $\eta/(sT\tau)$ to be small, we need $\eta/s\ll 1$. 
Note that ${\it Re}^{-1}=\eta/(sT\tau)$ is the inverse Reynolds number 
of the flow. In kinetic theory one finds 
\cite{Baym:1990uj,Arnold:2000dr}
\be 
\label{eta_s_w}
\frac{\eta}{s} \simeq \frac{9.2}{g^4\log(1/g)}\, , 
\ee 
where we have assumed that the quark-gluon plasma contains three
light quark flavors. The argument inside the logarithm has been 
determined \cite{Arnold:2003zc}, but the expansion in inverse powers
of $\log(1/g)$ converges very slowly. For numerical estimates we 
will assume $\log(1/g)\gsim 1$. Using $g\simeq 2$, which corresponds 
to $\alpha_s\simeq 0.3$, we conclude that at best $\eta/s\lsim 0.6$. 
This estimate implies that fluid dynamics is not likely to be 
quantitatively reliable in relativistic heavy ion collisions. 

 This pessimistic view was revised because of two discoveries, one 
experimental and the other theoretical. The first is the discovery of 
nearly ideal flow, indicative of very small viscous corrections, 
observed in the early RHIC data \cite{Adler:2003kt,Adams:2004bi}
and confirmed at the LHC \cite{Aamodt:2010pa}. The second is 
the theoretical realization that the strong coupling value
of $\eta/s$ in gauge theories with holographic duals is as small
as $\eta/s=1/(4\pi)$ \cite{Policastro:2001yc,Kovtun:2004de}. A
similar lower bound on $\eta/s$ was first suggested based on 
the quantum mechanical uncertainty relation \cite{Danielewicz:1984ww}.
Note that, if we reinstate all physical constants, the proposed 
bound takes the form $\eta/s=\hbar/(4\pi k_B)$, where $\hbar$ is 
Planck's constant, and $k_B$ is Boltzmann's constant.

 We will refer to fluids that approach $\eta/s=1/(4\pi)$ as ``nearly
perfect fluids''. Fluids of this type exhibit fluid dynamic behavior 
on time and distance scales as short as $t\sim l \sim T^{-1}$. The 
reason that hydrodynamics is successful as a theory of relativistic
heavy ion collisions is that the quark-gluon plasma is a nearly 
perfect fluid. Indeed, as we shall demonstrate, the best determinations
of $\eta/s$ at RHIC and the LHC are remarkably close to $1/(4\pi)$.

%%%%%%%%%%%%%%%%%%%%%%%%%%%%%%%%%%%%%%%%%%%%%%%%%%%%%%%%%%%%%%%%%%%%
\subsubsection{Beyond gradients: Hydrodynamic fluctuations}
\label{sec_hydro_flucs}
%%%%%%%%%%%%%%%%%%%%%%%%%%%%%%%%%%%%%%%%%%%%%%%%%%%%%%%%%%%%%%%%%%%%

  We have argued that fluid dynamics is a general effective theory 
that describes the long distance, late time response of a many body 
system perturbed away from thermal equilibrium. However, it is 
clear that in order to improve the fluid dynamic description it 
is not sufficient to include higher order gradients. Fluid dynamics
is a coarse grained description, and as we attempt to increase the 
resolution local fluctuations in the fluid dynamic description 
become more important. 
 
 The equations of fluid dynamics including fluctuations can be written 
as \cite{Landau:kin,Kapusta:2011gt,Young:2013fka,Murase:2013tma}
\be 
\label{hydro_fluc}
 \partial^\mu \left( T^{(0)}_{\mu\nu}+\pi_{\mu\nu}
  + \Xi_{\mu\nu} \right) = 0 \, , 
\ee
where $\Xi_{\mu\nu}$ is a stochastic term which satisfies
\begin{eqnarray}
\label{noise_cor}
\left\langle \Xi_{\mu\nu}(x) \Xi_{\alpha\beta}(x') \right\rangle 
  &=& \Big[2\eta T \left(\Delta_{\mu\alpha}\Delta_{\nu\beta}
                         +\Delta_{\mu\beta}\Delta_{\nu\alpha}\right)
  \nonumber \\
  & & \mbox{}\;\;\;
         -2\left(\zeta-\frac{2\eta}{3}\right)T 
                          \Delta_{\mu\nu}\Delta_{\alpha\beta}\Big]
   \delta(x-x')\, .
\end{eqnarray}
The structure of Eq.~(\ref{noise_cor}) is fixed by fluctuation
dissipation relations. As in Section~\ref{sec_npfl} it is easiest to 
understand the role of these terms near equilibrium. We consider the 
response function for $T_{xy}$ and study the role of fluctuations $\delta 
T_{xy}\simeq ({\cal E}+P)_0u_xu_y$. The leading term involves a pair of 
velocity correlators, which can be viewed as a Feynman diagram with two 
propagators for sound or shear modes. For example, the propagator of a 
shear mode is given by 
\be 
  \langle u_x u_y \rangle_{\omega k} =
     - \frac{2T}{{\cal E}+P}\, \frac{k_xk_y}{k^2}\, 
       \frac{\nu k^2}{\omega^2+\nu^2 k^4}\, .
\ee
In momentum space the convolution of the two propagators gives a loop 
integral. We find \cite{Kovtun:2011np}
\be 
\label{delta_G_R}
\delta G_R^{xyxy}(\omega,0) = -\frac{7T\Lambda^3}{90\pi^2}
 - i\omega\, \frac{17T\Lambda}{120\pi^2\nu}
 + (1+i)\omega^{3/2}T\, \frac{7+(3/2)^{3/2}}{240\pi\nu^{3/2}}
 + O(\omega^2)\, ,
\ee
where $\Lambda$ is a momentum space cutoff. Comparing this result 
with Eq.~(\ref{Kubo}) reveals a number of very interesting features: 

\begin{itemize}
\item Fluid dynamics behaves like a renormalizable effective theory. 
There are divergences, but the divergent terms can be absorbed 
into the low energy parameters, the pressure $P$ and the viscosity 
$\eta$. Non-analytic terms, like the $\omega^{3/2}$ term in 
Eq.~(\ref{delta_G_R}) are finite.

\item Fluid dynamics has an intrinsic resolution scale, given by 
the cell size in a simulation with fluctuating stresses, and
the parameters of fluid dynamics, the equation of state and 
the transport coefficients, evolve as a function of that scale. 

\item  Non-analytic corrections are smaller than the Navier-Stokes
term $-\omega\eta$, but larger than the second order term 
$\omega^2\tau_R\eta$. This means that a consistent second order 
calculation has to contain fluctuating forces. 

\item  The fluctuation contribution to the $i\omega$ term is inversely 
proportional to $\nu=\eta/(sT)$. This suggests that the physical viscosity 
cannot become arbitrarily small \cite{Kovtun:2011np}. The physical 
mechanism for this bound is related to the fact that shear viscosity
is a measure of momentum diffusion, and that the contribution of 
shear and sound modes to momentum diffusion can never vanish. Note 
that this bound is completely classical -- Planck's constant only
enters indirectly, via the equation of state in the quantum regime
\cite{Kovtun:2011np}.

\end{itemize}

\begin{figure}[t!]
\begin{center}
\includegraphics*[width=7.5cm,angle=-90]{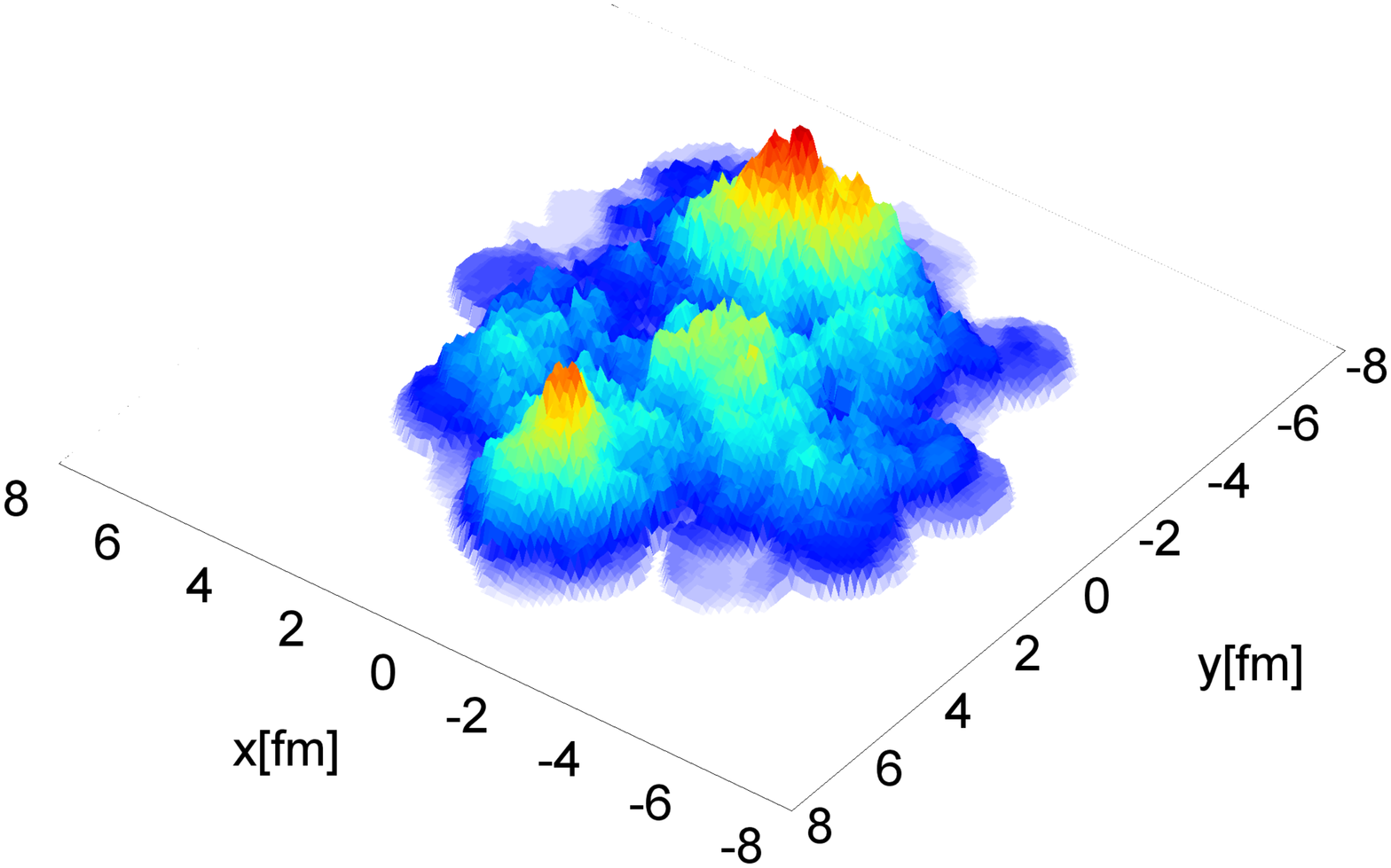}
\end{center}
\caption{\label{fig_init}
Initial energy density (arbitrary units) in a Au+Au collision at RHIC 
from the Monte-Carlo KLN model, see \cite{Schenke:2012wb,Kharzeev:2000ph}.
This model include the effects from the collision geometry, fluctuations
in the initial position of the nucleons inside the nucleus, and 
non-linear gluon field evolution. More sophisticated versions of the 
model also include quantum fluctuations of the gluon field. }
\end{figure}

  In calculations of flow observables the contribution of thermal 
fluctuations is likely to be small compared to the role of initial
state fluctuations (see Fig.~\ref{fig_init}) \cite{Kapusta:2011gt}.
However, the formalism described here is important in describing the 
dynamical evolution of fluctuations discussed in Section~\ref{sec:fluct},
in particular in the vicinity of a critical point, where the rate 
at which fluctuations can grow becomes a crucial concern.  

%%%%%%%%%%%%%%%%%%%%%%%%%%%%%%%%%%%%%%%%%%%%%%%%%%%%%%%%%%%%%%%%%%%%
\subsection{Collective expansion and transport properties of the 
quark-gluon plasma}
\label{sec_flow}
%%%%%%%%%%%%%%%%%%%%%%%%%%%%%%%%%%%%%%%%%%%%%%%%%%%%%%%%%%%%%%%%%%%%

 The experimental determination of transport properties of the 
quark-gluon plasma is mainly based on the comparison of flow measurements
at collider energies \cite{Voloshin:2008dg,Heinz:2013th} with 
dissipative fluid dynamics simulations \cite{Heinz:2009xj,Teaney:2009qa}.
Several observations support the assumption that heavy ion 
collisions create a locally thermalized system: 

\begin{itemize}

\item The total abundances of produced particles is described by 
a simple thermal model that depends on only two parameters, the 
temperature $T$ and the baryon chemical potential $\mu$ at freeze-out,
see Section~\ref{sec:PJ_thermal}.

\item In the regime of transverse momenta $p_\perp \lsim 2$ GeV the spectra
$dN/d^3p$ of produced particles follow a modified Boltzmann distribution
characterized by the freeze-out temperature and a collective radial
expansion velocity 
\cite{BraunMunzinger:1994xr,Heinz:2009xj,Schnedermann:1993ws,Adams:2003xp}.
Radial flow manifests itself in the fact that the spectra of heavy hadrons, 
which acquire a boost $p_\perp\sim m u_\perp$ from the collective radial
expansion with velocity $u_\perp$, have a larger apparent temperature 
than the spectra of light hadrons.

\item In non-central collisions the distribution of produced particles 
in the transverse plane shows a strong azimuthal anisotropy known as
elliptic flow \cite{Ollitrault:1992bk,Heinz:2009xj,Voloshin:2008dg}. 
Elliptic flow represents the hydrodynamic response of the quark-gluon 
plasma to energy density gradients in the initial state. These gradients 
are caused by a combination of geometric effects, related to the overlap
geometry, and fluctuation effects, related to the mechanism of the
initial energy deposition. 

\end{itemize}

 The quantitative analysis of the transverse flow pattern is based
on Fourier moments of the azimuthal particle distribution. We define 
the harmonic moments
\begin{eqnarray}
\label{v_2}
 \left. p_0\frac{dN}{d^3p}\right|_{p_z=0} &=& 
 \left. p_0\frac{dN}{\pi dp_T^2dp_z}\right|_{p_z=0}
 \Big( 1 + 2v_1(p_T)\cos(\phi-\Psi_1) \nonumber \\
 && \hspace{3.5cm}\mbox{}
         + 2v_2(p_T)\cos(2\phi-\Psi_2) +\ldots \Big) ,
\end{eqnarray}
where $p_z$ is the momentum in the longitudinal (beam) direction, $p_T$ 
is the transverse momentum, $\phi$ is the transverse angle relative 
to the impact parameter direction. The coefficient $v_2$ is known as 
elliptic flow, and the higher moments are termed triangular, quadrupolar, 
etc.~flow. The angles $\Psi_i$ account for the fact that the flow 
angles need not be aligned with the impact parameter plane, and are 
known as flow angles. Substantial elliptic flow, reaching about $v_2
(p_T\!=\!2\,{\rm GeV}) \simeq 20\%$ in semi-central collisions, was 
observed in the early RHIC data \cite{Adler:2003kt,Adams:2004bi} and 
confirmed at the LHC \cite{Aamodt:2010pa}. More recently, it was 
realized that fluctuations in the initial energy density generate 
substantial higher harmonics, including odd Fourier moments such as 
$v_3$ \cite{Alver:2010gr}, as well as fluctuations of the flow angles 
relative to the impact parameter plane \cite{Alver:2008zza}.

 Viscosity tends to equalize the radial flow velocity and suppress
elliptic flow and higher flow harmonics. An estimate of the relevant
scales can be obtained from simple scaling solutions of fluid dynamics. 
The simplest analytic solution was found by Bjorken, who considered a 
purely longitudinal expansion \cite{Bjorken:1982qr}. Bjorken assumed 
that the nuclei are infinitely extended in the transverse direction, 
and that the initial entropy density is independent of rapidity. This
implies that the subsequent evolution is invariant under boosts 
along the $z$ axis. The Bjorken solution provides a natural starting 
point for more detailed numerical and analytical studies 
\cite{Heinz:2009xj,Gubser:2010ui}. Bjorken flow is characterized 
by a flow profile of the form $u_\mu = \gamma(-1,0,0,v_z)= (-t/\tau,
0,0,z/\tau)$, where $\gamma=(1-v_z^2)^{1/2}$ is the boost factor and 
$\tau=(t^2-z^2)^{1/2}$ is the proper time. This velocity field solves 
the relativistic Navier-Stokes equation. Energy conservation then 
determines the evolution of the entropy density. We find
\be
\label{bj_ns}
-\frac{\tau}{s}\frac{ds}{d\tau} = 
  1 - \frac{4}{3}\frac{\eta}{sT\tau}\, ,
\ee
where we have neglected bulk viscosity. Using both the ideal equation of 
state, $s\sim T^3$, and ideal hydrodynamic evolution we obtain $T\sim 
1/\tau^{1/3}$. The validity of the gradient expansion requires that 
viscous corrections in Eq.~(\ref{bj_ns}) are small 
\cite{Danielewicz:1984ww}
\be
\label{DG}
\frac{\eta}{s} \ll \frac{3}{4}(T\tau)\, .
\ee
It is usually assumed that in the QGP $\eta/s$ is approximately constant. 
For the Bjorken solution $T\tau\sim \tau^{2/3}$ increases with time, and 
Eq.~(\ref{DG}) is most restrictive during the initial stages of the 
expansion. Using an equilibration time $\tau_0=1$ fm and an initial 
temperature $T_0=300$ MeV gives $\eta/s\lsim 0.6$. This confirms our 
earlier assertion that fluid dynamics can be applied to heavy ion 
collisions only if the QGP behaves as a nearly perfect fluid. 

%%%%%%%%%%%%%%%%%%%%%%%%%%%%%%%%%%%%%%%%%%%%%%%%%%%%%%%%%%%%%%%%%%%%
%%%%%%%%%%%%%%%%%%%%%%%%%%%%%%%%%%%%%%%%%%%%%%%%%%%%%%%%%%%%%%%%%%%%

 At late time the expansion becomes three dimensional and $T\tau$ 
is independent of time. The relevant degrees of freedom are hadronic 
resonances with interaction cross sections $\sigma$ that reflect hadronic 
sizes and are approximately independent of energy. In that case $\eta\sim 
T/\sigma$. Using $s\sim T^3$ and $T\sim 1/\tau$ we find that the viscous
correction $\eta/(sT\tau)$ increases with proper time as $\tau^2$. This 
result shows that fluid dynamics also breaks down at late times. At 
RHIC and LHC energies the duration of the fluid dynamic  phase is 
5-10 fm/c, depending on collision energy and geometry. 

 In heavy ion collisions we do not directly observe the final distribution
of energy and momentum. What is measured experimentally is the distribution
of hadrons. In principle one could imagine reconstructing azimuthal
harmonics of the stress tensor at freeze-out from the measured particle 
distribution, but doing so would require complete particle identification
as well as spatial and momentum information for the produced particles, 
and it has not been attempted. In any case, hadrons continue to interact
after the fluid freezes out, and some rearrangement of momentum takes 
place. This means that we need a prescription for converting hydrodynamic
variables to kinetic distribution functions. What is usually done is 
that we define a freeze-out hypersurface on which hydrodynamics is 
assumed to break down. In principle, this hypersurface is defined by 
a kinetic criterion, for example the condition that the mean free path 
of a typical hadron satisfies $l_{\it mfp}\gsim c_s/(\partial\cdot u)$. 
In practice, freeze-out is assumed to take place at constant temperature
or energy density. On the freeze-out surface the conserved densities in 
fluid dynamics are matched to kinetic theory \cite{Cooper:1974mv}. 

 In ideal fluid dynamics the distribution functions are Bose-Einstein
or Fermi-Dirac distributions characterized by the local temperature
and fluid velocity. Viscosity modifies the stress tensor, and via 
matching to kinetic theory this modification changes the distribution 
functions $f_p$. The value of $\eta/s$ constrains only the $p_iv_j$ 
moment of the distribution function. The full distribution function
can be reconstructed only if the collision term in the Boltzmann 
equation is specified. Using a simple model which assumes that 
collisions are fully specified by a single collision time (known as
the BGK model \cite{Bhatnagar:1954zz}) one obtains a very simple formula 
for the leading correction $\delta f_p$
\be
\label{del_f}
\delta f_p = \frac{1}{2T^3} \frac{\eta}{s}f_0(1\pm f_0)
   p_\alpha p_\beta  \sigma^{\alpha\beta} \, ,
\ee
where the $\pm$ sign refers to Bose/Fermi distributions. This result 
is a reasonable approximation to more microscopic theories 
\cite{Arnold:2000dr}. The shift in the distribution function leads to 
a modification of the single particle spectrum. In the case of the 
Bjorken expansion and at large $p_T$ we find
\be
\label{del_N}
 \frac{\delta (dN)}{dN_0} = \frac{1}{3\tau_fT_f}\frac{\eta}{s}
    \left( \frac{p_T}{T_f} \right)^2 \,,
\ee
where $dN_0$ is the number of particles emitted in ideal fluid dynamics, 
$\delta (dN)$ is the dissipative correction, and $\tau_f$ is the freeze-out 
time. Note that if it were not for nearly perfect fluidity, $\eta/s
\simeq 1/(4\pi)$, the prediction of spectra using fluid dynamics at 
RHIC and LHC would be completely hopeless. Even with a minimal viscosity 
corrections are of order 25\% at $p_T\sim 1$ GeV. As a consequence,
precision determinations of $\eta/s$ should be based on integrated 
observables that are dominated by low $p_T\lsim 1$ GeV.

  In a system with strong longitudinal expansion viscous corrections 
tend to equalize the momentum distribution by pushing particles to higher 
$p_T$. Because the single particle distribution enters into the denominator 
of $v_2$ this effect suppresses $v_2$ at large $p_T$. The effect from the 
numerator, dissipative corrections related to the $\cos(2\phi)$ component 
of the transverse flow, act in the same direction \cite{Teaney:2003kp}. 
The important observation is that corrections to the spectrum are controlled 
by the same parameter $\eta/(s\tau T)$ that governs the derivative expansion 
in fluid dynamics. This reflects the fact that in the regime in which 
kinetic theory can be matched to fluid dynamics, the expansion parameter
in kinetic theory, the Knudsen number 
\be 
{\it Kn} = \frac{l_{\it mfp}}{L}
\ee
with $L \sim c_s/(\partial\cdot u)$, is comparable to the expansion 
parameter in fluid dynamics,  ${\it Kn}\sim {\it Re}^{-1}$
\cite{Belenkij:1956cd}.

%%%%%%%%%%%%%%%%%%%%%%%%%%%%%%%%%%%%%%%%%%%%%%%%%%%%%%%%%%%%%%%%%%%%
\begin{figure}[t!]
\begin{center}
\includegraphics*[width=10cm]{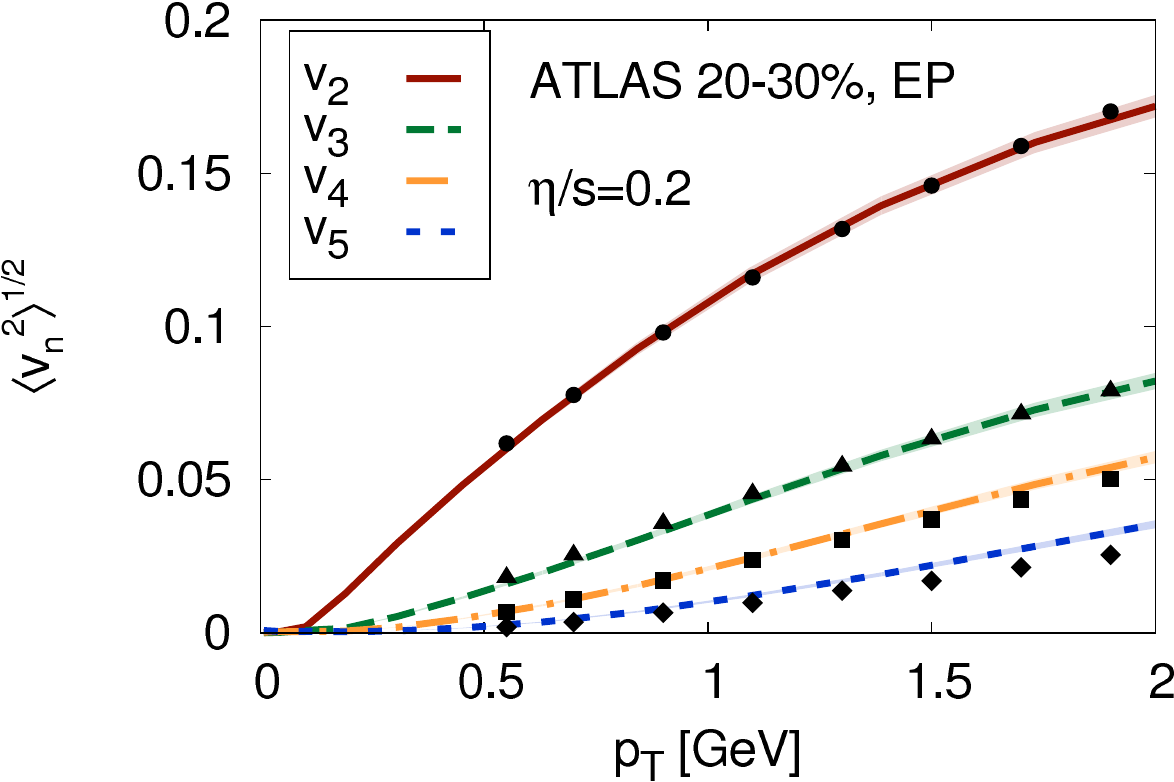}
\end{center}
\caption{\label{fig_vn}
Fourier coefficients $v_2,\ldots,v_5$ of the azimuthal charged
particle distribution as a function of the transverse momentum
$p_T$ measured in $Pb+Pb$ collisions at the LHC \cite{ATLAS:2012at}
The lines show a hydrodynamic analysis performed using $\eta/s=0.2$
\cite{Gale:2012rq}.}
\end{figure}
%%%%%%%%%%%%%%%%%%%%%%%%%%%%%%%%%%%%%%%%%%%%%%%%%%%%%%%%%%%%%%%%%%%%

 We obtain several simple predictions that have been confirmed
experimentally \cite{Heinz:2013th}: Dissipative corrections increase 
with $p_T$, they are larger in small systems that freeze out 
earlier, and they are larger for higher harmonics that are more 
sensitive to gradients of the radial flow profile. Quantitative 
studies that provide reliable measurements of $\eta/s$ together
with estimates of the uncertainties involved require a number 
of ingredients \cite{Gale:2013da}:

\begin{itemize}

\item An initial state model that incorporates the nuclear geometry 
and fluctuations in the initial energy deposition. The simplest 
possibility is a Monte-Carlo implementation of the Glauber model 
\cite{Miller:2007ri}, but some calculations also include a model of the 
color field generated by colliding nuclei, as well as the effects of 
quantum fluctuations and real time evolution of this field \cite{Gale:2012rq}. 
Alternatively, one may try to describe the pre-equilibrium stage using 
kinetic theory \cite{Martinez:2010sc,Petersen:2008dd} or the AdS/CFT 
correspondence \cite{vanderSchee:2013pia}. At the end of the initial 
stage the stress tensor is matched to fluid dynamics.

\item Second order dissipative fluid dynamics in 2+1 (boost invariant)
or 3+1 dimensions with a realistic equation of state (EOS). A realistic
EOS has to match lattice QCD results at high temperature, and a 
hadronic resonance gas below $T_c$ \cite{Huovinen:2009yb}. The
resonance gas EOS must allow for chemical non-equilibrium effects
below the chemical freeze-out temperature $T_{\it chem}\simeq T_c$. It 
is important to check the sensitivity to poorly constrained second 
order transport coefficients.

\item Kinetic freeze-out and a kinetic afterburner. At the kinetic
freeze-out temperature the fluid is converted to hadronic distribution
functions. Ideally, these distribution functions are evolved further
using a hadronic transport model \cite{Bass:2000ib,Hirano:2005xf}, but 
at a minimum one has to include feed-down from hadronic resonance decays.

\end{itemize}
 
 The first constraints on $\eta/s$ based on the RHIC data were derived 
in \cite{Teaney:2003kp}, and the effects of dissipation in the hadronic 
corona were studied in \cite{Teaney:2001av,Hirano:2005wx,Hirano:2005xf}.
Early estimates of $\eta/s$ were also obtained in \cite{Adare:2010de}
using the relationship between heavy quark diffusion and  momentum 
diffusion in the plasma obtained in kinetic theory \cite{Moore:2004tg}.
Determinations of $\eta/s$ at RHIC based on viscous fluid dynamics were 
obtained in \cite{Dusling:2007gi,Romatschke:2007mq,Song:2007ux}. A more 
recent analysis of LHC data is shown in Fig.~\ref{fig_vn} \cite{Gale:2012rq}. 
The authors found $\eta/s\simeq 0.2$ at the LHC, and $\eta/s\simeq 0.12$ 
from a similar analysis of RHIC data. Similar results were obtained by 
other authors. Song et al.~reported an average value of $\eta/s\simeq (0.2
-0.24)$ at the LHC and $\eta/s\simeq 0.16$ at RHIC \cite{Song:2011qa}. 
Luzum and Ollitrault tried to constrain the allowed range of $\eta/s$, 
obtaining $0.07\leq\eta/s\leq 0.43$ at RHIC \cite{Luzum:2012wu}. Given the 
complexity of the analysis, uncertainties are difficult to quantify. A 
survey of the main sources of error in the determination of $\eta/s$ can 
be found in \cite{Song:2008hj}. Interestingly, the extracted values 
of $\eta/s$ are lower at RHIC than they are at the LHC, as one would 
expect based on asymptotic freedom.  

%%%%%%%%%%%%%%%%%%%%%%%%%%%%%%%%%%%%%%%%%%%%%%%%%%%%%%%%%%%%%%%%%%%%
\subsection{Frontiers: Flow in $pA$?}
\label{sec_front}
%%%%%%%%%%%%%%%%%%%%%%%%%%%%%%%%%%%%%%%%%%%%%%%%%%%%%%%%%%%%%%%%%%%%

%%%%%%%%%%%%%%%%%%%%%%%%%%%%%%%%%%%%%%%%%%%%%%%%%%%%%%%%%%%%%%%%%%%%
\begin{figure}[t!]
\begin{center}
\includegraphics*[width=8.5cm]{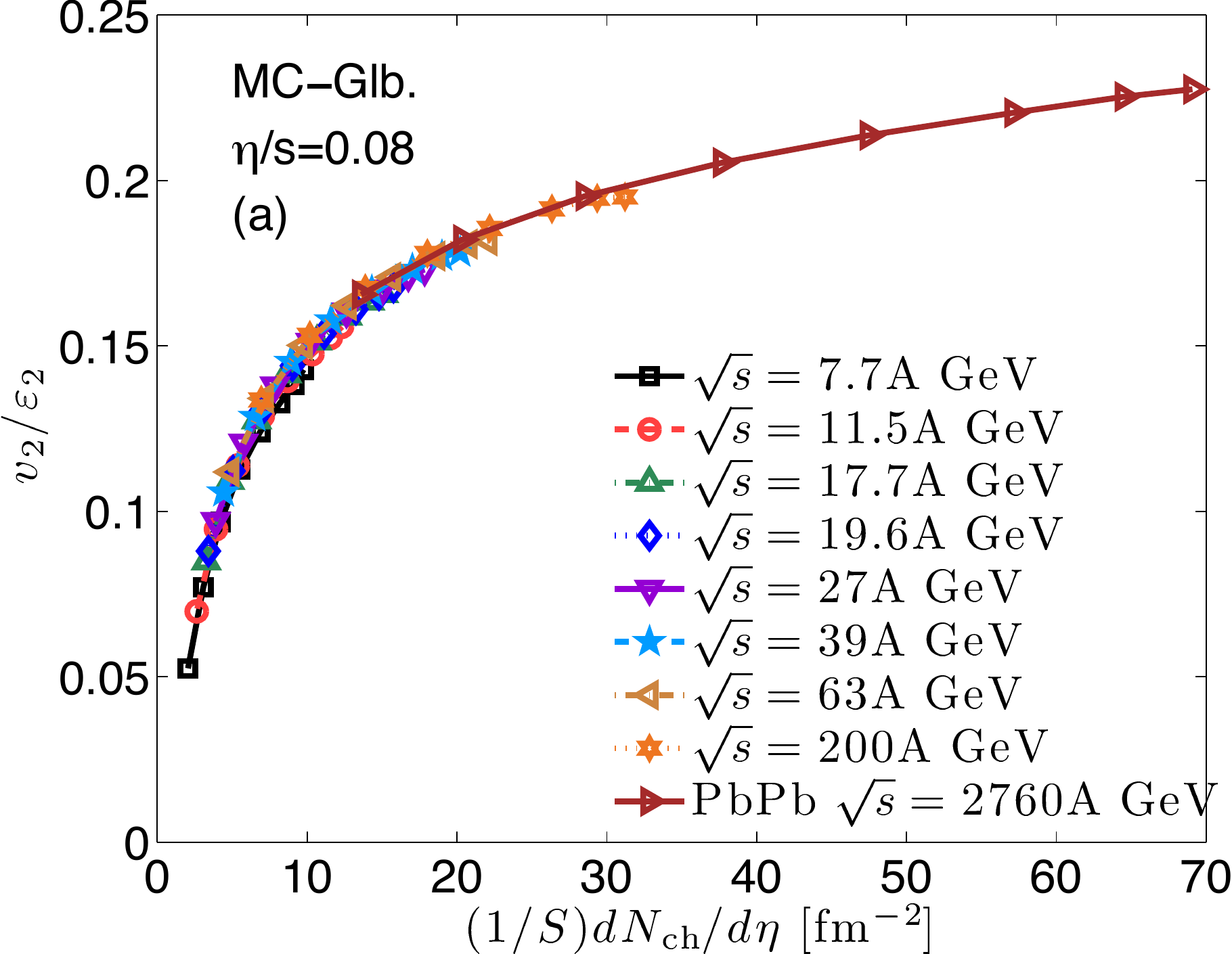}
\end{center}
\caption{\label{fig_dNdy_scal}
Eccentricity scaled elliptic flow $v_2$ plotted as a function of
the charged hadron multiplicity density $dN/dy$ divided by the 
nuclear overlap area $S$ for different collision energies, from
\cite{Shen:2012vn}. The figure shows results from a hydrodynamic 
simulation with MC-Glauber initial conditions and $\eta/s= 0.08$.}
\end{figure}
%%%%%%%%%%%%%%%%%%%%%%%%%%%%%%%%%%%%%%%%%%%%%%%%%%%%%%%%%%%%%%%%%%%%

 An important recent discovery is the observation of significant elliptic 
and triangular flow in high multiplicity p+Pb collisions at the LHC 
\cite{Chatrchyan:2013nka,Aad:2013fja,Abelev:2013wsa}. A particularly 
striking discovery is the mass ordering of $v_2(p_T)$ \cite{Abelev:2013wsa}, 
which is usually regarded as strong evidence for collective expansion 
\cite{Heinz:2009xj}. The result is surprising, because the proton nucleus 
collisions have generally been regarded as a control experiment in which 
dissipative corrections are too large for collective flow to develop. 
Indeed, several authors have shown that initial state effects
\cite{Dusling:2013oia,McLerran:2013oju}, or a simple free-streaming
expansion \cite{Romatschke:2015dha}, can give significant contributions
to flow observables in small systems.

 We should note, however, that the observed response to initial state 
fluctuations in nucleus-nucleus collisions indicates that the mean free 
path is very short, comparable to size of a single nucleon. The measured
flow in $pA$ may well be due to the same mechanism that generates flow
in $AA$, the hydrodynamic response to initial energy density gradients. 
Ultimately, only detailed studies of many particle correlations along
the lines of \cite{Khachatryan:2015waa} can disentangle the relative 
importance of collective and non-collective effects.   

%%%%%%%%%%%%%%%%%%%%%%%%%%%%%%%%%%%%%%%%%%%%%%%%%%%%%%%%%%%%%%%%%%%%
\begin{figure}[t!]
\begin{center}
\includegraphics*[width=7cm]{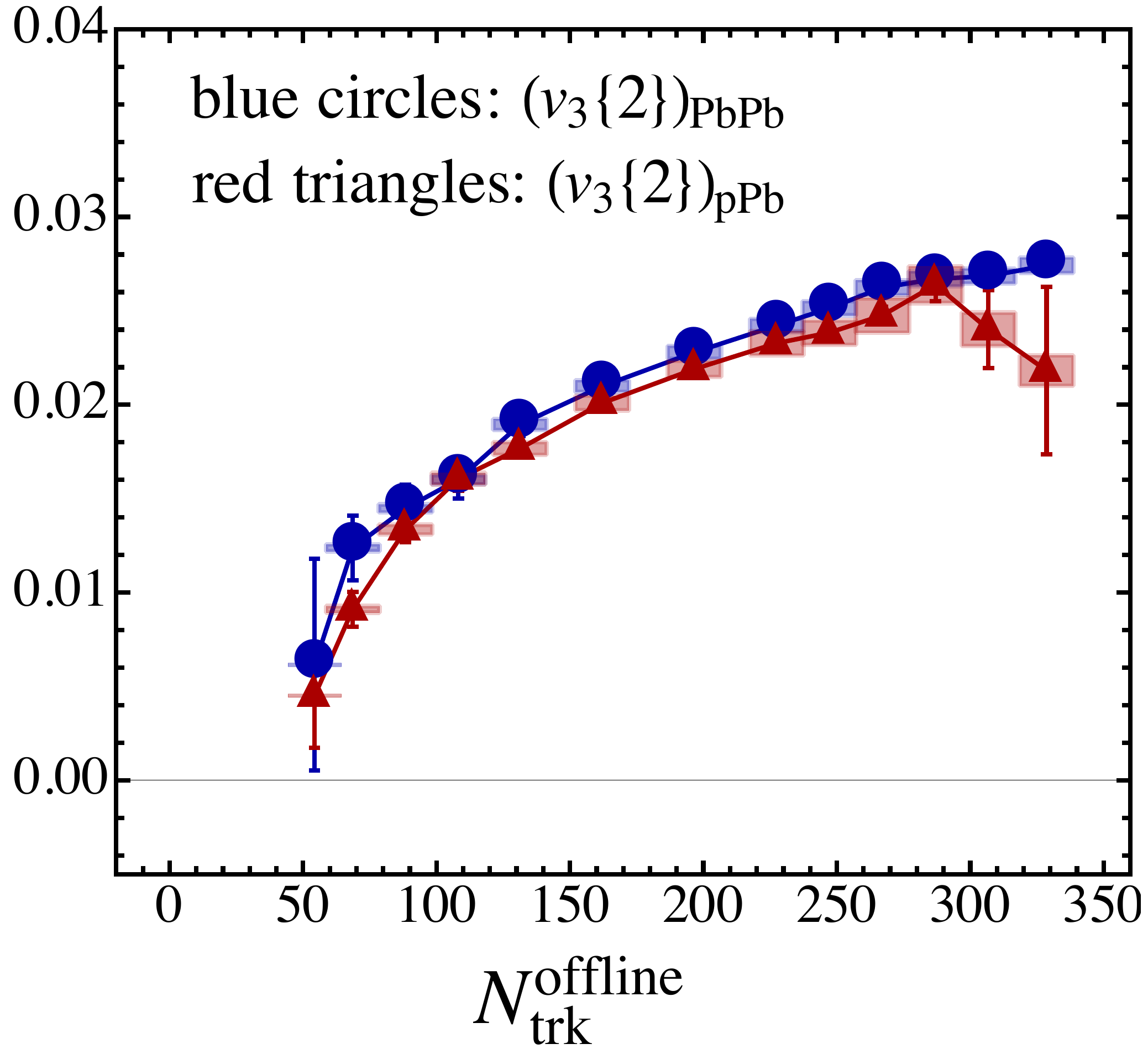}
\end{center}
\caption{\label{fig_v3_pA}
Integrated $v_3\{2\}$ for $PbPb$ and $pPb$ at the LHC as a function of 
the multiplicity. Here, $v_3\{2\}$ denotes the triangular flow extracted 
from two-particle cumulants. Data from \cite{Chatrchyan:2013nka},
analysis taken from \cite{Basar:2013hea}.}
\end{figure}
%%%%%%%%%%%%%%%%%%%%%%%%%%%%%%%%%%%%%%%%%%%%%%%%%%%%%%%%%%%%%%%%%%%%

 In this review we will concentrate on a simpler question, whether
there are simple scaling variables that can be used to compare 
experimental data from collisions at different beam energies, impact 
parameters, and nuclear mass numbers. An important step in this direction
was taken by Heiselberg and Levy, who studied elliptic flow in the dilute 
limit \cite{Heiselberg:1998es}. They showed that the contribution
from single elastic scattering events is 
\be 
 \frac{v_2}{\epsilon_2} \sim \frac{1}{S} \frac{dN}{dy} 
   \langle \sigma \rangle \, . 
\ee
where $S$ is the transverse overlap area, $dN/dy$ is the multiplicity 
per unit of rapidity, and $\epsilon_2$ is the initial deformation, 
defined by 
\be
\epsilon_2 = \frac{\langle y^2 -x^2\rangle}{\langle y^2+x^2 \rangle}\, .
\ee
Following the arguments given in the previous Section we expect that 
the parameter $(1/S)(dN/dy) \langle \sigma \rangle$ also appears in 
fluid dynamics. This is indeed the case, as we can see using the following 
argument \cite{Bhalerao:2005mm}. Consider a fireball of transverse size 
$\bar{R}\simeq \sqrt{R_x^2+R_y^2}$ which is undergoing Bjorken expansion 
in the longitudinal direction. The time scale for transverse expansion is 
$\tau=\bar{R}/c_s$, and the density at this time is $n\sim (\tau S)^{-1}
(dN/dy)$. This implies that the inverse Knudsen number is 
\be 
\frac{1}{{\it Kn}} = \frac{\bar{R}}{l_{\it mfp}} 
  = \bar{R}n\langle\sigma\rangle 
  = c_s \frac{1}{S}\frac{dN}{dy}\langle\sigma\rangle \, . 
\ee
Knudsen number scaling of $v_2/\epsilon_2$ was first studied by Voloshin 
and Poskanzer, see \cite{Voloshin:1999gs,Alt:2003ab}. The results compiled
in \cite{Alt:2003ab} demonstrate nice data collapse if different systems, 
centralities, and beam energies are plotted as a function of $(1/S)(dN/dy)$. 
The compilation also shows that $v_2/\epsilon_2$ rises almost linearly
with $(1/S)(dN/dy)$, and that the RHIC data at 200 GeV per nucleon
saturate the flow predicted by ideal hydrodynamics. A more recent 
analysis of data from the RHIC beam energy scan and $Pb+Pb$ collisions
at the LHC is shown in Fig.~\ref{fig_dNdy_scal} \cite{Shen:2012vn}.
There is some uncertainty related to different models for $\epsilon_2$.
Here, we show results based on the Monte Carlo Glauber model. In this 
case data collapse is excellent, but the results for MC-KLN model are 
not quite as good  \cite{Shen:2012vn}. We observe some curvature in 
$v_2/\epsilon_2$. This means that there are viscous effects even 
at LHC energy, and that there is no saturation of flow. 

 An important assumption in Fig.~\ref{fig_dNdy_scal} is that the 
effective cross section is not a function of the collision parameters.
At high temperature the quark-gluon plasma is scale invariant and
we expect $\langle\sigma\rangle \sim s^{-2/3}$. Then 
\be 
 \frac{1}{{\it Kn}} \sim 
   \left( c_s \frac{dN}{dy}\right)^{1/3} \, , 
\ee
and the overlap area does not appear in the estimate for the Knudsen
number. The choice of scaling variable makes a significant difference 
when comparing $pA$ and $AA$ collisions as shown in \cite{Basar:2013hea}.
Fig.~\ref{fig_v3_pA} shows the $p_T$ integrated triangular flow in 
$pPb$ and $PbPb$ collisions at the LHC, plotted as a function of 
the multiplicity. We observe excellent data overlap, suggesting that 
a common mechanism is at work, and that the conformal Knudsen scaling
holds at the highest energies.

%%%%%%%%%%%%%%%%%%%%%%%%%%%%%%%%%%%%%%%%%%%%%%%%%%%%%%%%%%%%%%%%%%%%
\subsection{Other frontiers, and puzzles}
\label{sec_front_other}
%%%%%%%%%%%%%%%%%%%%%%%%%%%%%%%%%%%%%%%%%%%%%%%%%%%%%%%%%%%%%%%%%%%%

 There are several other frontiers, and some puzzles, that 
are worth mentioning: 

\begin{itemize}

\item Flow even in $pp$? Azimuthal two-particle correlations
very similar to those produced by elliptic and triangular flow
in $AA$ collisions have also been observed in very high multiplicity
$pp$ collisions at $\sqrt{s}=7$ TeV \cite{Khachatryan:2010gv}
and $\sqrt{s}=13$ TeV \cite{Atlas-Conf:2015}. These observations
are intriguing, but it has not been checked whether these are 
true multi-particle correlation, as was done in the case of $pPb$
collisions \cite{Khachatryan:2015waa}.

\item Large photon elliptic flow: The photon $v_2(p_T)$ has been
measured at RHIC and LHC \cite{Adare:2011zr,Lohner:2012ct}, and the 
result is comparable (within sizable errors) to the elliptic flow 
of light hadrons. This is surprising, because photon emission is 
expected to be dominated by the early stages of the quark-gluon plasma 
evolution before a significant collective flow can develop 
\cite{Chatterjee:2005de}. The discrepancy between theory and
experiment is smaller at the LHC compared to RHIC, and it is further 
reduced by significant emission in the hadronic phase
\cite{Paquet:2015lta}.

\item Approximate beam energy independence of the charged particle
elliptic flow $v_2(p_T)$: The elliptic flow coefficient of charged 
particles has been measured over a large range of beam energies, from 
the low end of the beam energy scan at RHIC, $\sqrt{s_{NN}}=7.7$ GeV, 
to the initial LHC energy $\sqrt{s_{NN}}=2.76$ TeV
\cite{Aamodt:2010pa,Chatrchyan:2012ta,Adamczyk:2012ku}. For a given 
centrality the results are essentially beam energy independent. Within 
hydrodynamics this is somewhat surprising because many parameters, such 
as the lifetime of the system and $\eta/s$ are definitely changing. The 
observed universality could be accidental, because both the $v_2$ of 
identified particles, and the $p_T$ integrated $v_2$ do show beam 
energy dependence. 

\item Anomalous hydrodynamics: Several novel fluid dynamic effects 
have been discovered in recent years. An example is the chiral magnetic 
effect. Topological charge fluctuations in the initial state of a heavy 
ion collision, combined with the magnetic field generated by the highly 
charged ions, can manifest themselves in electric charge fluctuations 
in the final state  \cite{Kharzeev:2007jp}. This effect is now understood 
as part of a broader class of anomalous hydrodynamic effects 
\cite{Kharzeev:2010gr}. An interesting recent observation is a measurement 
of charge dependent elliptic flow at RHIC \cite{Adamczyk:2015eqo}, 
which could be interpreted as a manifestation of a new hydrodynamic
mode, a chiral magnetic wave \cite{Kharzeev:2010gd}.
\end{itemize}

%%%%%%%%%%%%%%%%%%%%%%%%%%%%%%%%%%%%%%%%%%%%%%%%%%%%%%%%%%%%%%%%%%%%%%%%%%%%%
\section{Correlations and Fluctuations}
\label{sec:fluct}
%%%%%%%%%%%%%%%%%%%%%%%%%%%%%%%%%%%%%%%%%%%%%%%%%%%%%%%%%%%%%%%%%%%%%%%%%%%%%

Fluctuations and correlations are important characteristics of any
physical system. They provide essential information about the
effective degrees of freedom and their possible  quasi-particle nature.

 In general, one can distinguish between several types of fluctuations. 
On the most fundamental level there are quantum fluctuations, which 
arise if we measure several non-commuting observables, or an observable
that does not commute with the Hamiltonian. In a system that thermalizes
we encounter thermal fluctuations. These reflect the fact that 
thermodynamics and hydrodynamics are coarse-grained descriptions, 
and that thermodynamic variables necessarily fluctuate in a finite 
sub-volume. An example is given by density fluctuations, which are 
controlled by the compressibility of the system. Finally, the dynamical 
evolution of a system may amplify small quantum or thermal fluctuations 
in the initial state. 

 In  heavy-ion collisions, we encounter fluctuations and correlations
related to the initial state of the system, fluctuations reflecting the
subsequent evolution of the systems, and trivial fluctuations induced 
by the experimental measurement process. Initial state fluctuations
are inhomgeneities in the initial energy and baryon number deposition,
see Fig.~\ref{fig_init}. These fluctuations are quite substantial,
and are reflected in higher harmonics of the radial flow field. If
the systems thermalizes and is described by fluid dynamics then 
we expect that fluctuations in the subsequent evolution are mostly 
thermal. Thermal fluctuations are typically small, suppressed by 
$1/\sqrt{N}$ where $N$ is the average number of particles in the 
volume considered. However, thermal fluctuations can become large
in the vicinity of a second order phase transition. This is the phenomenon
of critical opalescence. Finally, fluctuations related to the detectors
need to be understood, controlled and subtracted in order to access the 
dynamical fluctuations which tell as about the properties of the system.

A well known example for fluctuations in a physical system are those
of the cosmic microwave background first seen by the COBE satellite
\cite{Smoot:2007zz} and later refined by WMAP \cite{Spergel:2006hy}
and, most recently, by the Planck satellite \cite{Ade:2013zuv}. In
case of the cosmic microwave background the observed fluctuations are
at the level of $10^{-4 }$ with respect to the thermal background. In
addition a large dipole correlation due to the motion of earth through
the heat bath of the microwave background is observed.  In heavy ion
collision we are faced with a qualitatively similar situation. To leading
order the observed particles follow a thermal  distribution embedded
in a Hubble-like radial flow field. In addition, for non-central
collisions one observes a quadrupole correlation due to elliptic flow.
 
Experimentally fluctuations are most effectively studied by measuring
so-called event-by-event (E-by-E) fluctuations, where a given observable 
is measured on an event-by-event basis and its fluctuations are studied 
for the ensemble of events. Alternatively, one may analyze the appropriate 
multi-particle correlations measured over the same region in phase space 
\cite{Bialas:1999tv}.

\subsection{Fluctuations in a thermal system}
\label{sec:fluct:thermal}

As discussed in Section~\ref{sec:PJ_thermal} there is good evidence
that the system created in a ultrarelativistic heavy ion collision 
is, to a very good approximation, in thermal equilibrium. Therefore, 
let us start our discussion with thermal fluctuations. These are 
characterized by the appropriate cumulants of the partition function 
or, equivalently, by equal-time correlation functions which in turn 
correspond to the space-like (static) responses of the system.  

In the following we will concentrate on fluctuations or cumulants of 
conserved charges, such as baryon number and electric charge. Therefore, 
we will work within the grand-canonical ensemble, where the system is in
contact with an energy and ``charge'' reservoir. Consequently, the
energy and the various charges are only conserved on the average with 
their mean values being controlled by the temperature and the various
chemical potentials. As far as heavy ion reactions are concerned,
the grand canonical ensemble appears to be a good choice as long as
one only considers a sufficiently small subsystem of the entire final
state, and, 
as discussed in Section~\ref{sec:PJ_thermal}, the final state hadron
yields are very well described by a grand canonical thermal
system of hadrons.  

Fluctuations of  conserved charges are characterized by the cumulants
or susceptibilities of that charge. Given the partition function of
the system with conserved charges $Q_i$ 
\begin{align}
Z={\rm Tr} \left[ \exp \left(-\frac{H-\sum_{i} \mu_i Q_i}{T} \right) \right] 
\end{align}
the susceptibilities are defined as the derivatives with respect to
the appropriate chemical potentials. In case of three flavor QCD 
the conserved charges are the baryon number, strangeness and electric
charge, $\left( B,S,Q \right)$, and we have  
\begin{align}
\chi_{n_B,n_S,n_Q}^{B,S,Q} \equiv 
\frac{1}{V T^{3}} 
\frac{\partial^{n_B} }{\partial (\mu_B/T)^{n_B}} 
\frac{\partial^{n_S} }{\partial (\mu_j/T)^{n_S}} 
\frac{\partial^{n_Q} }{\partial (\mu_Q/T)^{n_Q}} \ln Z.
\label{eq:fluct:susz_genreal}
\end{align}
The above susceptibilities\footnote{Here we adopt the
normalization commonly used in the lattice QCD literature which
differs from other normalization e.g. in \cite{Koch:2008ia}.}
may also be expressed in terms of of
derivatives of the pressure $P= T/V \ln(Z)$
\begin{align}
\chi_{n_B,n_S,n_Q}^{B,S,Q} = 
\frac{\partial^{n_B} }{\partial (\mu_B/T)^{n_B}} 
\frac{\partial^{n_S} }{\partial (\mu_j/T)^{n_S}} 
\frac{\partial^{n_Q} }{\partial (\mu_Q/T)^{n_Q}} \left( \frac{P}{T^4}
  \right)
  \label{fluct:eq:cum_form_pressure}
\end{align}
Consequently, these susceptibilities also control the pressure at
small values of the various chemical potentials. For
example, at small baryon number chemical potential, $\mu_b/T < 1$,
the pressure may be expressed in terms of a Taylor series  
\begin{align}
  \frac{P\left( T,\mu_B \right)}{T^4} = 
  \frac{P\left( T, \mu_B=0 \right)}{T^4} + \sum_n c_n \, \left( \mu/T
  \right)^{n}
  \label{fluct:eq:pressure_mu}
\end{align}
where the expansion coefficients are given by the baryon-number
susceptibilities 
\begin{align}
  c_n=  \frac{ \chi_{n}^{B}}{n!} 
\end{align}
Due to the fermion sign problem, at present lattice QCD calculations
can only be reliably carried out at vanishing chemical
potentials. Therefore, the above Taylor expansion for the pressure is
employed in order to determine the QCD equation of state for small chemical
potentials \cite{Allton:2002zi,Gavai:2008zr,Borsanyi:2012cr}. Meanwhile
many susceptibilities at various orders and various combinations of
conserved charges have
been calculated in lattice QCD. In the following we will discuss a
selection of these results and their interpretation also in the
context of experiment.

%%%%%%%%%%%%%%%%%%%%%%%%%%%%%%%%%%%%%%%%%%%%%%%%%%%%%%%%%%%%%%%%%%%%%%%%%%%%%
\subsubsection{Example: Net Charge Fluctuations}
\label{sec:fluct:charge}
%%%%%%%%%%%%%%%%%%%%%%%%%%%%%%%%%%%%%%%%%%%%%%%%%%%%%%%%%%%%%%%%%%%%%%%%%%%%%

 To illustrate how fluctuations may be utilized to explore the relevant
degrees of freedom, let us briefly discuss the fluctuations of the
electric charge. In Refs. \cite{Jeon:2000wg,Asakawa:2000wh} it has
been realized that the electric charge of  particles contributes in square
to the fluctuations of the net-charge. Therefore, cumulants of the
net-charge are in principle sensitive to the fractional charge of 
quarks in a quark-gluon plasma. This can be easily seen by considering 
the variance of the net charge of a gas of uncorrelated
particles with charge $q$, 
\begin{eqnarray}
\ave{\left( \delta Q \right)^2} = q^2 \ave{\left( \delta N \right)^2}
  = q^2 \ave{N}, 
\end{eqnarray}
where in the last step we have, for simplicity, assumed that the particle 
number follows a Poisson distribution. Since the variance depends not 
only on the squared charge of the particles but also on the number of 
particles, it is advantageous to scale the charge variance by another 
extensive quantity, such as the entropy, $S$, so that the ratio
\begin{align}
\label{R-def}
R = \frac{\ave{\left( \delta Q \right)^2} }{S}
\end{align}
does not depend on the size of the system. A simple estimate using 
Boltzmann statistics gives \cite{Jeon:2000wg,Jeon:2003gk}  
\begin{align}
R_{QGP} = \frac{1}{24} 
\end{align}
for a two flavor quark-gluon plasma whereas for 
a gas of massless pions we get 
\begin{align}
R_{\pi} =  \frac{1}{6}.
\end{align}
In other words, due to the fractional charges of the quarks,
the charge fluctuations per entropy in a QGP is roughly a factor
four smaller than that in a pion gas at the same temperature. 
In reality the hadronic phase is made out of more than pions, and,
taking into account hadronic resonances, the charge variance per
entropy is reduced by about 30\% which still leaves roughly a factor
three difference between a hadronic system and a QGP. Incidentally, 
the fact that charges contribute in
square to fluctuations has been utilized to identify the
fractional charges in a quantum Hall system as well as the double
charge of cooper pairs in measurements of shot noise 
\cite{shot_quantum_hall,shot_cooper}.

%%%%%%%%%%%%%%%%%%%%%%%%%%%%%%%%%%%%%%%%%%%%%%%%%%%%%%%%%%%%%%%%%%%%%%%%%%%%%
\begin{figure}[t]
\begin{center}
\includegraphics[width=0.55\textwidth]{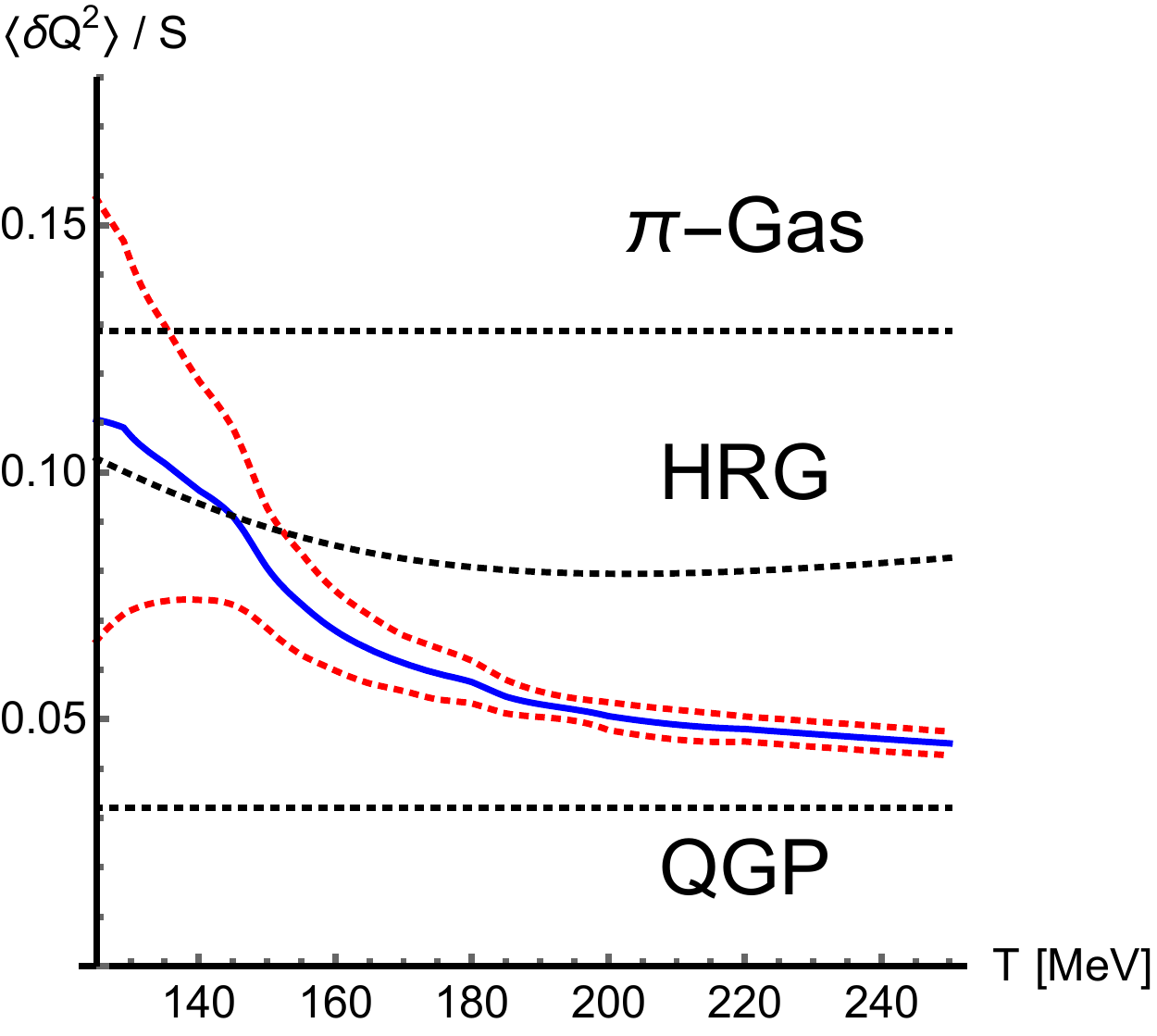} 
\end{center}
\caption{Net-charge variance per entropy, $R$, as a function of
temperature from 2+1 flavor lattice QCD with physical quark masses. 
The red-dashed line indicate the uncertainty. Results for $\ave{\left( 
\delta Q \right)^2}$ are from \cite{Borsanyi:2011sw} and the entropy 
is extracted from \cite{Borsanyi:2010cj}. The dashed horizontal lines 
indicate the results for a massless pion gas, a hadron gas as well 
as a non-interacting QGP with three flavors of massless quarks. }
\label{fig:fluct:charge_fluct_lattice}
\end{figure}
%%%%%%%%%%%%%%%%%%%%%%%%%%%%%%%%%%%%%%%%%%%%%%%%%%%%%%%%%%%%%%%%%%%%%%%%%%%%%

While our simple example is instructive, in reality one has to include
strange quarks and hadrons, quantum statistics, and possible
correlations among quarks or hadrons. Therefore, a realistic
calculation of $R$ will require lattice QCD methods. Since both the
variance of the net charge and the entropy are well defined
thermodynamic quantities this can be done, and in 
Fig.~\ref{fig:fluct:charge_fluct_lattice} we show the lattice QCD
result for the net-charge variance per entropy based on the results 
for the net-charge variance from \cite{Borsanyi:2011sw} and for the 
entropy density from \cite{Borsanyi:2010cj}. We also show the results 
for a free pion gas and a QGP with three flavors of mass-less quarks, 
both using the proper quantum statistics, as well as that for a hadron 
resonance gas. We see that the hadron resonance gas agrees well with the 
lattice results for temperatures up to $T \lesssim 160 \MeV$, which is 
close to the pseudo-critical temperature of $T_{pc}=154 \pm 9 \MeV$. 
For temperatures in the range of  $160 \MeV \lesssim T \lesssim 250 \MeV$ the 
lattice calculations are in between the prediction a resonance gas and that 
of a non-interacting QGP, indicating that some of the correlations 
leading to resonance formation are still present in the system. 
With few exceptions, 
this trend is seen for most quantities which have been calculated 
on the lattice, such as energy density, cumulant ratios etc.: Good 
agreement with the hadron resonance gas up to the critical temperature, 
followed by a rather smooth transition to a free QGP which takes 
place over a temperature interval of approximately $\Delta T \sim 
100 \MeV$, where the correlations slowly disappear.

%%%%%%%%%%%%%%%%%%%%%%%%%%%%%%%%%%%%%%%%%%%%%%%%%%%%%%%%%%%%%%%%%%%%%%%%%%%%%
\subsubsection{Correlations and mixed flavor susceptibilities}
%%%%%%%%%%%%%%%%%%%%%%%%%%%%%%%%%%%%%%%%%%%%%%%%%%%%%%%%%%%%%%%%%%%%%%%%%%%%%

\begin{figure}[t]
\begin{center}
\includegraphics[width=0.7 \textwidth]{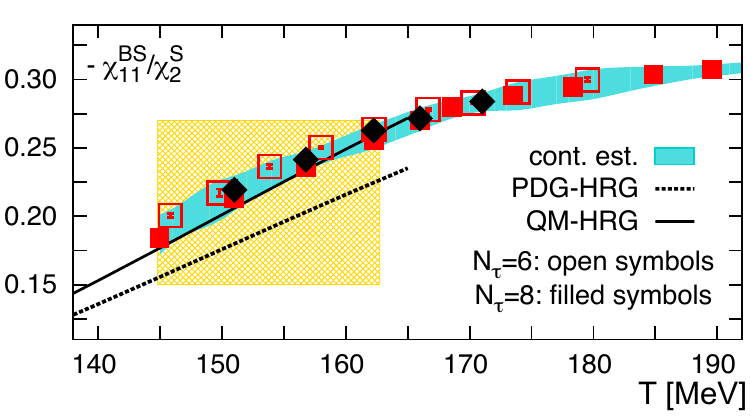}
%\hspace{0.05 \textwidth} 
%\includegraphics[width=0.45 \textwidth]{karsch_end_of_HRG.pdf}
\end{center}
\caption{Lattice QCD results
  for $-\frac{\Xi_{BS}^{11}}{\Xi_{S}^{2}}= \frac{1}{3} C_{BS} $ together with
  results from hadron resonance gas with (full line) and without
  (dashed line) extra strange
  mesons. Figure adapted from   \cite{Bazavov:2014xya}.} 
\label{fig:fluct:cbs}
\end{figure}

Some of these correlations, namely those between the various flavors,
can be explored explicitly by studying so called mixed flavor or 
``off-diagonal'' cumulants. One example is the co-variance between 
strangeness and baryon number, $\ave{\delta B \delta S} \sim
\chi_{1,1}^{B,S}$. Here $S$ refers to strangeness and not, as in the
previous section, to the entropy.
To illustrate the sensitivity of this 
co-variance to correlations among quarks, let us again compare  
a non-interacting QGP with a non-interacting hadron resonance gas (HRG). 
In the QGP strangeness is carried exclusively by baryons, namely the strange
quarks, whereas in a HRG strangeness can also reside in strange mesons. 
Therefore, baryon number and strangeness are more strongly correlated 
in a QGP than in a hadron gas, at least at low baryon number chemical
potential, where the mesons dominate. To quantify this observation, 
Ref.~\cite{Koch:2005vg} proposed the following quantity 
\begin{align}
C_{BS} \equiv - 3  \frac{\ave{\delta B \delta S}}{\ave{\delta S^2}} =
1 + \frac{\ave{\delta u \,\delta s } +\ave{\delta d \,\delta
    s}}{\ave{\delta s^2}},
\label{eq:fluct:CBS}
\end{align}
where we have expressed $C_{BS}$ also in terms of quark degrees of freedom, 
noting that the baryon number of a quark is $1/3$ and the strangeness of 
a s-quark is negative one. Here $(u,d,s)$ represent the net-number of 
up, down and strange quarks, i.e. the difference between up and anti-up 
quarks etc. For a non-interacting QGP, 
$\ave{\delta u \, \delta s} = \ave{\delta d \,\delta s} = 0$, so that
$C_{BS}=1$. 
For a gas of kaons and anti-kaons, on the other hand, where a light (up 
or down) quark is always correlated with a strange anti-quark (kaons) or 
vice versa (anti-kaons) $\ave{\delta u \, \delta s}<0$, resulting in
$C_{BS}<1$. Strange baryons, on the other hand, correlate light quarks
with strange quarks or light anti-quarks with strange anti-quarks, so
that $\ave{\delta u \, \delta s} >0$. Therefore, for sufficiently large
values of the baryon number chemical potential, $C_{BS}>1$ for a
hadron gas, whereas for a non-interacting QGP $C_{BS}=1$ for all
values of the chemical potential \cite{Koch:2005vg}. 
Since  $C_{BS}$ can be expressed in terms of susceptibilities,
$C_{BS}=-3 \frac{\chi_{BS}^{11}}{\chi_{S}^{2}}$, it can and has been calculated 
on the lattice with physical quark masses by two groups \cite{Borsanyi:2011sw,Bazavov:2012jq}. Both
calculations agree with each other, and both report a small, but
significant difference between the lattice results and that from the
hadron resonance gas. In \cite{Bazavov:2014xya} it has been argued that
this discrepancy may be removed by allowing for additional strange
hadrons, which are not in the tables of the Particle Data Group (PDG)
\cite{Agashe:2014kda}, but are predicted by various quark models. 
This is shown in Fig.~\ref{fig:fluct:cbs}, where
the lattice QCD results are compared with a hadron resonance gas based
on all the hadrons in the Review of Particles \cite{Agashe:2014kda}
(dotted line) and a hadron gas with {\em additional} strange hadrons
(full line). Whether or not this turns out to be the correct
explanation, this comparison demonstrates that these cumulant ratios
are a sensitive probe of the relevant microscopic degrees of freedom. 

%%%%%%%%%%%%%%%%%%%%%%%%%%%%%%%%%%%%%%%%%%%%%%%%%%%%%%%%%%%%%%%%%%%%%%%%%%
%%%%%%%%%%%%%%%%%%%%%%%%%%%%%%%%%%%%%%%%%%%%%%%%%%%%%%%%%%%%%%%%%%%%%%%%%%

%%%%%%%%%%%%%%%%%%%%%%%%%%%%%%%%%%%%%%%%%%%%%%%%%%%%%%%%%%%%%%%%%%%%%%%%%%
\begin{figure}[t]
\begin{center}
\includegraphics[width=0.65 \textwidth]{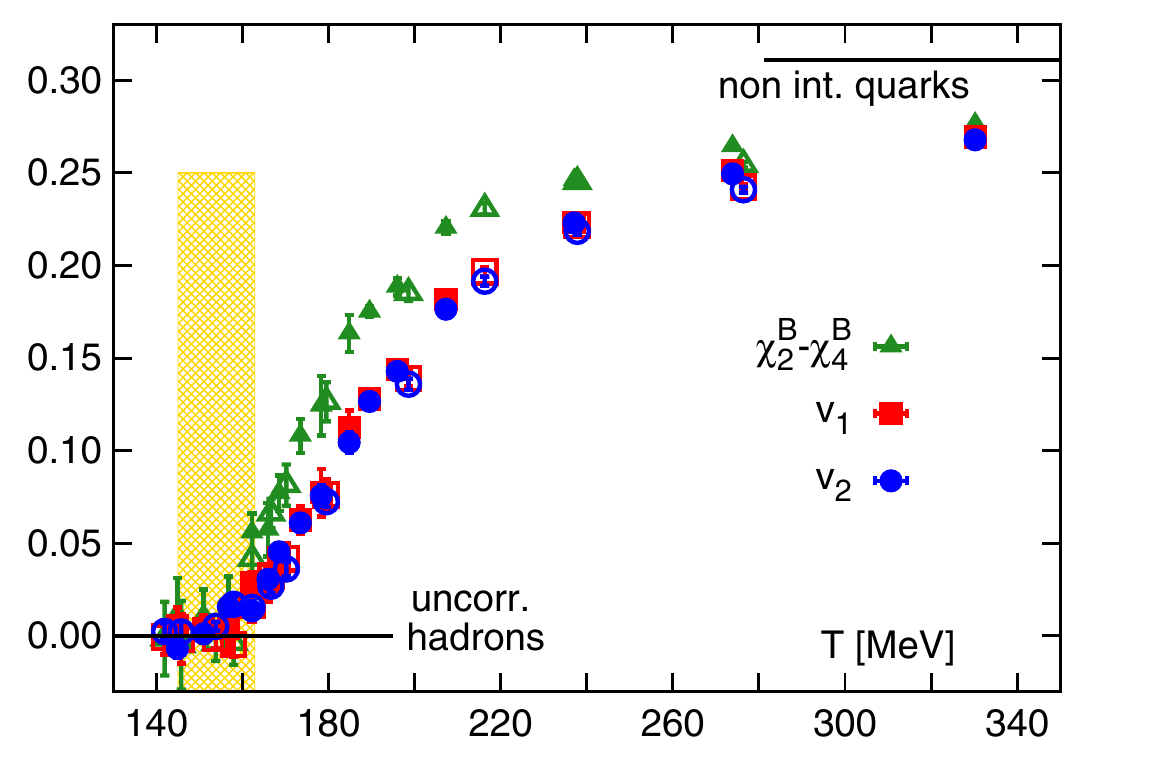}
\end{center}
\caption{Various combinations of higher
  order cumulants which demonstrate the melting of the hadrons. See
  text for details. Figure adapted from \cite{Bazavov:2013dta}.} 
\label{fig:fluct:cbs2}
\end{figure}
%%%%%%%%%%%%%%%%%%%%%%%%%%%%%%%%%%%%%%%%%%%%%%%%%%%%%%%%%%%%%%%%%%%%%%%%%%

To further explore at what temperature a hadronic description fails,
one can study even more involved combinations of cumulant ratios to
project out baryonic or mesonic states
\cite{Bazavov:2014xya,Bazavov:2014yba,Bazavov:2013dta}. 
For example, in a hadron resonance gas, which is well described 
in the Boltzmann approximation, the pressure may be written as 
\begin{align}
P/T^{4} &= 
\sum_{\mathrm{Baryons}\, i} \cosh(\hat{\mu}_B B_i + \hat{\mu}_S S_i +
\hat{\mu}_Q Q_i) \,f(g_i,m_i,T) \non
&+ \sum_{\mathrm{Mesons} \, j} \cosh( \hat{\mu}_S S_j +
\hat{\mu}_Q Q_j) f(g_j,m_j,T), 
\end{align}
where $\mh_{B,S,Q}=\frac{\mu_{B,s,Q}}{T}$ and
$f(g_{i},m_{i},T)=\frac{g_{i}}{2 \pi^{2}}
\frac{m^{2}}{T^{2}}K_{2}(m/T)$. 
Since the baryon number of all baryons in the hadron gas is 
$\pm 1$, the difference
\begin{align}
\chi_{2}^{B}-\chi_{4}^{B}= 0
\end{align}
vanishes in a hadron gas. For a non-interacting QGP, on the other
hand, all
quarks carry baryon number $B_{\mathrm{quark}}=\pm 1/3$, and the difference
between second and fourth order baryon-number cumulant is finite,
\begin{align}
\chi_{2}^{B}-\chi_{4}^{B}= \left(\frac{1}{9} -\frac{1}{81}\right)
  P_{\mathrm{quarks}}/T^{4} >0.
\end{align}
The same holds for other, more complicated, combinations, involving 
strange particles, such as \cite{Bazavov:2013dta}
\begin{align}
  v_{1}&=\chi_{31}^{BS}-\chi_{11}^{BS}\non
  v_{2}&=\frac{1}{3}\left( \chi_{2}^{S}-\chi_{4}^{S} \right)
  -2  \chi_{13}^{BS} -4 \chi_{22}^{BS}-2 \chi_{31}^{BS}
\end{align}
Here, $v_{1}$ and $v_{2}$ represent combinations of cumulants, and,
thus, should not be confused with the moments of the azimuthal
distribution discussed in previous sections, which are commonly
denoted by $v_{n}$ as well. 
In Fig.~\ref{fig:fluct:cbs2} we show the results from
lattice QCD for these various combinations of cumulants as a function
of temperature. All start deviating from the HRG value of zero at
about the same temperature, indicating that both light and strange
hadrons seem to loose their identity at temperatures above $T
\gtrsim 150 \MeV$, which coincides with the pseudo-critical
temperature of the QCD transition. A similar exercise has also been 
carried out for charmed hadrons \cite{Bazavov:2014yba} with the 
somewhat surprising result that even hadrons with open charm seem 
to ``melt'' at the same temperature of $T\sim 150 \MeV$. 

To summarize this Section, we have demonstrated that the cumulants of
conserved charges contain useful information about the correlations
and relevant degrees of freedom of QCD matter. Since they are amenable  
to lattice QCD methods, the insights derived from such
studies are rather model independent.

%%%%%%%%%%%%%%%%%%%%%%%%%%%%%%%%%%%%%%%%%%%%%%%%%%%%%%%%%%%%%%%%%%%%%%%%%%%
\subsection{Non-equilibrium and Experimental considerations}
\label{sec:fluct:exp_issues}
%%%%%%%%%%%%%%%%%%%%%%%%%%%%%%%%%%%%%%%%%%%%%%%%%%%%%%%%%%%%%%%%%%%%%%%%%%%

Given the wealth of information which can be extracted from cumulants
of conserved charges and the fact that they can be determined model
independently, it would be very desirable to measure these cumulants
in heavy ion collisions. However, a heavy ion collision is a highly dynamical
process whereas lattice QCD deals with a static system in global
equilibrium. In addition, real experiments have limitations in
acceptance etc, which are difficult to map onto a lattice QCD
calculation. Consequently a direct comparison of experiment with
lattice QCD results for fluctuation observables is a non-trivial
task. In the following we will discuss various
issues which need to be understood and addressed in order for such a 
comparison to be meaningful.

\begin{itemize}

\item {\bf Dynamical evolution:} 
So far our discussion assumed that the system is static and in global
thermal equilibrium. However, even if fluid dynamics is applicable 
the system is at best in local thermal equilibrium, and in viscous fluid 
dynamics local equilibrium is never complete. The difference between local 
and global thermal equilibrium is an important aspect of the evolution of 
fluctuations of conserved charges, because the amount of conserved charge 
in a given comoving volume can only change by diffusion, and the rate of 
diffusion is limited by causality \cite{Shuryak:2000pd}.
 This observation is central to the use 
of the variable $R$ defined in Eq.~(\ref{R-def}) to detect the presence of 
quark-gluon plasma. If we consider a sufficiently large rapidity window 
$\Delta y$ then the value of $R$ is frozen in during the QGP phase, and 
cannot change in the subsequent hadronic stage. Of course, if $\Delta y$ 
is chosen too large, then $R$ never equilibrates, and reflects properties 
of the initial state. 

 This observation can be made more quantitative using the theory of 
fluctuating hydrodynamics outlined in Section~\ref{sec_hydro_flucs}.
However, so far most theoretical studies have focused on schematic 
models, see, for example \cite{Kitazawa:2013bta}. More importantly, 
there are other physical and experimental considerations that affect 
the choice of $\Delta y$, which we will discuss next.

\item {\bf Global charge conservation:} Obviously, baryon number, 
 electric charge and
  strangeness are conserved globally, i.e. if we detected all
  particles, none of the conserved charges would fluctuate. In
  contrast, lattice QCD calculations are carried out in the grand
  canonical ensemble, which allows for exchange of conserved charges
  with the heat bath. Consequently, charges are conserved only on the
  average and, thus, do fluctuate due to the exchange with the
  heat-bath. These exchanges and thus the fluctuations depend on 
  the correlations between
  particles and, as demonstrated above, on the magnitude of the
  charges of the individual particles.
  Therefore, in order to compare with lattice QCD, one has to mimic a
  grand canonical ensemble in experiment. This can be done by
  analyzing only a subset of the particles in the final
  state. However, even in this case, corrections due to global charge
  conservation are present. These corrections increase with the order
  of the cumulant \cite{Bzdak:2012an} and need to be taken into account 
  as discussed in
  \cite{Kitazawa:2013bta,Koch:2001zn,Aziz:2004qu,Kitazawa:2012at}. 

\item {\bf Finite acceptance:} All real experiments do have a finite
  acceptance, i.e. they are not able to cover all of phase space. In
  addition, most experiments are unable to detect neutrons, which do
  carry baryon number. However, due to rapid isospin exchange
  processes, the lack of neutron detection may be successfully modeled
  by a binomial distribution \cite{Kitazawa:2012at}. While it is
  desirable to study only a subset of particles, in order to mimic a
  grand canonical ensemble, it is mandatory to have sufficient
  coverage in phase space in order to capture all correlations.
  We  note, that at the
  lowest beam energy, $\sqrt{s}=7.7 \GeV$, STAR finds a rather significant
  dependence on the width of the rapidity window for the fourth order
  net-proton cumulant \cite{Luo:2015ewa}. 
\item {\bf Efficiency corrections:} A real world experiment detects a 
  given particle only with a probability, commonly referred to as 
  efficiency  $\epsilon$, which is
  smaller than one, $\epsilon<1$. However, this does not mean that in
  every event one detects the same fraction of produced
  particles. Consequently, the number of measured particles fluctuates
  even if the number of produced particles does not. In other words
  the finite detection efficiency gives rise to fluctuations, which
  need to be removed or unfolded before a comparison with any theoretical
  calculation. If the efficiency follows a binomial distribution,
  analytic formulas for the relation between measured and true
  cumulants can be derived
  \cite{Bzdak:2012ab,Bzdak:2013pha,Luo:2014rea}. 
  These have been applied to the most recent analysis by the STAR
  collaboration.  
\item {\bf Dynamical fluctuations:} A heavy ion collision is a highly
  dynamical process and the initial conditions as well as the time
  evolution may easily give rise to additional
  fluctuations. Especially at lower energies, $\sqrt{s}\lesssim 30
  \GeV$, the incoming nuclei are stopped effectively and deposit
  baryon  number and electric charge in the mid-rapidity
  region. Clearly the amount of baryon number deposited will vary from
  event to event, resulting in fluctuations of the baryon number at
  mid-rapidity, which are not necessarily the same as those of a
  thermal system. This potential source of background needs to be
  understood and removed, especially at low energies where one uses
  higher cumulants of the net proton distribution in order to find
  signals for a possible QCD critical point. Not only does the number
  of baryon and charges fluctuate due to the collision dynamics, so
  does the size of the system. And while ratios of cumulants do not
  depend on the average system size, they are affected by event by
  event fluctuation of the system size. This has been studied in 
  \cite{Skokov:2012ds} and it was found that only for the very most central
  collisions these fluctuations are suppressed. Therefore, any
  measurement of the centrality dependence of cumulant ratios needs to
  be interpreted with care. Alternatively, one can devise observables,
  which are not sensitive to size fluctuation \cite{Koch:2008ia,Jeon:2003gk,Gazdzicki:2013ana,Sangaline:2015bma}.
\end{itemize}

The first three points deserve some additional discussion, as they
pose contradictory demands on the measurement \cite{Koch:2008ia}. 
In order to minimize
corrections from global charge conservation, one wants to keep the
acceptance window $\Delta$, say in rapidity, as small as possible. On the 
other hand, in order to capture the physics, the acceptance window needs 
to be sufficiently wide to catch the correlation among the particles. 
Therefore, if $\sigma$ is the correlation length in rapidity
and $\Delta_{charge}$ the range over which all the charges are distributed, 
then $\Delta/\Delta_{charge}\ll 1$ in order to minimize the effects of
charge conservation, and $\sigma/\Delta\ll 1$ in order to capture the
physics. 

%%%%%%%%%%%%%%%%%%%%%%%%%%%%%%%%%%%%%%%%%%%%%%%%%%%%%%%%%%%%%%%%%%%%%%%%%%%%%%
\begin{figure}[t]
\begin{center}
\includegraphics[width=0.65 \textwidth]{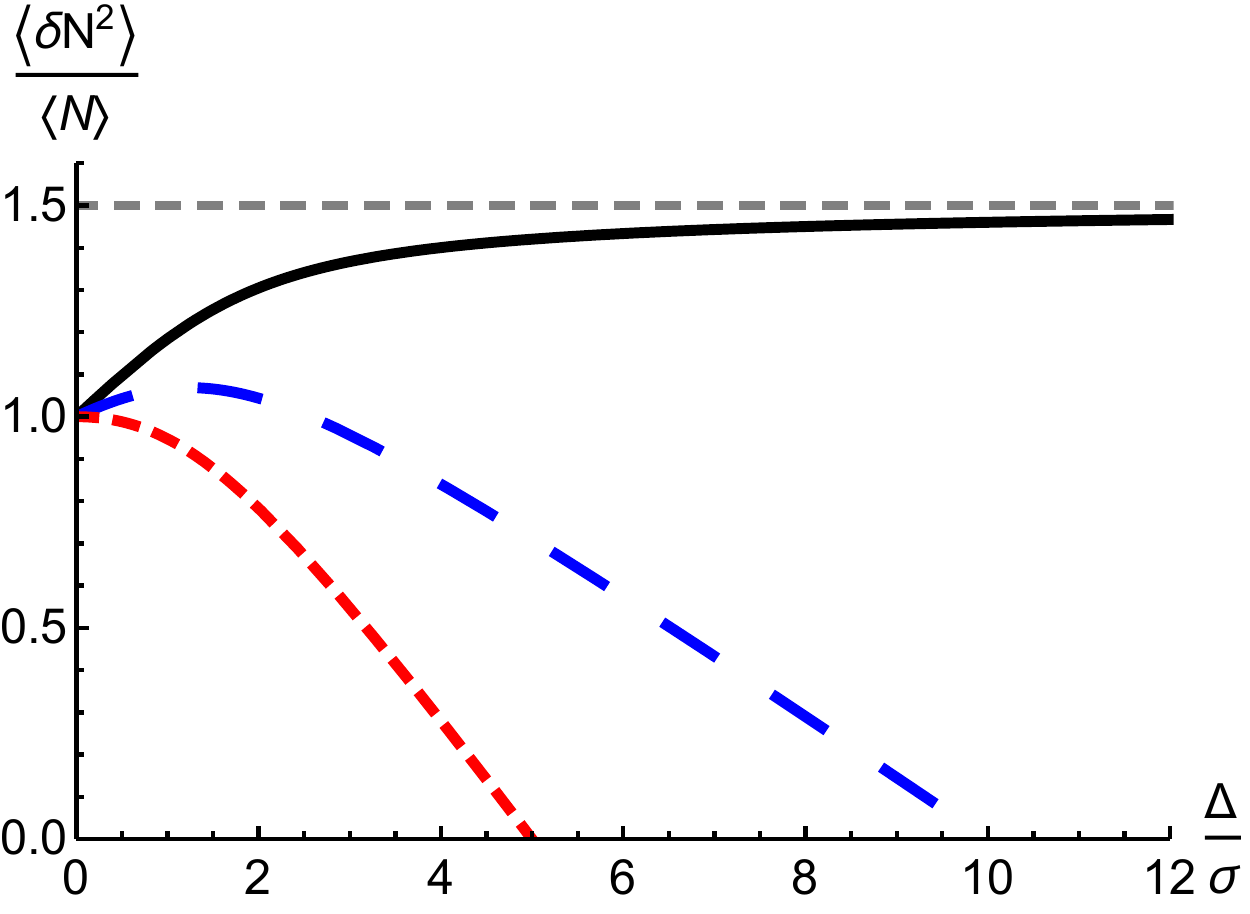}
\end{center}
\caption{Observed scaled variance as a function of the acceptance
  window in units of the correlation length. The full (black) line
  corresponds to an infinite system where global charge conservation
  can be ignored. The long-dashed (blue) and short-dashed (red) line
  correspond the situation where the 
  where the charge is conserved within $(-10\sigma,10\sigma)$ and
  $(-5\sigma,5\sigma)$, respectively.} 
\label{fig:fluct:schematic_correlation}
\end{figure}
%%%%%%%%%%%%%%%%%%%%%%%%%%%%%%%%%%%%%%%%%%%%%%%%%%%%%%%%%%%%%%%%%%%%%%%%%%%%%%

To illustrate this point, let us consider the following schematic model. Let
us define a two-particle correlation function (in rapidity $y$) 
\begin{align}
\ave{n(y_{1})\left( n(y_{2}) -\delta\left( y_{1}-y_{2} \right)
  \right))}=\ave{n \left( y_{1} \right)}\ave{n\left( y_{2}
  \right)}\left( 1+ C\left( y_{1},y_{2} \right) \right)  
\end{align}
Then the (acceptance dependent) scaled variance of the particle number 
is given by
\begin{align}
  \frac{\ave{\delta N^{2}}}{\ave{N}}=1 + \int_{-\Delta/2}^{\Delta/2}
  dy_{1}\,dy_{2}\,C\left( y_{1},y_{2} \right) 
\label{fluct:eq:schematic_correlate}
\end{align}
where the acceptance in rapidity is given by $-\Delta/2<y<\Delta/2$. Using 
a simple Gaussian for the correlation function 
\begin{align}
  C\left( y_{1},y_{2} \right) = 
C_{0}\exp\left( -\frac{\left(y_{1}-y_{2}\right)^{2}}{2\sigma^{2}}\right) 
\end{align}
in Fig.~\ref{fig:fluct:schematic_correlation}
we show the scaled variance as a function of the
size of the acceptance window in units of the correlation length
$\Delta/\sigma$. The black line is simply the expression of
Eq.~\ref{fluct:eq:schematic_correlate}, where we have ignored any
effects due to global charge conservation,
i.e. $\Delta_{charge}\rightarrow \infty$. The red and blue dashed
lines represent the situation where were the total charge is
distributed over a range of $\Delta_{charge}/\sigma \le 5 $ and
$\Delta_{charge}/\sigma \le 10 $, respectively. Here we used the leading order
formulas of \cite{Bleicher:2000ek} to account for charge conservation
noting that a more sophisticated treatment a la \cite{Sakaida:2014pya}
would not change the picture qualitatively. Lattice QCD and model
calculations, on the other hand would give the asymptotic value 
indicated by the dashed gray line, which we have chosen to be
$\frac{\ave{\delta N^{2}}}{\ave{N}}=1.5$. 
The obvious lesson from this exercise is that a comparison of a
measurement at one single acceptance window $\Delta$ with any model
calculation is rather meaningless. Instead, one needs to measure the
cumulants for various values of $\Delta$, and remove the effect of charge
conservation. If the subsequent results trend towards an asymptotic
value for large $\Delta$, it is this value which should be compared
with model and lattice calculations. Such a program has been carried by
the ALICE collaboration in order to extract the aforementioned charge
fluctuations \cite{Abelev:2012pv}.  

%%%%%%%%%%%%%%%%%%%%%%%%%%%%%%%%%%%%%%%%%%%%%%%%%%%%%%%%%%%%%%%%%%%%%%%%%%%
\subsection{Freeze-out conditions}
\label{fluct:sec:freeze}
%%%%%%%%%%%%%%%%%%%%%%%%%%%%%%%%%%%%%%%%%%%%%%%%%%%%%%%%%%%%%%%%%%%%%%%%%%%

As we have discussed in Section~\ref{sec:PJ_thermal} the hadron
resonance gas is very successful in describing the (chemical)
freeze out conditions of a heavy ion collision. In addition, with a few
exceptions such as $C_{BS}$, lattice
QCD calculations for the various cumulants agree very well with
the HRG prediction for temperatures up to $T \sim 150 \MeV$. However,
since the abundance of hadrons, such as the pion number, is not a well
defined concept in an interacting thermal field theory, it would be
desirable to extract the freeze out conditions by  direct comparison of
lattice QCD calculations with experimental data, thus, avoiding the,
albeit successful, HRG model as an intermediate step. As first
suggested in \cite{Karsch:2012wm,Bazavov:2012vg} this goal can be
achieved by 
comparing ratios of cumulants of conserved charges. 
The cumulants of the distribution of conserved charges are
well defined in thermal field theory and they can, in principle, be
measured in experiment, although the issues raised in the previous
Section need to be resolved for such a comparison to be meaningful.

Since the cumulants depend both on the temperature and baryon number
chemical potential, the two main parameters characterizing the chemical
freeze out, a comparison with experiment should be able to
constrain both of them. Following the specific strategy proposed in
\cite{Karsch:2012wm,Bazavov:2012vg} one first 
extracts the dependence of the cumulants on the chemical potential
$\mh_{X}=(\mh_{B},\mh_{Q},\mh_{S})$. Since the cumulants are
derivatives of the pressure, Eq.~\eqref{fluct:eq:cum_form_pressure}, 
all that is needed is the pressure at finite chemical potential, which
is given by the Taylor expansion, Eq.~\eqref{fluct:eq:pressure_mu}.
For example, to leading order in the chemical potentials,
the first order cumulant or the mean value of the net charge,
$M_Q$, is given by 
\begin{align}
  M_{Q}(T,\mh_{B},\mh_{Q},\mh_{S}) &= \frac{1}{T^{4}}\frac{\partial }{\partial
  \mh_{Q}} P(T,\mh_{B},\mh_{Q},\mh_{S}) \non
  &= \frac{1}{T^{4}}\frac{\partial }{\partial  \mh_{Q}} \left( P(T,0) + \frac{\partial
  P }{\partial \mh_{B}} \mh_{B}+ \frac{\partial P}{\partial \mh_{Q}}
  \mh_{Q} + \frac{\partial P}{\partial \mh_{S}} \mh_{S}\right)\non
  &= \chi_{11}^{BQ}\mh_{B}+\chi_{2}^{Q}\mh_{Q} + \chi_{11}^{QS}
    \mh_{S}.
\label{fluct:eq:mean_charge_mu}
\end{align}
and similar for the mean net baryon number $M_{B}$ and net strangeness
$M_{S}$. Here, the cumulants in the last line are evaluated at vanishing 
chemical potential and thus are accessible to lattice QCD methods.
Similar expressions can be derived for other, higher order
cumulants. 

\begin{figure}[t]
\begin{center}
\includegraphics[width=0.48 \textwidth]{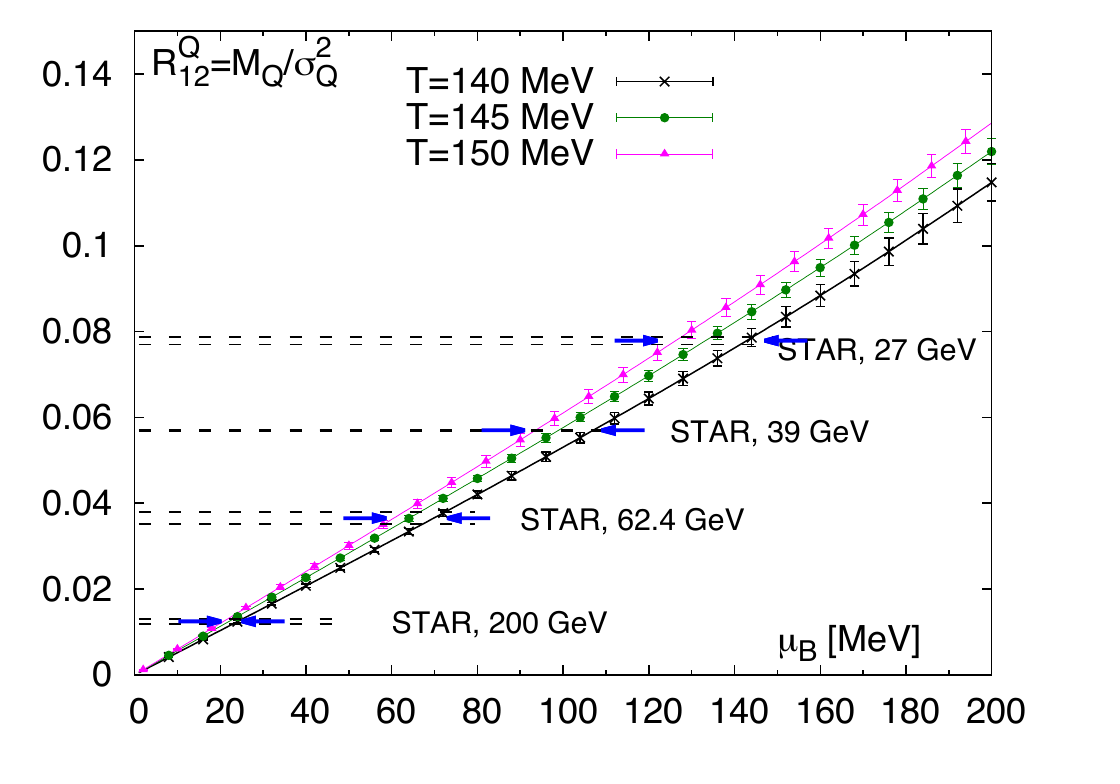}
\includegraphics[width=0.48 \textwidth]{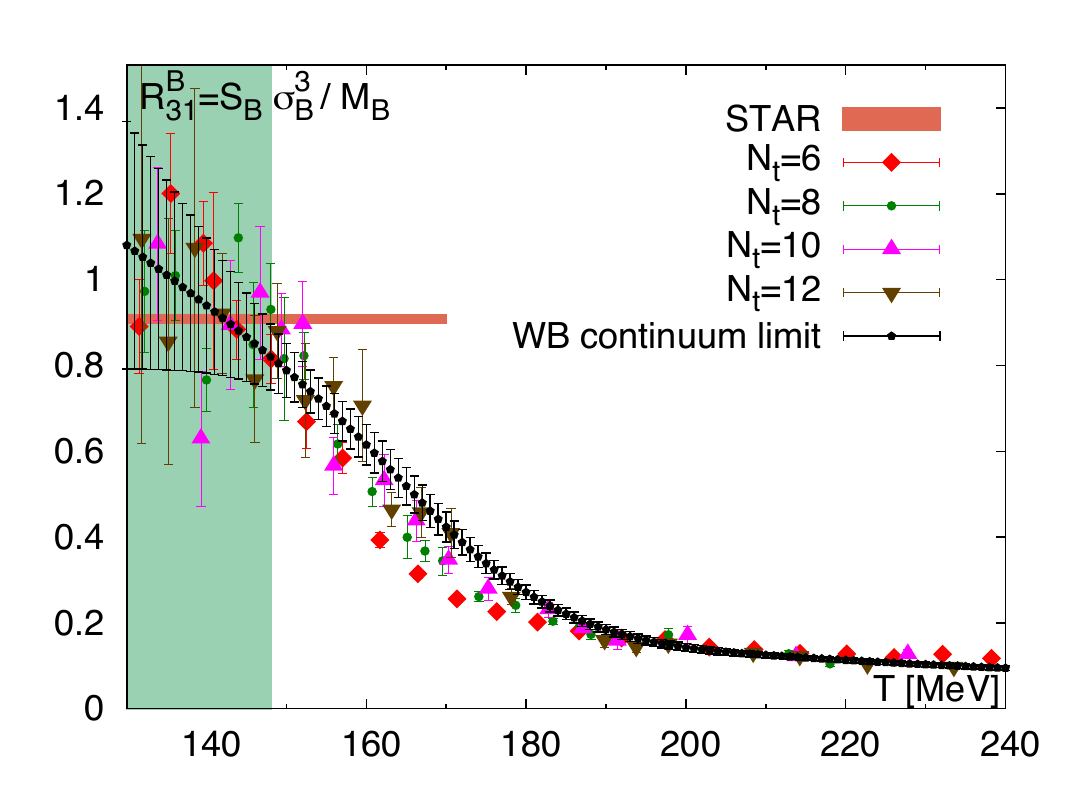}
\end{center}
\caption{Comparison of lattice results with STAR data
\cite{Adamczyk:2013dal,Adamczyk:2014fia} for $R_{12}^{Q}$ (left
panel) and $R_{32}^{B} $ (right panel). Figures adapted from 
\cite{Borsanyi:2014ewa}.} 
\label{fig:fluct:freeze_out}
\end{figure}

\begin{figure}[t]
\begin{center}
\includegraphics[width=0.7 \textwidth]{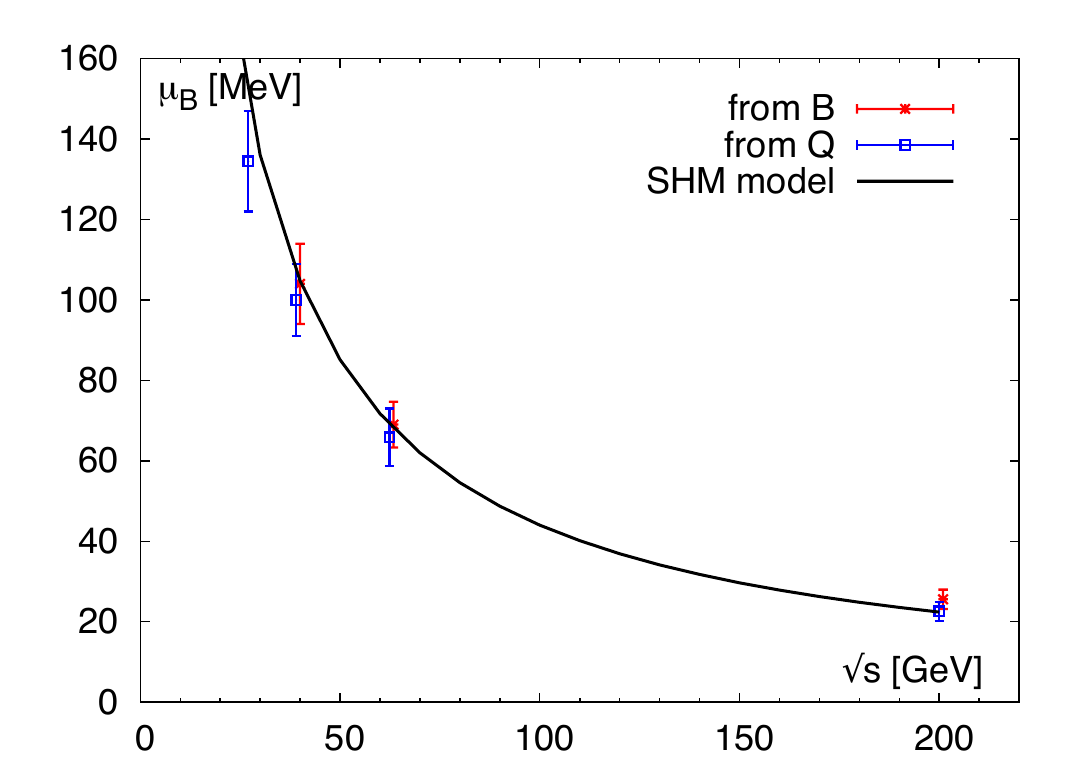}
\end{center}
\caption{Comparison of lattice results with STAR data
\cite{Adamczyk:2013dal,Adamczyk:2014fia} for charge and baryon 
number fluctuations. In this figure we show the  extracted freeze 
out chemical potential in comparison with the HRG model (full line). 
Figures adapted from \cite{Borsanyi:2014ewa}.} 
\label{fig:fluct:freeze_out2}
\end{figure}

In a heavy ion collision, no net strangeness is produced, i.e. $M_{S}=0$. 
Also, the ratio of electric charge over baryon number, $r=M_{Q}/M_{B}$,
can be determined by experiment and is likely close to that of the
incoming nuclei, $r \simeq 0.4$. These two constraints relate  
the charge and strangeness chemical potential to the baryon number
chemical potential. Consequently, the cumulants depend only on the
temperature and baryon number chemical potential, which then can
be extracted from the comparison with experiment. 
In \cite{Bazavov:2012vg}, the authors proposed to use the ratios of
mean over variance and that of third order cumulant over variance
for such a comparison
\begin{align}
  R_{12}^{X}=\frac{M_{X}\left( T,\mh_{B} \right)}{\sigma_{X}\left(
  T,\mh_{B} \right)},\,\,\,\,\,\,\, 
  R_{32}^{X}=\frac{\chi_{3}^{X}\left( T,\mh_{B} \right)}{\sigma_{X}\left(
  T,\mh_{B} \right)}.
\end{align}
Here $X$ stands for charge or net baryon number. The first ratio,
$R_{12}$ depends strongly on the baryon number chemical potential
$\mh_{B}$ whereas the $R_{32}$ has only a mild dependence on the
$\mh_{B}$, as can seen in the Boltzmann limit where
$R_{12}\sim \sinh(\mh_{B})/\cosh(\mh_{B})\sim \mh_{B}$ and $R_{32}\sim
\cosh(\mh_{B})/\cosh(\mh_{B})=1$. Therefore, $R_{12}$ 
constrains the chemical potential and $R_{32}$ the
temperature. This is shown in Figs.~\ref{fig:fluct:freeze_out} and \ref{fig:fluct:freeze_out2}  where
we show the result of \cite{Borsanyi:2013hza,Borsanyi:2014ewa}, who used 
the methods
of \cite{Bazavov:2012vg} together with their own lattice
calculations in order to compare with the data of the STAR collaboration 
\cite{Adamczyk:2013dal,Adamczyk:2014fia}. 
While the chemical potential can be determined rather well from
$R_{12}^{Q}$ (left panel of Fig.~\ref{fig:fluct:freeze_out}) and it agrees well with the HRG model (full
line in Fig.~\ref{fig:fluct:freeze_out2}), the large errors in the lattice
calculations for $R_{32}^{B}$ allow only for an upper limit of the
freeze-out temperature, $T_{f}<148 \MeV$ (right panel of Fig.~\ref{fig:fluct:freeze_out} ). The 
freeze out temperature may be determined better if one additionally assumes
that electric charge and  baryon number freeze out
at the same temperature. In this case the double ratio
$R_{12}^{Q}/R_{12}^{B}$ constrains the freeze out temperature to
within $T_{f}=144\pm 6 \MeV$ \cite{Borsanyi:2014ewa}. 
Overall the extracted freeze out parameters from the comparison of
cumulant ratios agree remarkably well with those obtained from the HRG
analysis. 

%%%%%%%%%%%%%%%%%%%%%%%%%%%%%%%%%%%%%%%%%%%%%%%%%%%%%%%%%%%%%%%%%%%%%%%%%%%%%%
%%%%%%%%%%%%%%%%%%%%%%%%%%%%%%%%%%%%%%%%%%%%%%%%%%%%%%%%%%%%%%%%%%%%%%%%%%%%%%

With regards to our discussion in the previous Section, we note that
the measured cumulants have not been corrected for global charge and 
baryon number
conservation and have been obtained for a fixed acceptance window. 
In addition, the experiment measures the net proton
cumulants which are compared with net baryon cumulants from the
lattice. Therefore, the fact that the freeze out parameters agree
rather well with those obtained from and HRG analysis, is somewhat
surprising. 
However, if the system at freeze-out approximately follows Poisson statistics,
i.e. correlations are negligible, as is the case in the HRG model,
then the lack of neutron detection and the effect of the acceptance
window cancel in the ratios and we recover the HRG values. A recent
comparison of data taken at the LHC by the ALICE collaboration with
lattice cumulants seems to support this possibility 
\cite{Braun-Munzinger:2014lba}. In this paper, the authors assumed
that the net-charges follow a Skellam distribution, i.e. absence of
any correlations. The various cumulants are then determined by
combinations of mean values, and, using the particle yields measured by
ALICE, they found a very good agreement with the lattice calculation.
However, only a careful analysis of the
experimental data along the lines discussed above will be able to
verify, if indeed the particles at freeze out follow Poisson statistics.

%%%%%%%%%%%%%%%%%%%%%%%%%%%%%%%%%%%%%%%%%%%%%%%%%%%%%%%%%%%%%%%%%%%%%%%%%%%%%%
%%%%%%%%%%%%%%%%%%%%%%%%%%%%%%%%%%%%%%%%%%%%%%%%%%%%%%%%%%%%%%%%%%%%%%%%%%%%%%

%%%%%%%%%%%%%%%%%%%%%%%%%%%%%%%%%%%%%%%%%%%%%%%%%%%%%%%%%%%%%%%%%%%%%%%%%%%%%
\subsection{Fluctuations and the QCD Phase diagram}
\label{fluct:sec:critical}
%%%%%%%%%%%%%%%%%%%%%%%%%%%%%%%%%%%%%%%%%%%%%%%%%%%%%%%%%%%%%%%%%%%%%%%%%%%%%

 The phase diagram of QCD has potentially a very rich structure, 
see Fig.~\ref{fig_qcd_phase}. As discussed in Section~\ref{sec_EOS}
one of the more remarkable aspects of the phase diagram is the 
possible existence of a critical endpoint of the chiral transition. 
The idea about a possible co-existence region and critical point has
spawned a dedicated program at the Relativistic Heavy Ion Collider
(RHIC) at Brookhaven National Laboratory, where the so called beam
energy scan (BES) tries to scan the phase diagram by colliding nuclei
over the whole range of collision energies available at RHIC. As
discussed in Section~\ref{sec:PJ_thermal}, by lowering the beam energy one
increases the baryon number chemical potential of the system created
in these collisions, and the energy range of RHIC allows to scan the
phase diagram over the range of $0\lesssim \mu_{B} \lesssim 400 \MeV$
\cite{Odyniec:2013aaa}.

The measurement of fluctuations, in particular cumulants of the net baryon
number, play a central role in the experimental search for a possible 
critical point. At the critical point, where we have a second order 
phase transition, the correlation length $\xi$ diverges for a static 
system of infinite size, resulting in diverging cumulants of the baryon number
\cite{Stephanov:1998dy,Stephanov:1999zu}. However, the system created
in a heavy ion collision has finite size as well as a finite lifetime
which limits the maximum correlation length, and, as shown in
\cite{Berdnikov:1999ph}, due to critical slowing down, it is
the finite lifetime which restricts the  correlation length to $\xi
\simeq 2$ fm. Thus, instead on diverging susceptibilities, 
one will only observe moderate enhancements. These enhancements are 
larger for higher order cumulants, as pointed out in
\cite{Stephanov:2008qz} where it was shown that the fourth order
cumulant scales with the seventh power of the correlation length
$\chi_{4}^{B}\sim \xi^{7}$, while the second order scales like the
square $\chi_{B}\sim \xi^{2}$. Consequently the ratio, $R_{42}=
\chi_{4}^{B}/\chi_{2}^{B} \sim \xi^{5}$  grows with a large
power of $\xi$.

 More recently is was realized that in the region around the critical 
point $R_{42}$ may be enhanced or reduced, and that a typical trajectory 
of freeze-out points as a function of beam energy would suggest the 
following scenario: With decreasing beam energy $R_{42}$ is initially 
below the Poisson baseline $R_{42}^{\mathrm{Poisson}}=1$, then rises above 
it, and finally returns to the baseline value $R_{42}=1$ as kinetic
freeze-out occurs outside the critical region in the hadronic phase 
\cite{Skokov:2010uh,Stephanov:2011pb}. A first measurement by the STAR 
collaboration of the beam energy dependence of various cumulant ratios 
\cite{Adamczyk:2013dal} indeed showed a small decrease below the baseline 
for energies $\sqrt{s}<20 \GeV$. However, this observation may not 
be significant because a similar reduction is also seen in hadronic 
event-generators, and thus may simply be an effect of baryon number 
conservation. On the other hand, recent preliminary data, which also 
include  protons at higher transverse momentum, show a small dip 
followed by a large increase at the lowest beam energy $\sqrt{s}= 7.7 
\GeV$ \cite{Luo:2015ewa}. Although the data have significant statistical 
errors, these results are rather intriguing and beg for a measurement 
at even lower energies. While statistics will be improved during the 
second phase of the RHIC beam energy scan, a measurement at energies 
below $\sqrt{s}= 7.7 \GeV$ will have to be carried out at another
facility. One should also point out that the enhancement of $R_{42}$
observed in the preliminary STAR data is due to protons with
transverse momentum $p_{t}> 800 \MeV$ and that the enhancement
increases with increased rapidity coverage. While the latter is to be
expected following our discussion in Section~\ref{sec:fluct:exp_issues}, the
former requires more scrutiny, since naively one would expect critical
fluctuations to arise from low momentum modes.

 Finally, we should remark that a first order transition will also 
give rise to enhanced fluctuation of the baryon number. Especially if 
one enters the mechanically unstable spinodal region, as it is possible 
in a dynamic process such as a heavy ion collision. The associated 
instabilities give rise to the rapid formation of blobs of high 
density matter \cite{Randrup:2009gp}, which should reflect itself
in enhanced fluctuations of the baryon number. While the formation of
such blobs of matter has been demonstrated in model calculations 
\cite{Steinheimer:2012gc,Steinheimer:2013xxa}, their effect on baryon
number cumulants has not yet been investigated quantitatively.

 Aside from  exploring the phase diagram at large baryon density, fluctuation 
measurements also allow to experimentally address the nature of the 
cross-over transition at vanishing baryon number chemical potential. 
It is commonly believed that the  cross-over transition is a remnant 
of the $O(4)$ chiral transition which in the limit of vanishing quark 
or pion masses would be of second order \cite{Pisarski:1983ms}.  
In this case, one can make quite general predictions about the baryon 
cumulants at vanishing chemical potential, as discussed in some detail 
e.g. in \cite{Friman:2011pf}. Essential for this to work is the fact 
that the so-called pseudo critical line (black dashed line in 
Fig.~\ref{fig_qcd_phase}) defined by the inflection point of the chiral 
condensate $\ave{\bar{\psi}\psi}$ \cite{Aoki:2009sc,Bazavov:2011nk}, has 
a finite curvature, $T_{pc}(\mu_{B}) = T_{pc}(\mu_{B}=0) \left[ 1 -
  \kappa \,\mh^{2}_{B} \right]$ with $\kappa = 0.0066$
\cite{Endrodi:2011gv,Kaczmarek:2011zz}. Consequently, close to the
pseudo-critical line one may write the free energy as a function of an
effective reduced ``temperature'' $t=\left[(T-T_{c})/T_{c} + \kappa \mh_{B}^{2}\right]$
\begin{align}
  F(T,\mu_{B}) = F(t).
\end{align}
As a result, derivatives with respect to the chemical potential can be
can be related with derivatives with respect to the temperature. Since
the effective temperature has quadratic dependence on $\mh_{B}$, the
$n$-th derivative with respect to $\mh_{B}$  corresponds to a
derivative of order $n/2$ with respect to the temperature. 
For example,
\begin{align}
  \frac{\partial^{2} }{\partial \mh_{B}^{2}}F = 2 T \kappa
  \frac{\partial }{\partial T} F.
\end{align}
With this relation one can then apply the results of $O(4)$ scaling
theory directly to the baryon number cumulants, which in turn can be
tested by experiment. One of the predictions is, for example, that the
ratio of the sixth over the second order cumulant is negative right
below the freeze out temperature, $R_{62}^{B}(T=T_{f})<0 $, which in
principle is accessible to experiments at the LHC. For a detailed
discussion we refer to \cite{Friman:2011pf}.

%%%%%%%%%%%%%%%%%%%%%%%%%%%%%%%%%%%%%%%%%%%%%%%%%%%%%%%%%%%%%%%%%%%%
\section{Hadrons in a hot and dense medium}
\label{sec_hadrons}
%%%%%%%%%%%%%%%%%%%%%%%%%%%%%%%%%%%%%%%%%%%%%%%%%%%%%%%%%%%%%%%%%%%%

%%%%%%%%%%%%%%%%%%%%%%%%%%%%%%%%%%%%%%%%%%%%%%%%%%%%%%%%%%%%%%%%%%%%
\subsection{Chiral effective theory}
\label{sec_efts}
%%%%%%%%%%%%%%%%%%%%%%%%%%%%%%%%%%%%%%%%%%%%%%%%%%%%%%%%%%%%%%%%%%%%

 Penetrating probes like photons and dileptons can be used
to test the quasi-particle structure of hot and dense matter. 
In the following we will concentrate on the regime below the 
phase transition, where the relevant degrees of freedom are hadrons. 
At low temperature and small baryon density the dominant degrees 
of freedom are pions. Pions are Goldstone bosons associated with 
the spontaneously broken chiral symmetry, and their interactions
are constrained by the underlying symmetry. 

 The Goldstone boson fields can be parameterized by unitary matrices 
$\Sigma = \exp(i\lambda^a\phi^a/f_\pi)$ where $\lambda^a$ are the 
Gell-Mann matrices for $SU(3)$ flavor and $f_\pi=93$ MeV is the pion 
decay constant. For example, $\pi^0=\phi^3$ and $\pi^\pm=(\phi_1\pm 
i\phi_2)/2$ describe the neutral and charged pion. Other components
of $\phi^a$ describe the neutral and charged kaons, as well as the 
eta. The eta prime acquires a large mass because of the axial anomaly,
and is not a Goldstone boson. 

 At low energy the effective Lagrangian for $\Sigma$ can be organized 
as an expansion in the number of derivatives of $\Sigma$. This is the
case because higher derivative terms describe interactions that scale 
as either the momentum or the energy of the Goldstone boson. Since
Goldstone bosons are approximately massless, the energy is of the 
same order of magnitude as the momentum. We will see that the expansion
parameter is $p/(4\pi f_\pi)$. At leading order in $(\partial/f_\pi)$ 
there is only one structure which is consistent with chiral symmetry, 
Lorentz invariance and C,P,T. This is the Lagrangian of the non-linear 
sigma model. In the presence of a small explicit symmetry breaking 
term we have
\be
\label{l_chpt}
{\mathcal L} = \frac{f_\pi^2}{4} {\rm Tr}\left[
 \partial_\mu\Sigma\partial^\mu\Sigma^\dagger\right] 
  +\left[ B {\rm Tr}(M\Sigma^\dagger) + h.c. \right]
+ \ldots. 
\ee
Here, $M={\rm diag}(m_u,m_d,m_s)$ is the quark mass matrix and $B$, the 
coefficient of the symmetry breaking term, is a low energy constant that 
we will fix below. In order to show that the parameter $f_\pi$ is related 
to the pion decay amplitude we have to gauge the non-linear sigma model. 
This is achieved by introducing the gauge covariant derivative $\nabla_\mu
\Sigma = \partial_\mu\Sigma+ig_w W_\mu\Sigma$ where $W_\mu$ is the charged 
weak gauge boson and $g_w$ is the weak coupling constant. The gauged 
non-linear sigma model gives a pion-$W$ boson interaction ${\mathcal L}
=g_w f_\pi W^\pm_\mu \partial^\mu \pi^\mp$. This term leads to an amplitude
for the decay $\pi^\pm\to W^\pm\to e^\pm\nu_e$ or $\pi^\pm\to W^\pm\to\mu^\pm
\nu_\mu$ which is proportional to $g_wf_\pi q_\mu$, where $q_\mu$ is the 
momentum of the pion. This result agrees with the standard definition of 
$f_\pi$ in terms of the pion-weak axial current matrix element. In the ground 
state $\Sigma=1$ and the ground state energy is $E_{vac}=-2B{\rm Tr}[M]$. 
Using the relation $\langle\bar{q}q\rangle = \partial E_{vac}/(\partial m)$ 
we find $\langle\bar{q}q\rangle=-2B$. Fluctuations around $\Sigma=1$ 
determine the Goldstone boson masses. The pion mass satisfies the 
Gell-Mann-Oaks-Renner relation (GMOR) \cite{GellMann:1968rz}
\be
\label{GMOR}
m_\pi^2 f_\pi^2 = (m_u+m_d)\langle\bar{q}q\rangle
\ee
and analogous relations exist for the kaon and eta masses. 

 Corrections to this result arise from higher derivative corrections to 
the effective Lagrangian Eq.~(\ref{l_chpt}), and from loop corrections 
computed using the leading order vertices. Expanding out Eq.~(\ref{l_chpt})
to fourth order in $\phi^a$ gives interaction terms of the form ${\cal L}
\sim f_\pi^{-2}(\phi^a\partial_\mu\phi^a)^2$. This means that the tree level 
meson-meson interaction scales as $q^2/f_\pi^2$, and that the low energy 
interaction is indeed weak. Computing loop diagrams gives corrections to 
this result that contain additional factors of $q/(4\pi f_\pi)$. The 
numerical factor $1/(4\pi)$ arises from the phase space in loop integrals, 
and clearly helps in obtaining meaningful results for pion momenta up to 
a few hundred MeV. 

 Higher order gradient corrections involve terms like ${\cal L} \sim 
c_4 {\rm Tr}[(\partial_\mu\Sigma\partial^\mu\Sigma^\dagger)^2]$. On dimensional
grounds $c_4$ is suppressed by a factor $f_\pi^{-2}$ relative to the 
leading order interaction. Since $c_4$ acts as a counter-term that 
can be used to absorb the scale dependent pieces of one-loop terms a 
more accurate estimate is $c_4f_\pi^{-2}\sim (4\pi f_\pi)^{-2}$. Alternatively, 
we can view higher order gradient terms as arising from integrating 
out resonances like the rho meson and the $K^*$. This suggests
$c_4f_\pi^{-2}\sim m_V^{-2}$, where $V=\rho,K^*,\ldots$. We note that 
$m_V$ is numerically close to the scale $4\pi f_\pi \simeq 1160$ MeV, 
implying that the values of $f_\pi$ and the vector meson masses in 
QCD are natural. 

%%%%%%%%%%%%%%%%%%%%%%%%%%%%%%%%%%%%%%%%%%%%%%%%%%%%%%%%%%%%%%%%%%%%
\subsection{Chiral effective theory at finite temperature}
\label{sec_eft_T}
%%%%%%%%%%%%%%%%%%%%%%%%%%%%%%%%%%%%%%%%%%%%%%%%%%%%%%%%%%%%%%%%%%%%

 As a first application of the chiral Lagrangian we study the 
dependence of the chiral condensate on the temperature. Like the 
GMOR relation this result can be extracted from thermodynamic
properties. At finite $T$ we have to consider the free energy
$F=E-TS$ instead of the energy $E$. At leading order in $T/f_{\pi}$,
we can ignore interactions between pions. The free energy of an 
ideal pion gas is 
\be
\label{z_pi}
F = (N_f^2-1)T\int \frac{d^3p}{(2\pi)^3} \log
  \left( 1- e^{-E_\pi/T} \right), 
\ee
where $E_\pi=\sqrt{p^2+m_\pi^2}$. The quark condensate is 
$\langle\bar{q}q\rangle = (N_f)^{-1}\partial F/\partial m$. 
Equation (\ref{z_pi}) depends on the quark mass only through the 
pion mass. Using the Gell-Mann-Oakes-Renner relation (\ref{GMOR})
we find \cite{Gasser:1986vb}
\be
\label{qbarq_T}
\langle\bar{q}q\rangle_T = 
\langle\bar{q}q\rangle_0 \left\{ 1-\frac{N_f^2-1}{3N_f}
\left(\frac{T^2}{4f_\pi^2}\right)+\ldots\right\}.
\ee
This result shows that there is a tendency towards chiral symmetry 
restoration already at low temperature, and that the relevant scale is 
set by $T\sim 2f_\pi \sim 180$ MeV. There is a nice physical interpretation 
of the result given in Eq.~(\ref{qbarq_T}). The chiral condensate in 
vacuum is negative, but the pion matrix elements of $m_q\bar{q}q$ is 
positive. A finite density of thermal pions therefore reduces the 
vacuum condensate, or, to quote Gerry Brown, ``they act as a
  vacuum cleaner'' 

 Another simple application of chiral perturbation theory involves
the vector and axial-vector correlation functions. The correlator 
of the vector current can be accessed using di-lepton measurements, 
see Section~\ref{sec:dilepton}, and the difference between the vector and
axial-vector correlators is a measure of chiral symmetry breaking. 
The correlation functions are defined by 
\be 
\label{Pi_VA}
\Pi_{\mu\nu}^{V,A}(q) = -i \int d^4x\, \Theta(x_0)e^{iq\cdot x}
  \langle [j_\mu^{V,A}(x),j_\nu^{V,A}(0)]\rangle
\ee
where $j_\mu^V=\bar{q}\frac{\tau^a}{2}\gamma_\mu q$ and $j_\mu^A=\bar{q}
\frac{\tau^a}{2}\gamma_\mu\gamma_5 q$ are the vector and axial-vector 
currents, and we have suppressed the isospin indices $a,b$ on the 
correlation function. The functions in Eq.~(\ref{Pi_VA}) are 
retarded correlation functions. As usual, the spectral function
is determined by the imaginary part of the retarded correlator. 
A set of sum rules for the vector and axial-vector spectral functions
was derived by Weinberg. We can split the correlators into 
transverse and longitudinal parts
\be 
 \Pi_{\mu\nu}(q)=\Pi_T(q^2)P_{\mu\nu}^T + \Pi_L(q^2)P^L_{\mu\nu}\, ,
\ee
with $P_{\mu\nu}^T=q_\mu q_\nu/q^2-g_{\mu\nu}$ and $P_{\mu\nu}^L=q_\mu q_\nu
/q^2$. We define $\rho^{V,A}(s)=\frac{1}{\pi}{\rm Im}\Pi_T^{V,A}(s)$. Then 
the Weinberg sum rules in the chiral limit read \cite{Weinberg:1967kj}
\begin{eqnarray}
\int_0^\infty  \frac{ds}{s} \left( \rho_V (s)- \rho_A (s)\right)
 &=& f_\pi^2 \; , \\
\int_0^\infty  \,ds\, \left(\rho_V(s)-\rho_A(s)\right)
 &=& 0 \; .
\end{eqnarray}
Additional sum rules, and corrections due to finite quark masses can 
be studied using the operator product expansion \cite{Narison:2007spa}.
The sum rules provide an explicit relation between the difference
of the spectra in the vector and axial-vector channels and 
spontaneous chiral symmetry breaking, controlled by $f_\pi$. 
The Weinberg sum rules can be extended to finite temperature provided
we interpret the sum rules as integrals over energy at fixed
momentum, and separate out the pion contribution more carefully
\cite{Kapusta:1993hq}. At lowest order in the chiral expansion 
one can show that the finite temperature correlators are related 
to the $T=0$ functions by the simple mixing relation 
\cite{Dey:1990ba}
\be 
\label{VA_mix}
\Pi^{V,A}_{\mu\nu}(q) = (1-\epsilon) \Pi^{V,A}_{\mu\nu\, ,0} 
   + \epsilon \Pi^{A,V}_{\mu\nu\, , 0}(q)\, , 
\ee
with $\epsilon=T^2/(6f_\pi^2)$ and the subscript $0$ refers to the 
$T=0$ result. This formula, too, has a simple 
interpretation. In a thermal medium the vector current can couple
to thermal pions and mix with the axial current, and vice versa. 

 In order to make more quantitative statements about the vector
current spectral function we have to understand the coupling of 
the vector current to hadronic states, in particular the $\rho$ meson. 
The $\rho$ meson is not a Goldstone boson, and these calculations inevitably
involve model assumptions. A successful scheme for constructing 
effective Lagrangians for vector mesons is the massive Yang-Mills
scheme \cite{Meissner:1987ge}. Here, we assume that the $\rho$ and
$a_1$ mesons are vector and axial-vector gauge fields associated 
with the $SU(2)_L\times SU(2)_R$ symmetry of the non-linear sigma
model. The gauge symmetry is broken by a Higgs field that gives
masses to the $\rho$ and $a_1$. A different scheme, known as the 
hidden local symmetry scheme, was introduced by Bando, Kugo, and
Yamawaki \cite{Bando:1987br}, and elaborated by many others, see
\cite{Halasz:1997xc}.

 In the massive Yang-Mills scheme the leading interaction between
pions, rho mesons, and the $a_1$ is given by 
\bea 
{\cal L} &=& \frac{1}{2}m_\rho^2 \vec{\rho}^{\,2}_\mu
   + \frac{1}{2}m_{a_1}^2 \vec{a}^{\,2}_{1\mu}
   + g^2 f_\pi \vec{\pi}\times\vec{\rho}^{\,\mu}\cdot\vec{a}_{1,\mu}
\nonumber \\
    && \mbox{}\;\;  
   +  g_{\rho\pi\pi} \left( \vec{\rho}^{\,2}_\mu \vec{\pi}^2
         - \vec{\rho}^{\,\mu}\cdot\vec{\pi} 
           \vec{\rho}_{\mu}\cdot\vec{\pi} \right)
   +  g_{\rho\pi\pi} \vec{\rho}^{\,\mu}\cdot
          \left(\vec{\pi} \times\partial^\mu\vec{\pi}\right) + \ldots\, , 
\eea
where $2g_{\rho\pi\pi}^2=g^2$ and the Higgs mechanism leads to the 
mass formula $m_{a_1}^2=m_{\rho}^2+g^2f_\pi^2$. The massive Yang-Mills
Lagrangian exhibits vector meson dominance, that means the coupling
of the isospin-one component of the photon to the vector current 
is saturated by the rho meson. As a result, the calculation of the 
vector spectral function can be reduced to computing the self-energy 
of the rho meson \cite{Rapp:2009yu}. The main contributions
at $T=0$ are the $\pi\pi$ and $\pi a_1$ intermediate states. Both 
of these receive corrections at finite temperature due to thermal
pion states. At low temperature these thermal effects reproduce the
mixing formula in Eq.~(\ref{VA_mix}).

 In a dense hadron gas higher resonances become important. There are,
in particular, many nucleon resonances $N^*$ that have a strong coupling 
to $N\rho$ \cite{Peters:1997va}. In the $\rho$ self energy this corresponds 
to intermediate states of the form $N^*N^{-1}$. In vacuum excited-nucleon 
anti-nucleon intermediate states are strongly suppressed, but in a medium 
with non-zero baryon density we find excited-nucleon nucleon-hole
states that make significant contributions. Also, at finite temperature
there is a non-zero thermal population of baryons and anti-baryons.
The net effect of these contributions is a significant broadening 
of the $\rho$, together with extra strength at low invariant mass 
\cite{Rapp:2009yu}. 

 Similar effects have been studied in nuclear physics for a long
time. In finite nuclei and nuclear matter pions (or states with
the quantum numbers of pions) can mix with nucleon-hole and 
delta-hole pairs \cite{Brown:1975di}. The p-wave $\pi N\Delta$
interaction is quite strong, and these effects can lead to 
a softening of the pion dispersion relation, which is a precursor 
of a possible pion condensed phase in cold dense nuclear matter. 
The Delta-hole mechanism is not restricted to cold nuclear phases, 
and possible effects in heavy ion collisions in the fixed 
target regime where studied in \cite{Bertsch:1988xu,Koch:1992zi}.

 Brown and Rho suggested that, because of the chiral and scale 
symmetries of the QCD Lagrangian, the complicated many-body dynamics
of quarks and gluons can be summarized in terms of simple scaling laws
for the effective masses of hadrons \cite{Brown:1991kk}. They 
proposed that 
\be 
\label{BR_scal}
 \frac{m_\rho^*}{m_\rho}=\frac{m_{a_1}^*}{m_{a_1}}
  = \left( \frac{\langle\bar{q}q\rangle_{\rho,T}}
                {\langle\bar{q}q\rangle_0}\right)^{1/3}\,. 
\ee
where $m_\rho^*$ and $m_{a_1}^*$ are the in-medium masses. In this 
scenario chiral symmetry is restored because all hadrons become
massless at the critical temperature. This is different from 
the picture discussed above, where chiral symmetry is restored 
because of the effects of mixing between chiral partner channels, 
combined with the melting of hadronic resonances. The Brown-Rho
scenario was investigated in great detail by analyzing the 
spectra of dileptons emitted in heavy ion collisions at different
energies, see Section~\ref{sec:dilepton}.

%%%%%%%%%%%%%%%%%%%%%%%%%%%%%%%%%%%%%%%%%%%%%%%%%%%%%%%%%%%%%%%%%%%%
\subsection{Quasiparticles in the quark-gluon plasma}
\label{sec_qp_qgp}
%%%%%%%%%%%%%%%%%%%%%%%%%%%%%%%%%%%%%%%%%%%%%%%%%%%%%%%%%%%%%%%%%%%%

  We can also identify quasi-particles in the high temperature 
phase. In Section~\ref{sec_EOS} we noted that in the quark-gluon 
plasma the color Coulomb interaction is screened at a distance 
$r\sim m_D^{-1}$,  where 
\be
\label{m_D}
m_D^2 = g^2 \left[  \left(1+\frac{N_f}{6}\right)\, T^2
            + \frac{N_f}{2\pi^2}\, \mu^2 \right]\, ,
\ee
is called the Debye mass. In perturbation theory the static magnetic 
interaction is unscreened \cite{Baym:1990uj}, but non-static magnetic 
interactions are dynamically screened at a distance $r\sim (m_D^2
\omega)^{-1/3}$. This phenomenon, known as Landau damping, is due to 
the coupling of gluons to particle-hole (or particle-anti-particle)
pairs, and also play a role in electromagnetic plasmas. 
Unlike classical plasmas the QCD plasma has a non-perturbative 
static magnetic screening mass $m_M\sim g^2 T$. This is the scale 
that determines the non-perturbative contributions to the pressure. 
Modes below the magnetic screening scale contribute
\be 
\label{P_mag}
 P \sim T \int^{m_M} d^3q \sim g^6T^4\, ,
\ee
which implies that the ``last'' perturbatively calculable 
contribution to the QGP pressure is $O(g^6\log(g))$ \cite{Kajantie:2002wa}.

The electric screening scale also determines the properties 
of collective gluonic modes, see \cite{Kapusta_Book}. For momenta 
$q\gg gT$ there are two transverse modes with dispersion relation 
$\omega\simeq q$. For momenta $q< gT$ there are two transverse and 
one longitudinal mode. The longitudinal mode is sometimes called 
the plasmon. The energy of all three modes approaches $\omega=
\omega_p=m_D/\sqrt{3}$ as $q\to 0$. The quantity $\omega_p$ is known 
as the plasma frequency. The gluon (and plasmon) decay constant 
in the limit $q\to 0$ is \cite{Braaten:1990it}
\be
\label{gam_plas}
\gamma = 6.64 \frac{g^2N_c T}{24\pi} \, ,
\ee
which confirms that the quasi-particle width is parametrically
small compared to their energy. Numerically, the ratio of the 
plasmon width to the plasmon energy is $\gamma/\omega_p \simeq 
g/2.2$, which is of order one even in a weakly coupled plasma. 
However, as discussed in Section~\ref{sec_lQCD}, models based 
on screened quasi-particles are successful in describing the 
thermodynamics of QCD down to $T\sim 2T_c$. The calculation of 
the collisional width of quasi-particles with momenta of order 
$T$ is a complicated, non-perturbative problem, but the width 
remains parametrically small, $\gamma\sim g^2\log(1/g)T$ 
\cite{Blaizot:1999fq}.

 Quarks also form collective modes by coupling to gluons. 
Perturbative interactions do not contribute to chiral symmetry
breaking mass terms that connect left and right-handed. However,
perturbative interaction can generate an effective energy, 
and lead to mixing between chirality and helicity eigenstates. 
It is standard to define an effective fermion ``mass''
\be 
\label{m_f}
 m_f^2 = \frac{g^2}{6} \left[ T^2 +\frac{\mu^2}{\pi^2} \right].
\ee
There are two fermionic branches, one with helicity equal 
to chirality, and another, called the plasmino, in which 
the two quantities are opposite. Both modes satisfy $\omega
\sim m_f$ for small momenta. At large momenta the plasmino
disappears, and the standard quark mode satisfies $\omega
\simeq q+m_f^2/q$. 

 At very low momenta the only propagating mode is the 
hydrodynamic sound modes discussed in Section~\ref{sec_hydro}. 
Sound is damped by shear and bulk viscosity. The width of 
a sound mode with energy $\omega=c_sq$ is given by $\gamma
=[c_s\Gamma_s q^3]^{1/2}$ with 
\be 
\Gamma_s= \frac{1}{sT}\left(\frac{4\eta}{3}+\zeta \right)\, . 
\ee
Using the perturbative result for $\eta/s$ given in Eq.~(\ref{eta_s_w})
we conclude that $\Gamma_s\sim 1/(g^4T)$, and only very low 
momentum sound modes with $q\lsim g^4T$ can propagate. In a 
nearly perfect fluid, on the other hand, modes with $q\sim T$ 
are propagating. Numerically, the width over energy of a sound 
mode is $\gamma/\omega \simeq [2(\eta/s)(\omega/T)]^{1/2}$. For
$\eta/s\simeq 1$ only modes with $\omega\lsim 0.5 T$ are 
propagating, but in a nearly perfect fluid we find a much
less restrictive bound,  $\omega\lsim 6 T$.

%%%%%%%%%%%%%%%%%%%%%%%%%%%%%%%%%%%%%%%%%%%%%%%%%%%%%%%%%%%%%%%%%%%%
\subsection{Quasiparticles in dense quark matter}
\label{sec_qp_qm}
%%%%%%%%%%%%%%%%%%%%%%%%%%%%%%%%%%%%%%%%%%%%%%%%%%%%%%%%%%%%%%%%%%%%

 Finally, quasi-particle properties can be studied in the 
regime of very high baryon density. Above the critical temperature
for color superconductivity we have quasi-quarks and quasi-gluons with 
the effective masses given in Eq.~(\ref{m_D}) and (\ref{m_f}). 
The perturbative expansion has some unusual features, known as
non-Fermi liquid behavior \cite{Schafer:2004zf}, but quasi-particles 
are well-defined and have small widths. At $T_c$ quarks acquire 
a gap, and gluons acquire magnetic screening masses by a QCD 
analog of the Meissner effect. The situation is simplest in 
the CFL phase, where all quarks and gluons acquire a gap. The
gap in the quark sector was determined by Son \cite{Son:1998uk},
\be
\label{gap_oge}
\Delta_{\it CFL} \simeq 2^{4/3}\Lambda_{BCS}
   \exp\left(-\frac{\pi^2+4}{8}\right)
   \exp\left(-\frac{3\pi^2}{\sqrt{2}g}\right), 
\ee
with $\Lambda_{BCS}=256\pi^4 (2/N_f)^{5/2}g^{-5}\mu$ \cite{Schafer:2003jn}.
The nine different color-flavor combinations organize themselves into an 
octet with gap $\Delta_{\it CFL}$, and a singlet with gap $2\Delta_{\it CFL}$
\cite{Alford:1998mk}.

 The CFL order parameter breaks chiral symmetry, and for energies 
below the gap the propagating modes are Goldstone bosons. Based on 
symmetry arguments, the effective Lagrangian has the same structure
as the Lagrangian of chiral perturbation theory, Eq.~(\ref{l_chpt}),
except that Lorentz-invariance is no longer a symmetry. We have
\cite{Casalbuoni:1999wu}
\be
\label{l_cheft_cfl}
{\mathcal L}_{eff} &=& \frac{f_\pi^2}{4} {\rm Tr}\left[
 \nabla_0\Sigma\nabla_0\Sigma^\dagger - v_\pi^2
 \partial_i\Sigma\partial_i\Sigma^\dagger \right] \, 
\ee
where the speed of Goldstone modes is $v_\pi^2\simeq 1/3$ \cite{Son:1999cm}.
If quark masses are taken into account then the Goldstone bosons 
acquire small masses, $m_\pi,m_K\ll \Delta$. It interesting to consider 
the properties of gapped quasi-quarks in more detail. We already noted
that quarks are organized into an octet and a singlet of the $SU(3)$
flavor group. We also find that quasi-particles have integer electric
and baryon charges. These arise from the diquark polarization cloud
that surrounds a single quark, and the phenomenon can be described 
as quark-hadron complementary \cite{Schafer:1998ef}. The effective 
Lagrangian for fermions in the CFL phase can be written as 
\cite{Kryjevski:2004jw}
\bea 
\label{l_bar}
{\mathcal L} &=&  
 {\rm Tr}\left(N^\dagger iv^\mu D_\mu N\right) 
 - D{\rm Tr} \left(N^\dagger v^\mu\gamma_5 
               \left\{ {\mathcal A}_\mu,N\right\}\right)
 - F{\rm Tr} \left(N^\dagger v^\mu\gamma_5 
               \left[ {\mathcal A}_\mu,N\right]\right)
  \nonumber \\
 & &  \mbox{} + \frac{\Delta}{2} \left\{ 
     \left( {\rm Tr}\left(N_LN_L \right) 
   - \left[ {\rm Tr}\left(N_L\right)\right]^2 \right)  
   - (L\leftrightarrow R) + h.c.  \right\}.
\eea
$N_{L,R}$ are left and right handed baryon fields in the adjoint 
representation of flavor $SU(3)$. We can think of $N$ as describing 
a quark which is surrounded by a diquark cloud, $N_L \sim q_L\langle 
q_L q_L\rangle$. The covariant derivative of the nucleon field is 
given by $D_\mu N=\partial_\mu N +i[{\mathcal V}_\mu,N]$, and 
${\mathcal V}_\mu$ and ${\mathcal A}_\mu$ are the pionic (and kaonic)
vector and axial-vector currents. The coupling constants $D,F$ 
control the axial-vector couplings in flavor $SU(3)$, and the 
last line in Eq.~(\ref{l_bar}) is a Majorana mass term that 
describes the flavor structure of the gap. What is interesting 
about Eq.~(\ref{l_bar}) is that, except for the gap, this Lagrangian
has the same structure as the flavor $SU(3)$ chiral Lagrangian for 
baryons at zero temperature and density, which illustrates the 
possibility of a continuous phase transition between (hyperonic)
nuclear matter and strange quark matter \cite{Schafer:1998ef}.

%%%%%%%%%%%%%%%%%%%%%%%%%%%%%%%%%%%%%%%%%%%%%%%%%%%%%%%%%%%%%%%%%%%%%%%%%
\subsection{Landau Fermi liquid theory}
\label{sec_fl}
%%%%%%%%%%%%%%%%%%%%%%%%%%%%%%%%%%%%%%%%%%%%%%%%%%%%%%%%%%%%%%%%%%%%%%%%%

 An important issue related to the modification of quasi-particle
properties in a hot and dense medium is the consistent treatment
of hadronic transport properties. This problem does not arise
in the context of fluid dynamics, where the only relevant properties
are the equation of state and a small set of transport coefficients, 
but it is an issue in hadronic transport models. The simplest 
example of a theory that provides a consistent treatment of single 
particle and transport properties is the Landau theory of Fermi 
liquids, which is directly applicable to cold and dense systems of 
baryons. Consider a cold Fermi system in which the low energy 
excitations are spin 1/2 quasi-particles. Landau proposed to define 
a distribution function $f_p=f_p^0+\delta f_p$ for the quasi-particles. 
Here, $f_p^0$ is the ground state distribution function, and $\delta f_p
\ll f_p^0$ is a correction. The energy density can be written as 
\cite{Landau:kin,Brown:1971zza,Brown:1972,Baym:1991}
\be
 {\cal E} = {\cal E}_0 
  + \int d\Gamma_p\, \frac{\delta{\cal E}}{\delta f_p}\delta f_p
  + \frac{1}{2} \int\int d\Gamma_p d\Gamma_{p'}
    \frac{\delta^2{\cal E}}{\delta f_p\delta f_{p'}} 
    \delta f_p\delta f_{p'} + \ldots\, , 
\ee
with $d\Gamma_p=d^3p/(2\pi)^3$. Functional derivatives of ${\cal E}$ 
with respect to $f_p$ define the quasi-particle energy $E_p$ and the 
effective interaction $t_{pp'}$
\be 
\label{ep_flt}
E_p = \frac{\delta{\cal E}}{\delta f_p} \hspace{0.1\hsize}
t_{pp'}=\frac{\delta^2{\cal E}}{\delta f_p\delta f_{p'}} \, . 
\ee
Note that, in general, $E_p$ is a non-trivial function of the 
distribution function $f_p(x,t)$. This implies, in particular, that 
the particles have density and temperature dependent effective masses. 
Near the Fermi surface we can write $E_p=v_F(|\vec{p}|-p_F)$, where 
$v_F$ is the Fermi velocity, $p_F$ is the Fermi momentum, and $m^*=
p_F/v_F$ is the effective mass. We can decompose $t_{pp'}=F_{pp'}+G_{pp'}
\vec{\sigma}_1\cdot\vec{\sigma}_2$. On the Fermi surface the effective 
interaction is only a function of the scattering angle and we can 
expand the angular dependence as
\be 
F_{pp'} = \sum_l F_l\, P_l\left(\cos\theta_{\vec{p}\cdot\vec{p}'}\right) \, ,
\ee
where $P_l(x)$ is a Legendre polynomial, and $G_{pp'}$ can be expanded in 
an analogous fashion. The coefficients $F_l$ and $G_l$, which control
the properties of quasi-particles, are called Landau parameters. There
are a number of interesting connections between single-particle and
collective properties. For example, the effective mass is 
\be
m^*= m \left( 1 +\frac{F_1}{3}\right)\, ,
\ee
and the speed of sound is given by
\be
c_s^2=\frac{v_F^2}{3}\frac{1+F_0}{1+F_1/3}\, , 
\ee
where $v_F$ is the Fermi velocity defined above. 
 
The distribution function satisfies a Boltzmann equation
\be
\Big( \partial_t + \vec{v}_p\cdot\vec{\nabla}_x + 
    \vec{F}_p\cdot\vec{\nabla}_p\Big) f_p(x,t) = C[f_p]
\ee
where $\vec{v}_p=\vec{\nabla}_p E_p$ is the quasi-particle velocity, 
$\vec{F}_p=-\vec{\nabla}_x E_p$ is an effective force, and $C[f_p]$ 
is the collision term. Conserved currents can be defined in terms of
$f_p$ and the single particle properties $E_p$ and $v_p$. For example, 
we can write the mass density $\rho$ and mass current $\vec{\jmath}$ as
\be 
\rho =\int d\Gamma_p\, mf_p\, ,\hspace{0.3cm}
\vec{\jmath} =\int d\Gamma_p\, m\vec{v}_pf_p\, , 
\ee
where $d\Gamma_p=d^3p/(2\pi)^3$. The Boltzmann equation implies that the 
current is conserved, $\partial_0 \rho+\vec{\nabla}\cdot\vec{\jmath}=0$. 
Since $E_p$ and $v_p$ are functionals of the particle distribution $f_p(x,t)$, 
the validity of conservation laws is non-trivial. In the framework of Landau 
Fermi liquid theory, conservation laws follow from the condition that $E_p$
can be derived from an energy density functional, see Eq.~(\ref{ep_flt}).
The equation of momentum conservation is $\partial_0\pi_i+\nabla_jT_{ij}=0$, 
where
\be
\label{T_ij_kin}
T_{ij}\left(\vec{x},t\right) &=& 
   \int d\Gamma_p\, p_iv_j  f_p\left(\vec{x},t\right)
    + \delta_{ij} \left( \int d\Gamma_p \, E_p f_p\left(\vec{x},t\right)
       - {\cal E}\left(\vec{x},t\right)\right) ,
\ee
Similar expressions hold in relativistic theories, see \cite{Baym:1975va}
and \cite{Jeon:1995zm}. A difficulty in constructing quasi-particle 
models of equilibrium and non-equilibrium properties is to find an 
explicit expression for ${\cal E}[f_p]$. This problem can be 
avoided by focusing on the enthalpy
\be 
\label{enth_kin}
{\cal E}+P = \int d\Gamma_p\, \left(\frac{1}{3}\vec{v}\cdot\vec{p}
 + E_p \right)f_p(\vec{x},t)\, ,
\ee
which can be expressed directly in terms of quasi-particle properties
$E_p$ and $v_p=\nabla_p E_p$. Equation (\ref{enth_kin}) can be used
in connection with any of the quasi-particle theories discussed earlier 
in this Section in order to construct a consistent kinetic and thermodynamic 
model. Indeed, enthalpy functionals are also at the center of many 
quasi-particle models that explore the more difficult regime near the 
QCD phase transition, see \cite{Blaizot:1999ip,Peshier:2004bv}.

%%%%%%%%%%%%%%%%%%%%%%%%%%%%%%%%%%%%%%%%%%%%%%%%%%%%%%%%%%%%%%%%%%%%%%%%%%%%
\section{Dilepton Production}
\label{sec:dilepton}
%%%%%%%%%%%%%%%%%%%%%%%%%%%%%%%%%%%%%%%%%%%%%%%%%%%%%%%%%%%%%%%%%%%%%%%%%%%%

The measurement of dileptons, i.e. lepton anti-lepton pairs, such 
as $(e^{+}e^{-})$ or $(\mu^{+}\mu^{-})$ in heavy ion collisions provides 
insight into the early, dense phase of the system. Dileptons, which
originate from the decay of time-like virtual photons, only interact
electromagnetically and thus, contrary to hadrons, do not suffer from final state
interaction. Compared to real photons, dileptons
offer a larger kinematic range since they are not restricted to the
light cone, $E=p$.

The first measurement of electron positron pairs in a heavy ion
collision was carried out by the DLS collaboration at the BEVALAC, 
where invariant mass spectra in proton-proton, Carbon-Carbon and 
Ca+Ca collisions at beam energies up to  $2\, A\GeV$ 
\cite{Roche:1989jx,Porter:1997rc} were measured. The motivation 
for these measurements was to gather information about the early
phase of the system. Soon thereafter it was also realized 
that these measurements may  be sensitive to in-medium properties 
of pions \cite{Gale:1987ki}.  At high energies, in connection with
the search for a QGP, thermal dileptons were proposed as signature
of an equilibrated plasma \cite{Shuryak:1978ij,Shuryak:1980tp,McLerran:1984ay,Kajantie:1986cu,Kajantie:1989yr}.

Since vector mesons such as $\rho$, $\omega$, and $\phi$ have an
exclusive decay channel into lepton pairs, dileptons are an
excellent probe to study the in-medium properties of vector mesons, 
or, more generally, the in-medium spectral function of the vector-correlator 
of the strong interaction as discussed in Section~\ref{sec_efts}.
This possibility received considerable attention due to the influential 
paper by Brown and Rho \cite{Brown:1991kk}, where it was conjectured 
that the mass of the $\rho$-meson scales with the chiral condensate 
$\ave{\bar{\psi}\psi}$, the order parameter of chiral symmetry restoration.   

The thermal dilepton production rate is given by 
\cite{McLerran:1984ay,Gale:1990pn,Rapp:1999ej,vanHees:2007th}
\begin{align}
  \frac{dR}{dM d^{4}q} = - \frac{\alpha^{2}}{3\pi^{2}} \frac{L\left(
  M^{2} \right)}{M^{2}} \mathrm{Im} \Pi^{\mu}_{em,\mu}\left( M,q;\mu_{B},T
  \right) \,\, f^{B}\left( q0;T \right)
\end{align}
where $\alpha$ is the fine structure constant, $f^{B}(q_{0};T)$
the Bose-Einstein distribution function. The lepton phase space factor
\begin{align}
  L(M)=\left( 1+\frac{2m_{l}^{2}}{M^{2}} \sqrt{1-\frac{4m_{l}^{2}}{M^{2}}} \right)
\end{align}
is unity except for small invariant masses $M\lesssim 2 m_{l}$. 
The retarded (electromagnetic) current-current correlator
\begin{align}
  \Pi^{\mu,nu}_{em} = i \int d^{4}x e^{i q x}\Theta(x^{0})\ave{\left[
  J^{\mu}_{\mathrm{em}}, J^{\nu}_{\mathrm{em}} \right]}
\end{align}
is proportional to the isospin $(0,0)$ component of the vector correlator, 
Eq.~\eqref{Pi_VA}. Therefore, the dilepton rate is proportional to the time
integrated, thermally weighted correlation function of the (chiral) conserved 
vector correlator $\Pi_{V}(q)$ which enters in the Weinberg sum-rules, as 
discussed in Section~\ref{sec_efts}. Consequently a dilepton measurement 
may provide constraints on the dynamics of chiral restoration at finite 
density and temperature.   

 According to the conjecture of Brown and Rho \cite{Brown:1991kk}, where
the above correlator is saturated by a $\rho$-meson at a reduced mass, the 
dilepton invariant mass spectrum for high energy heavy ion collisions 
should exhibit more strength at masses below the $\rho/\omega$ peak. And 
indeed, the first measurement of dilepton invariant mass spectra by the 
CERES collaboration at the CERN SPS showed such an  enhancement, first 
in $\mathrm{S+Au}$ \cite{Agakishiev:1995xb} and later in $\mathrm{Pb+Pb}$ 
collisions \cite{Agakishiev:1997au}. However, as discussed in Section~\ref{sec_efts}, the alternative view of a broadened spectral function 
via mixing of the $\rho$-meson with hadronic states, predominantly 
excited baryons \cite{Peters:1997va}, could also explain the first 
CERES data \cite{Rapp:1997fs}. For a detailed review see \cite{Rapp:1999ej}. 

\begin{figure}[t]
\begin{center}
\includegraphics[width=0.6 \textwidth]{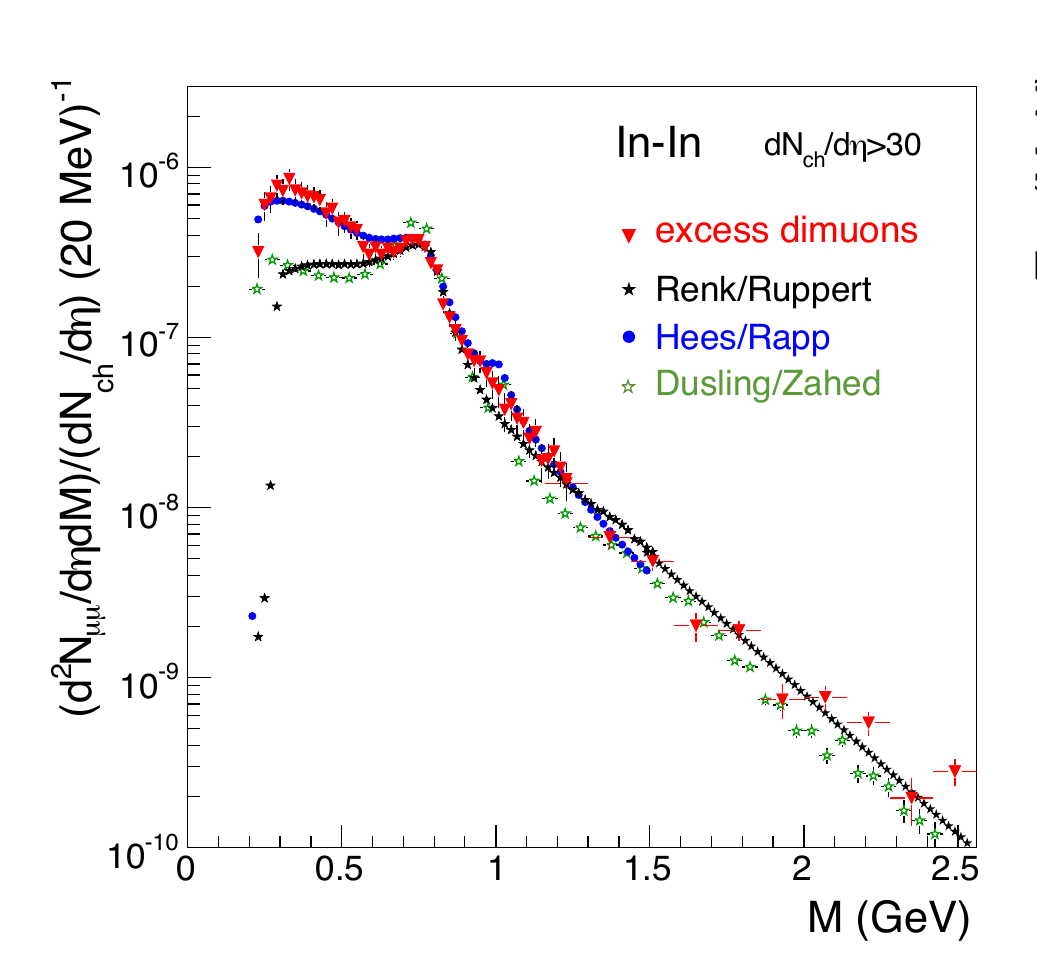}
\end{center}
\caption{Invariant mass spectrum of excess dileptons from the NA60
  experiment \cite{Arnaldi:2008fw,Arnaldi:2008er}. Also shown are theoretical
  calculations by Rupert
  et al, \cite{Ruppert:2007cr}, Rapp
  and van Hees \cite{vanHees:2007th}, as well as Dusling and Zahed
  \cite{Dusling:2007kh}. Figure adapted from
  \cite{Specht:2010xu}.}
\label{fig:dilep:na60_dndm}
\end{figure}

 The definitive resolution between these alternative explanations was later 
provided by the NA60 experiment, which measured  di-muon invariant mass 
spectra of unprecedented quality \cite{Arnaldi:2006jq,Arnaldi:2008fw}. This 
measurement clearly ruled out the originally conjectured scaling of the 
$\rho$-meson mass while the picture of a broadened in-medium 
$\rho$-meson spectral function \cite{Rapp:1997fs,Rapp:1999ej,Eletsky:2001bb}
prevailed. The acceptance corrected invariant mass spectrum of excess 
dileptons measured by the NA60 collaboration \cite{Arnaldi:2006jq,Arnaldi:2008fw,Arnaldi:2008er}
is shown in Fig.~\ref{fig:dilep:na60_dndm} together with theoretical
calculations by various groups
\cite{vanHees:2007th,Ruppert:2007cr,Dusling:2007kh}. The excess spectrum 
is obtained by subtracting the contributions from long lived sources, 
such as $\eta$ and $\omega$-Dalitz decays, direct decays of $\omega$ 
and $\phi$, as well as open charm and Drell-Yan pairs \cite{Arnaldi:2008er}. 
The excellent resolution and high statistics of the NA60 experiment allowed 
to measure these sources directly and to remove them in a model 
independent fashion. In the low mass region, $M_{inv}< 0.7 \GeV$, the 
model calculations shown in Fig.~\ref{fig:dilep:na60_dndm} diverge and 
the result of van Hees and Rapp \cite{vanHees:2007th} is closest to the 
data (see also \cite{Endres:2014zua}). While all calculations shown take 
into account in-medium broadening of the $\rho$-meson, those of Renk and 
Ruppert \cite{Ruppert:2007cr} and Dusling and Zahed \cite{Dusling:2007kh}
consider only the leading order correction in density and temperature,
while van Hees and Rapp re-sum the self-energy corrections  and thus
take higher order effects into account.

%%%%%%%%%%%%%%%%%%%%%%%%%%%%%%%%%%%%%%%%%%%%%%%%%%%%%%%%%%%%%%%%%%%%%%%%%%%%
%%%%%%%%%%%%%%%%%%%%%%%%%%%%%%%%%%%%%%%%%%%%%%%%%%%%%%%%%%%%%%%%%%%%%%%%%%%%

\begin{figure}[t]
\begin{center}
\includegraphics[width=0.5 \textwidth]{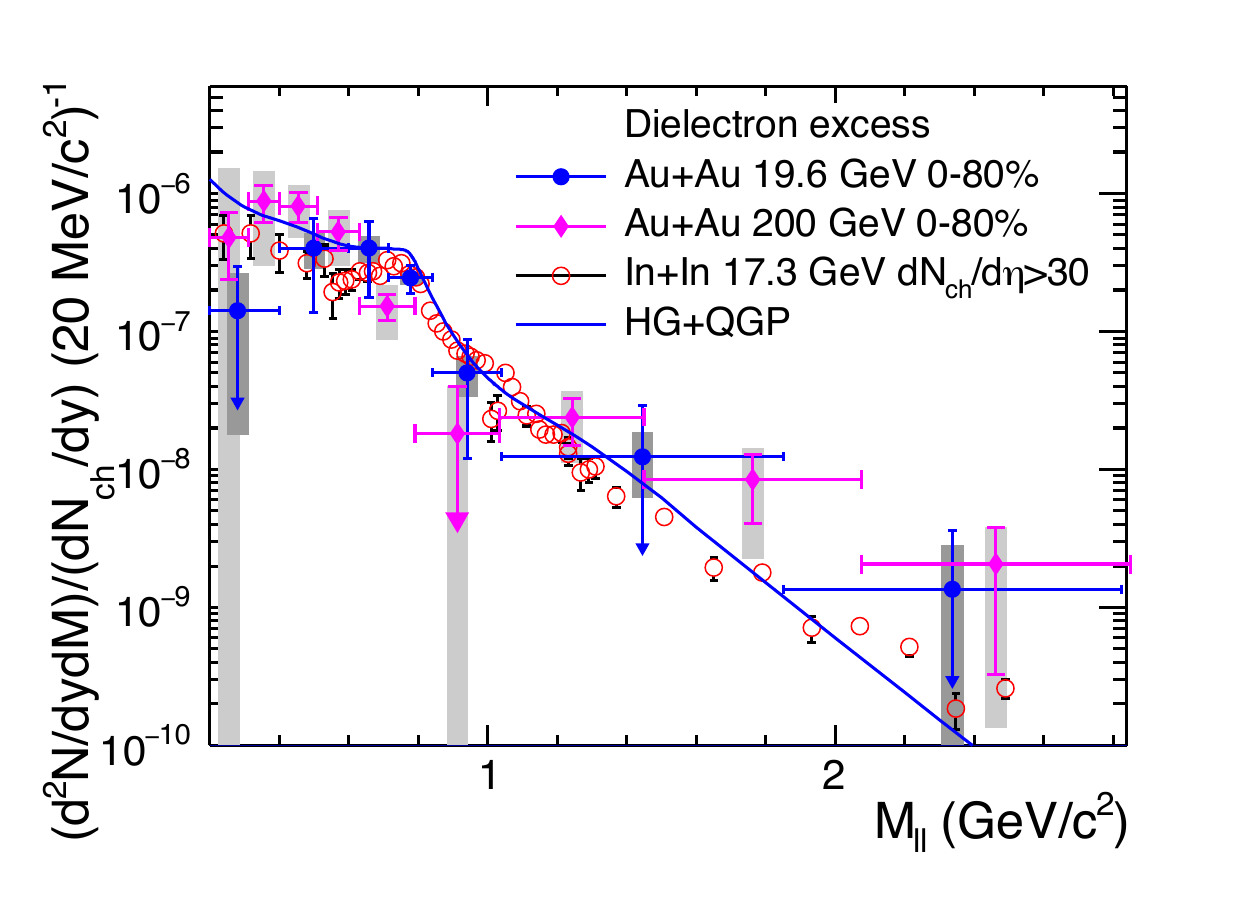}
\includegraphics[width=0.4 \textwidth]{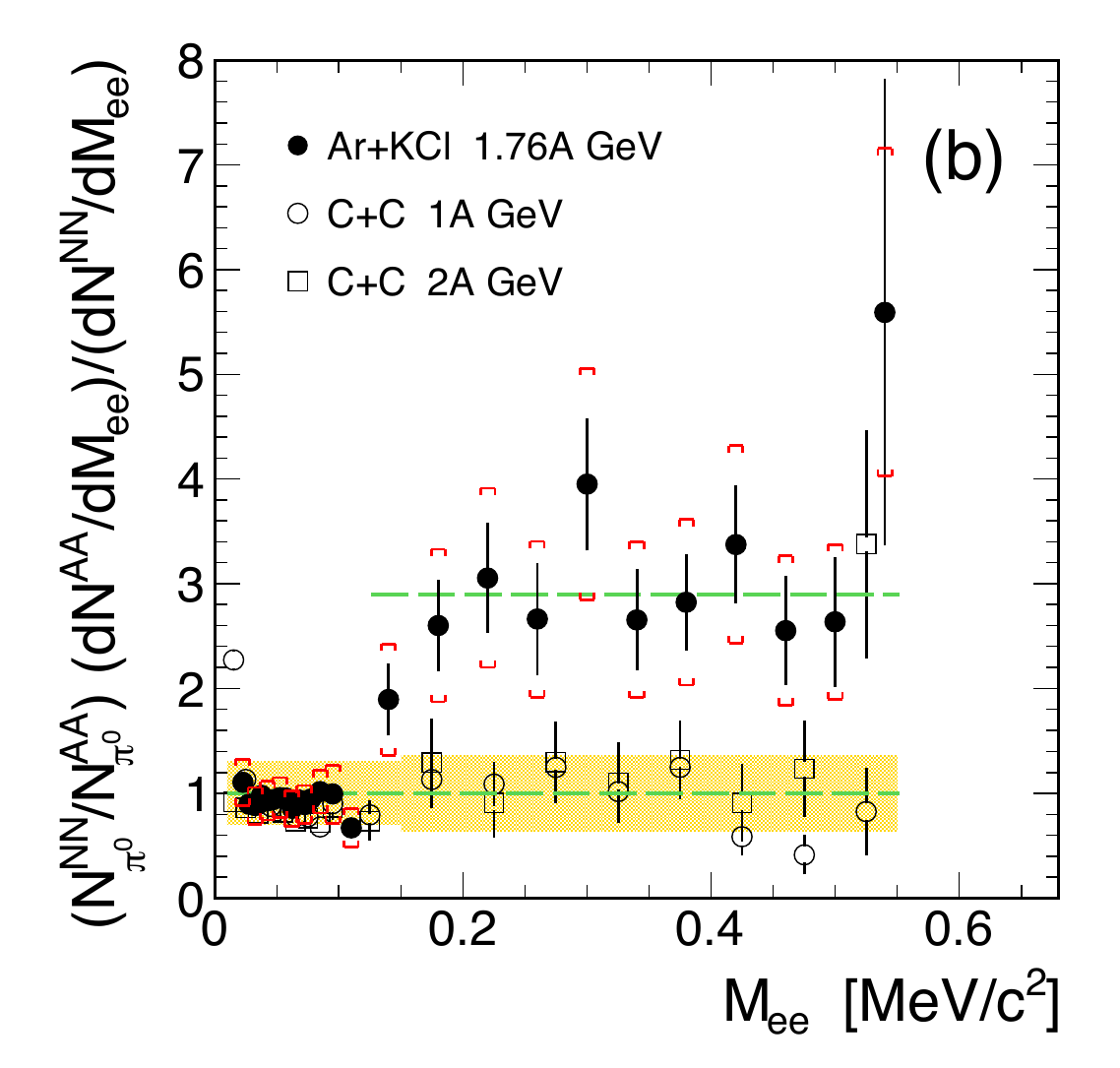}
\end{center}
\caption{Left panel: Comparison of excess dileptons measured by NA60
  for $\mathrm{In}+\mathrm{In}$ at $\sqrt{17.3} \GeV $ with those by STAR
  for  $\mathrm{Au}+\mathrm{Au}$ at $\sqrt{19.6} \GeV$ and $\sqrt{200}
  \GeV$. Figure adapted from \cite{Adamczyk:2015mmx}. Right panel: Ratio of dilepton excess
   for $\mathrm{Ar}+\mathrm{KCl}$ over $\mathrm{C}+\mathrm{C}$. Figure
   adapted from \cite{Agakishiev:2011vf}.}
\label{fig:dilep:star_hades}
\end{figure}

Meanwhile, low mass dilepton spectra have also been measured  at much
lower energies $\sqrt{s}=2.24 \GeV $ by the HADES collaborations
\cite{Agakichiev:2006tg,Agakishiev:2007ts,Agakishiev:2009yf,Agakishiev:2011vf} 
and at higher energies (up to $ \sqrt{s}=200 \GeV$) by the STAR collaboration
\cite{Adamczyk:2013caa,Adamczyk:2015lme,Adamczyk:2015mmx} and the
PHENIX collaboration \cite{Adare:2009qk,Adare:2015ila}. 
In both cases one finds a qualitatively similar enhancement in the mass 
region below the $\rho$ meson. At low energies, HADES finds that the 
dilepton spectrum for the small system of $\mathrm{C}+\mathrm{C}$ can be 
explained in terms of nucleon-nucleon scattering \cite{Agakishiev:2007ts}, 
whereas the heavier system of $\mathrm{Ar}+\mathrm{KCl}$ shows a clear
enhancement \cite{Agakishiev:2011vf}, as shown in the right panel of 
Fig.~\ref{fig:dilep:star_hades}. The STAR experiment at RHIC
has measured the dilepton spectra for several beam energies ranging
from $\sqrt{s}=19.6 \GeV $ to $\sqrt{s}=200 \GeV$ and observes an
enhancement below the $\rho$ in all cases \cite{Xu:2014jsa}. As shown
in the left panel of Fig.~\ref{fig:dilep:star_hades}, the
excess seen by STAR agrees, within the comparatively large errors, 
with that observed by NA60 even for the high energy collisions at 
$\sqrt{s}=200 \GeV$. One of the reasons for this mild energy dependence 
of the excess is that the density of hadrons hardly changes, as can be 
seen by the nearly constant freeze out temperature as discussed in 
Section~\ref{sec:PJ_thermal}. Furthermore, the important contribution 
from baryons does not depend on the net baryon density but rather on 
the sum of baryons and anti-baryons, which in the thermal model remains 
nearly constant for top SPS energies ($\sqrt{s} = 17.3 \GeV$) and higher 
\cite{Randrup:2006nr}. Below top SPS energies, on the other hand, the 
density of baryons and thus that of baryons and anti-baryons 
increases as the incoming nuclei are stopped more
effectively. Consequently, one  
expects further enhancement of the excess below the $\rho$-mass, and 
indeed data taken by the CERES collaboration at $\sqrt{s}=8.75 \GeV$ 
confirm this expectation, albeit with large statistical errors 
\cite{Adamova:2002kf}. Thus, the measurement of dilepton invariant 
mass spectra in the energy range of $5 \GeV \lesssim \sqrt{s} \lesssim 8 
\GeV$ would be of great interest as this is the region of highest baryon 
plus anti-baryon density.

It is worth pointing out that the model of Rapp, Wambach and van Hees,
is able to reproduce the observed excess for all beam energies even at
the low energies where the fireball consists mostly of baryon
resonances \cite{Endres:2015fna}. In this sense it is fair to say that
the question of the low mass enhancement is settled and that it is the
various hadronic resonances and their interaction and mixing which are 
the origin of the observed excess. To which extent this can be related
to the fundamental question of chiral symmetry is still an open
question which, at present, can only be addressed within models
\cite{Hohler:2013eba}. Experimentally, this would require the
measurement of the axial correlator. If this is feasible
e.g.  via  $\gamma$-$\pi$ correlation remains to be seen.

%%%%%%%%%%%%%%%%%%%%%%%%%%%%%%%%%%%%%%%%%%%%%%%%%%%%%%%%%%%%%%%%%%%%%%%%%%%%
%%%%%%%%%%%%%%%%%%%%%%%%%%%%%%%%%%%%%%%%%%%%%%%%%%%%%%%%%%%%%%%%%%%%%%%%%%%%

 While the low mass dilepton spectrum  provides information on 
the chiral dynamics at finite temperature and density, the intermediate 
mass region between $1 \GeV \leq M_{inv} \lesssim 3 \GeV$ may provide
access to the thermal radiation from the quark-gluon plasma, as argued 
many years ago \cite{Shuryak:1978ij,Shuryak:1980tp,McLerran:1984ay,Kajantie:1986cu,Kajantie:1989yr}.
The major sources of background in the intermediate mass region are
lepton pairs from Drell-Yan production and semi-leptonic decays of
correlated open charm pairs. The latter turns out to be especially
difficult to cope with at the higher energies at RHIC and LHC, where
open charm is abundantly produced. This background is less severe at
SPS energies and the NA60 collaboration, which also measured open
charm, could successfully remove it together with the Drell-Yan pairs. 
The resulting spectrum, Fig.~\ref{fig:dilep:na60_dndm}, exhibits an 
almost perfect exponential fall off, which is consistent with thermal 
radiation with a temperature of $T = 205 \pm 12 \MeV$
\cite{Arnaldi:2008er,Specht:2010xu}. Since dileptons are emitted
during the entire evolution of the fireball, this temperature should
be considered as a lower limit of the temperature reached in these
collisions. Given the thermal nature of the spectrum 
and the fact that lattice QCD
predicts a transition temperature of $T_{c}\simeq 150 \MeV$ (see
Section~\ref{sec_lQCD}), one cannot but conclude that the intermediate
mass dileptons observed by NA60 originate from a QGP. This argument 
finds support in the model calculations shown in
Fig.~\ref{fig:dilep:na60_dndm}, which find the spectrum above
$M>1\GeV$ to be dominated by thermal radiation from the QGP. This was
recently confirmed in \cite{Rapp:2014hha}, where the authors augmented 
their calculation shown in Fig.~\ref{fig:dilep:na60_dndm} by a realistic 
treatment of the QGP based on the latest lattice results.

We note that the PHENIX collaboration reported a temperature of
$T_{\gamma} = 239 +- 26 \MeV $
\cite{Adare:2008ab,Afanasiev:2012dg,Adare:2014fwh} and ALICE an even
higher value of $T_{\gamma} = 297 \pm 43 \MeV $ \cite{Adam:2015lda}
from a measurement of the low transverse momentum direct photon
spectrum.
However, the photon momentum spectrum is subject to blue shift and, thus, 
the actual temperature may be considerably lower \cite{Paquet:2015lta,vanHees:2011vb}, 
potentially close to the transition temperature. The dilepton invariant 
mass spectrum, on the other hand, is unaffected by such a blue shift, and, 
therefore, appears to be the preferred method to determine the 
temperature of the QGP. 

Since the dileptons in the intermediate mass region seem to originate
predominantly from the QGP, it would be desirable to measure the
elliptic flow or azimuthal asymmetry of these lepton pairs. The
prevailing paradigm asserts that the observed elliptic flow is
generated predominantly in the QGP. If this is correct, it also should
be cleanly visible with intermediate mass dileptons. To which extent
such a measurement is feasible remains to be seen, and it is
encouraging that first attempts to extract azimuthal asymmetries of
dileptons, albeit at lower invariant mass, have successfully been
carried out \cite{Adamczyk:2014lpa}.

%%%%%%%%%%%%%%%%%%%%%%%%%%%%%%%%%%%%%%%%%%%%%%%%%%%%%%%%%%%%%%%%%%%%%%%%%%%%%
\section{Hadrons with heavy quarks} 
\label{sect:quarkonium}
%%%%%%%%%%%%%%%%%%%%%%%%%%%%%%%%%%%%%%%%%%%%%%%%%%%%%%%%%%%%%%%%%%%%%%%%%%%%%

We showed in Section~\ref{sec:PJ_thermal} that the thermal model
provides a successful description of the production of hadrons
composed of light quarks in ultrarelativistic nuclear collisions.
Given this result, it makes sense to ask whether a similar approach
can be used in the heavy quark sector. It was realized some time ago
\cite{BraunMunzinger:2000dv} that, because of the large charm quark
mass ($m_c \approx 1.2$ GeV), for temperatures $T$ reached
realistically in a Pb--Pb collision at LHC energy, thermal production
of charm quarks is strongly suppressed compared to the number of charm
quarks produced in initial, hard collisions. This clearly implies that
chemical equilibration is not achieved for charm quarks and certainly
not for beauty quarks. However, thermal equilibration of heavy quarks
may well take place in the hot and dense fireball. Indeed, there are
strong indications for significant rescattering of charm quarks after
they are formed very early, less than 0.1 fm/c after the start of the
collision. In the QGP, the energy loss of high energy heavy quarks 
is similar to that of light quarks or gluons. The measurements from the ALICE
collaboration at the LHC with fully reconstructed D mesons
\cite{ALICE:2012ab} demonstrate this over a large range of transverse
momenta. These measurements corroborate the earlier observations at
RHIC, where inclusive single electron data provided first indications
for energy loss of charm quarks at relatively low values of transverse
momentum \cite{Adare:2010de,Adare:2006nq}. Further evidence for
thermalization of charm quarks comes from flow studies. Measurements
of elliptic flow of heavy quarks at the LHC \cite{Abelev:2014ipa} and
at RHIC \cite{Adare:2006nq,Adamczyk:2014yew} demonstrate clearly that
heavy quarks also participate in the collective expansion of the hot
fireball. The simultaneous description of energy loss and flow imposes
further constraints on theoretical models \cite{He:2014cla}.

Nearly 30 years ago, charmonium production and, in particular, the
possible dissociation of charmonia in relativistic nuclear collisions
was proposed as a unique signature for a dense, deconfined medium
\cite{Matsui:1986dk}. Further theoretical development considering the
J$/\psi$ meson and all its excited states led to the prediction of
``sequential melting'' at different temperatures
\cite{Karsch:1990wi,Digal:2001ue,Karsch:2005nk} of charmonia with a
hierarchy determined by size and binding energies of the various
excited states.  Measurements first at the CERN SPS and later at RHIC
provided first evidence for such a dissociation mechanism, although
many puzzling aspects remained. In particular, the J/$\psi$
dissociation or suppression patterns observed at SPS and RHIC energies
nearly coincide, although the energy density in the hot fireball is
significantly increased in nuclear collisions at RHIC as compared to
top SPS energy. Furthermore, the suppression measured at RHIC is
smallest at mid-rapidity and suppression is stronger at forward and
backward rapidities. This is opposite to what is expected in scenarios
where the suppression increases with increasing energy density
\cite{Andronic:2007bi}, as the energy density peaks at mid-rapidity. A
recent review and compilation of data and their possible
interpretation in the original Matsui-Satz dissociation scenario can
be found in Ref. \cite{Kluberg:2009wc}.

Lattice QCD calculations, discussed in Section~\ref{sec_lQCD}, have
been used to shed light on possible melting effects in the plasma (for
a review see \cite{Mocsy:2013syh}). Indeed, evidence for sequential
melting was found in some of these calculations.  However, one should
remember that lattice QCD calculations can only be used to describe
completely equilibrated systems, and contain none of the dynamical
effects present in a nucleus-nucleus collision at high energy.
Investigating various scenarios of the possible influence of a thermal
medium on charmonium production two of us proposed an entirely new
approach. In this statistical hadronization model
\cite{BraunMunzinger:2000px}, the charm quarks which are all produced
in initial hard collisions (see above) thermalize in the QGP. At
chemical freeze-out, which for small enough values of $\mu_b$ (say
below 100 MeV) closely coincides with hadronization, as discussed in
Section~\ref{sec:PJ_thermal}, these thermalized charm quarks form
hadrons with heavy quarks; their yields are determined by thermal weights
equivalent to those used in the description of hadrons containing
light valence quarks \cite{Andronic:2006ky,BraunMunzinger:2000px},
with an additional charm fugacity factor described below. A more
detailed discussion of this new approach, together with predictions
made before data from the LHC became available, can be found in
\cite{BraunMunzinger:2007zz,BraunMunzinger:2009ih,Andronic:2011yq}.

The main new feature in this statistical hadronization model is that
all charmonia are formed late in the collision phase, i.e at the phase
boundary (hadronization). Since the charm quarks are explicitly not in
chemical equilibrium as they are produced in early, hard collisions, a
charm fugacity factor g$_c$ has to be introduced into the thermal
formulation, as described in
\cite{BraunMunzinger:2000px,BraunMunzinger:2000ep,Andronic:2003zv}. As
a consequence, all yields for open charm hadrons are multiplied with
g$_c$ while charmonia are enhanced by a factor g$_c^2$. Since g$_c$ is
(nearly) proportional to N$_c$, the number of charm quarks, this
implies a strong enhancement of charmonia with increasing collision
energy. This N$_c^2$ dependence implies that a charm quark in the
fireball may combine with any anti-charm quark to form a J$/\psi$ or
other charmonium state, as long as the charm and anti-charm quarks can
be causally connected. For this reason, in practice, only charm quarks
within a rapidity interval of $\Delta y < 1$ are considered for
charmonium formation \cite{Andronic:2006ky}. This also implies that
observation of J/$\psi$ production enhanced according to g$_c^2$ is a
clear sign for the presence of deconfined charm quarks. In view of all
the above considerations, little suppression or even enhancement of
charmonia was predicted for LHC energy, see, e.g.,
\cite{BraunMunzinger:2000px,Andronic:2003zv,BraunMunzinger:2007zz}.

Shortly after the proposal of \cite{BraunMunzinger:2000px}, an
alternative idea for production of charmonia was put forward via the
kinetic recombination of charm and anti-charm quarks in the QGP
\cite{Thews:2000rj}.  In this approach which has been taken up by
several groups
\cite{Liu:2009nb,Zhao:2011cv,Emerick:2011xu,Zhou:2014kka} continuous
dissociation and generation of charmonia is modeled by rate equations
and takes place during the lifetime of the QGP. Also in this approach,
the final yield of charmonia scales $\propto N_c^2$ and charmonium
enhancement is expected as N$_c$ becomes very large.

Before describing the recent data from the LHC we note that, for
historical reasons, both approaches are dubbed somewhat colloquially
in the literature as ``charmonium regeneration'' models.

%%%%%%%%%%%%%%%%%%%%%%%%%%%%%%%%%%%%%%%%%%%%%%%%%%%%%%%%%%%%%%%%%%%%%%%%%%%%%%
\begin{figure}[t]
\begin{tabular}{lr} \begin{minipage}{.49\textwidth}
\hspace{-0.3cm}\includegraphics[width=1.02\textwidth]{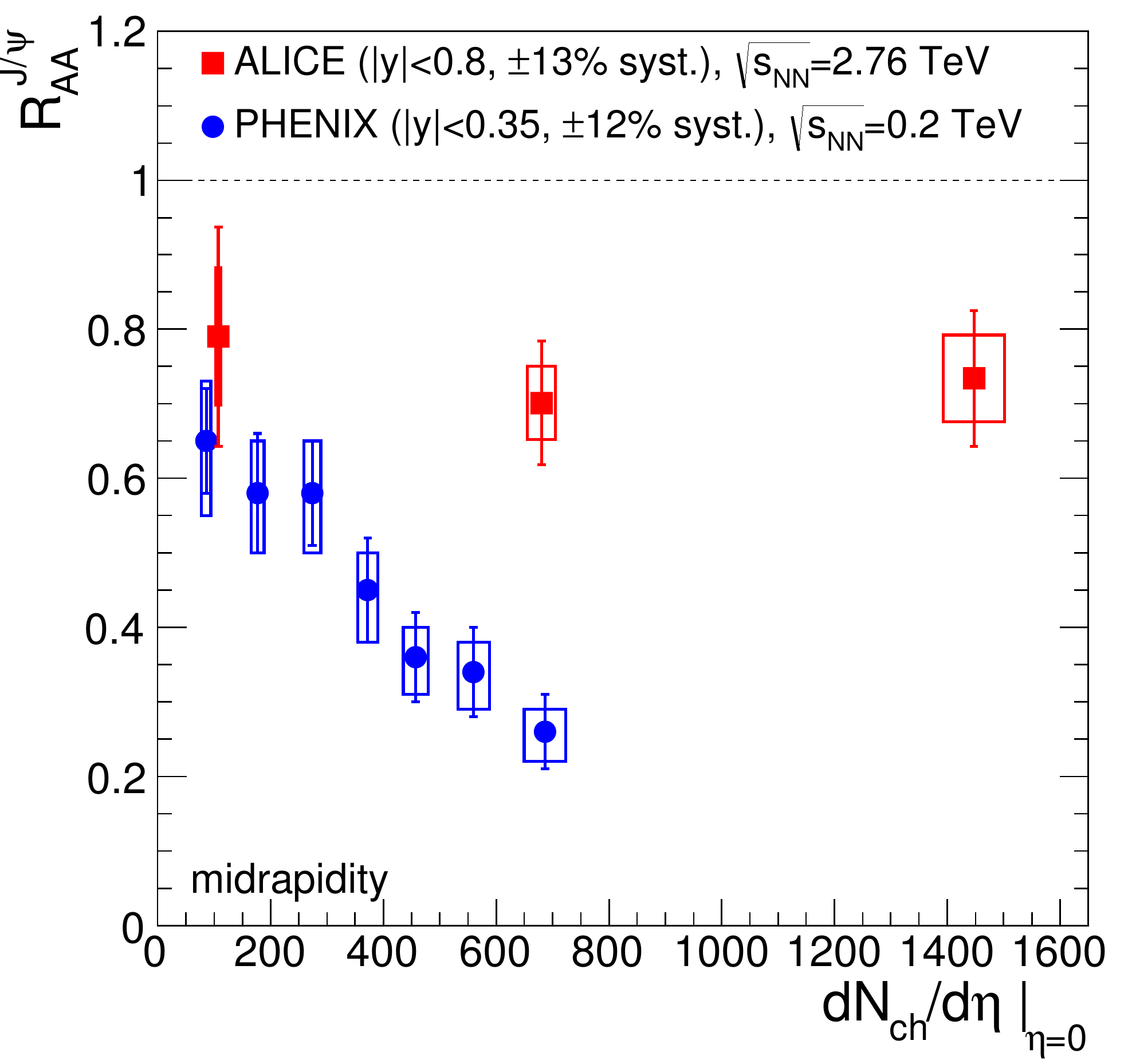}
\end{minipage} & \begin{minipage}{.49\textwidth}
\hspace{-0.3cm}\includegraphics[width=1.02\textwidth]{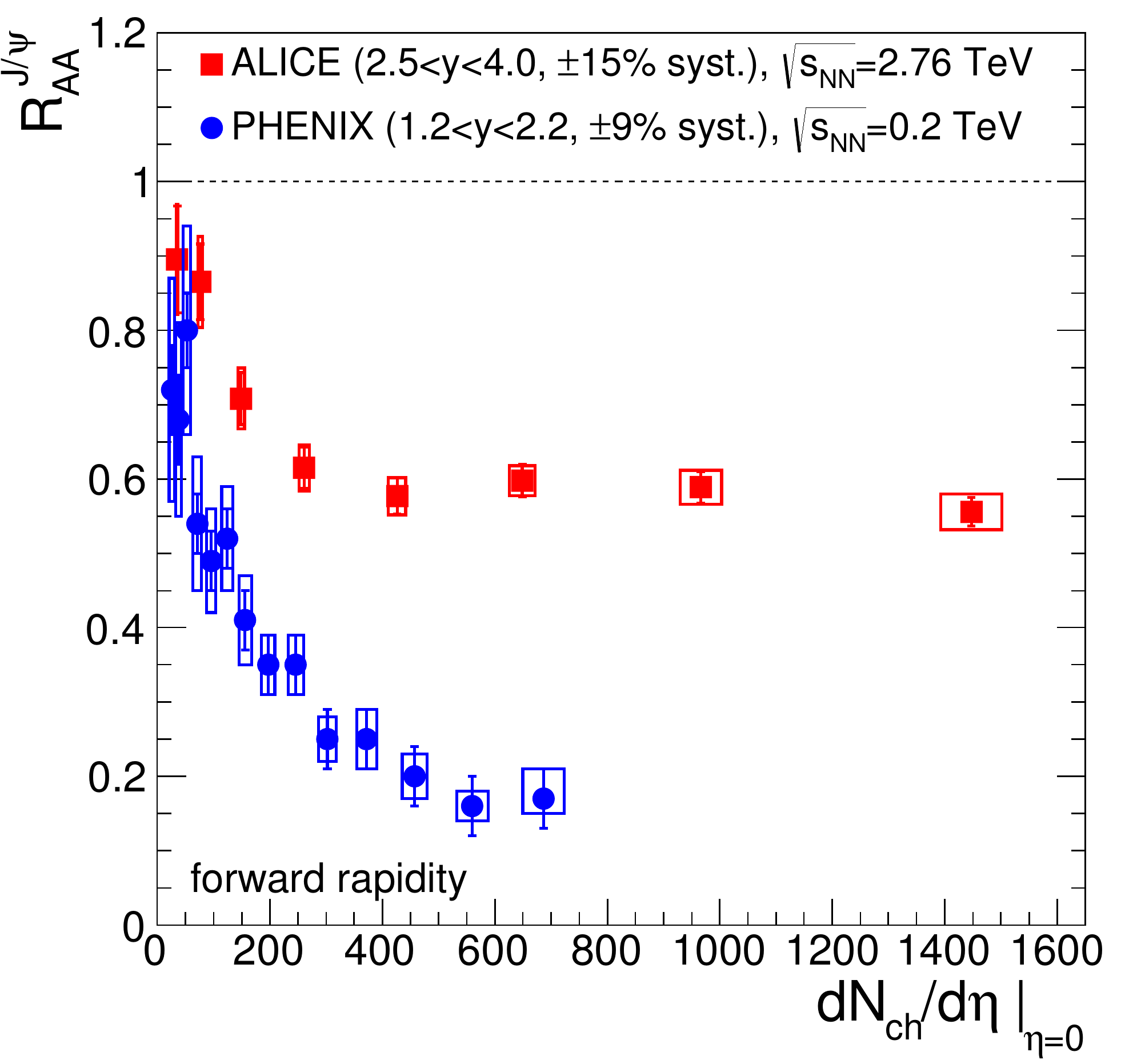}
\end{minipage}\end{tabular}
\caption{Dependence of the nuclear modification factor
  $R_{\mathrm{AA}}$ for inclusive J/$\psi$ production on the
  multiplicity density (at $\eta$=0).  The left panel shows data at
  mid-rapidity, the right panel at forward rapidity. The data are
  integrated over $p_T$ and are from the PHENIX collaboration
  \cite{Adare:2006ns} at RHIC and the ALICE collaboration
  \cite{Abelev:2013ila} at the LHC. Figure taken from
  \cite{Abelev:2013ila}. }
\label{fig:raa_jpsi1}
\end{figure}
%%%%%%%%%%%%%%%%%%%%%%%%%%%%%%%%%%%%%%%%%%%%%%%%%%%%%%%%%%%%%%%%%%%%%%%%%%%%%%

Since the center of mass energy increase between RHIC and LHC is more
than a factor of 20, the corresponding increase in N$_c$ was estimated
to be about an order of magnitude, while the energy density of the
fireball should increase by about a factor of 2-3. Consequently, the
measurement of the centrality dependence of $R_{\mathrm{AA}}$ for
inclusive J/$\psi$ production in Pb--Pb collisions at the LHC was
expected to provide important information on the question of strong
suppression due to Debye screening vs. regeneration.

%%%%%%%%%%%%%%%%%%%%%%%%%%%%%%%%%%%%%%%%%%%%%%%%%%%%%%%%%%%%%%%%%%%%%%%%%%%%%%
\begin{figure}[t]
\centering
\includegraphics[width=.61\textwidth,height=.6\textwidth]{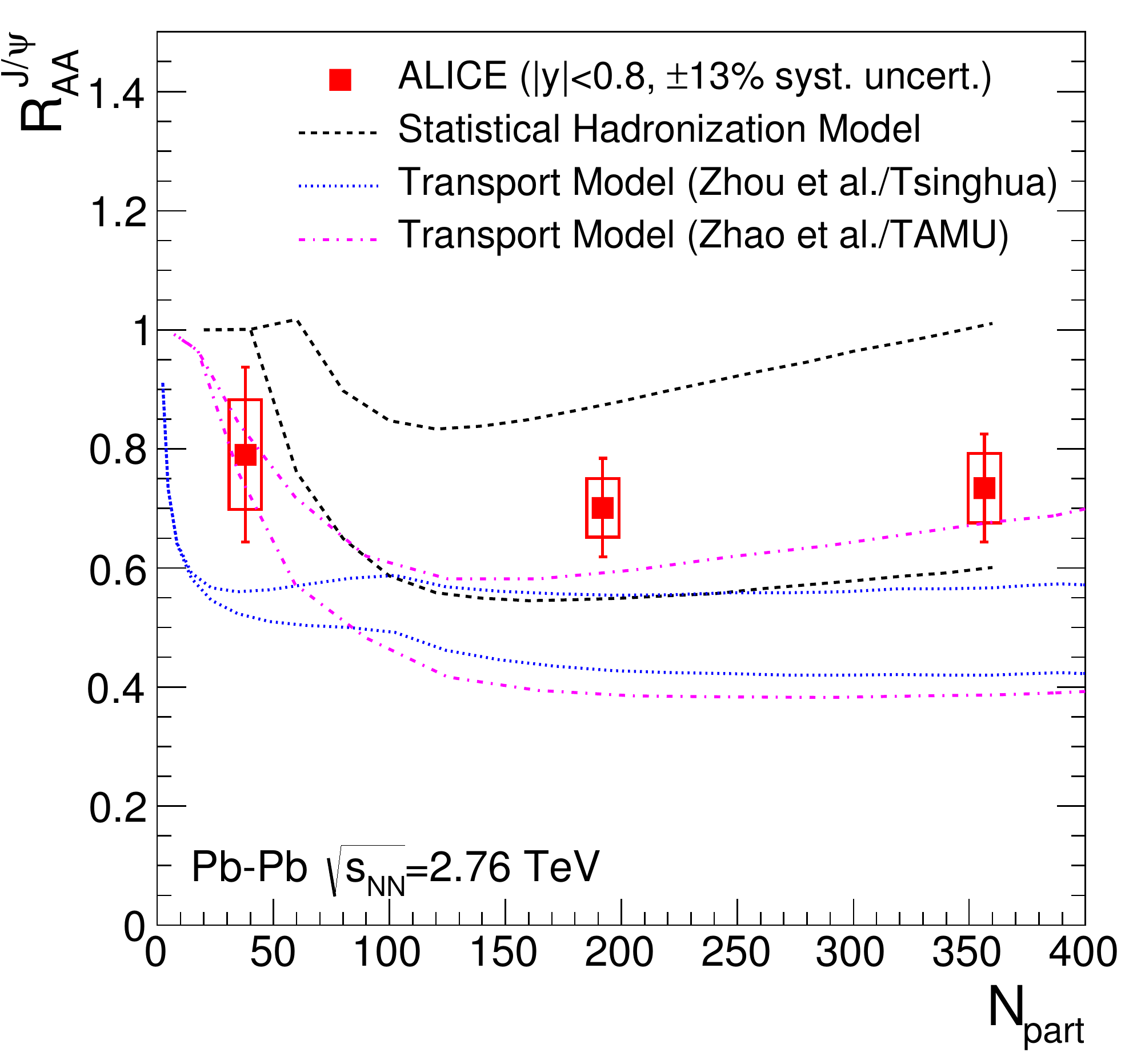}
\caption{Centrality dependence of the nuclear modification factor for
  inclusive J$/\psi$ production at the LHC. The data at mid-rapidity
  \cite{Abelev:2013ila} are compared to model calculations: the
  statistical hadronization model \cite{Andronic:2011yq} and transport
  models of the TAMU \cite{Zhao:2011cv,Zhao:2012gc} and Tsinghua
  \cite{Liu:2009nb,Zhou:2014kka} groups. The band of the statistical
  hadronization model reflects the uncertainty in the charm production
  cross section: The lower and upper black dashed lines correspond to
  d$\sigma_{c\bar{c}}$/dy = 0.3 and 0.4 mb. Figure taken from
  \cite{Abelev:2013ila}.}
\label{fig:raa_jpsi2} 
\end{figure}
%%%%%%%%%%%%%%%%%%%%%%%%%%%%%%%%%%%%%%%%%%%%%%%%%%%%%%%%%%%%%%%%%%%%%%%%%%%%%%

Indeed, the first data from the LHC on the centrality dependence of
$p_{T}$ integrated J/$\psi$ yields \cite{Abelev:2012rv} exhibited, for
forward rapidities, values of the nuclear modification factor
$R_{\mathrm{AA}}$ which significantly exceeded those measured at RHIC
energies, providing first qualitative evidence for the charmonium
regeneration scenario. Soon higher statistics data including
measurements at mid-rapidity \cite{Abelev:2013ila} confirmed this, see
Fig.~\ref{fig:raa_jpsi1}. Note that the quantity on the horizontal
axis on this plot, the charged particle pseudo-rapidity density, is
essentially proportional to the energy density. For the most central
collisions, the energy density between RHIC and LHC increases by more
than a factor of 2, but $R_{\mathrm{AA}}$ does not decrease at the
higher energy density, but rather increases by nearly a factor of 3.

Further evidence for the regeneration mechanism comes from the fact
that the data are well described by both the statistical hadronization
model \cite{Andronic:2011yq} and by transport models
\cite{Liu:2009nb,Zhao:2011cv}. This is demonstrated in
Fig.~\ref{fig:raa_jpsi2}, although admittedly the uncertainties in the
models are still quite large. The main uncertainty is due to the fact
that the charm production cross section has not yet been measured for
Pb--Pb collisions at the LHC and, when extrapolating from the pp cross
section the uncertainty in the nuclear parton distributions is
significant.

%%%%%%%%%%%%%%%%%%%%%%%%%%%%%%%%%%%%%%%%%%%%%%%%%%%%%%%%%%%%%%%%%%%%%%%%%%%%%%
\begin{figure}[t]
\centering\includegraphics[width=.62\textwidth]{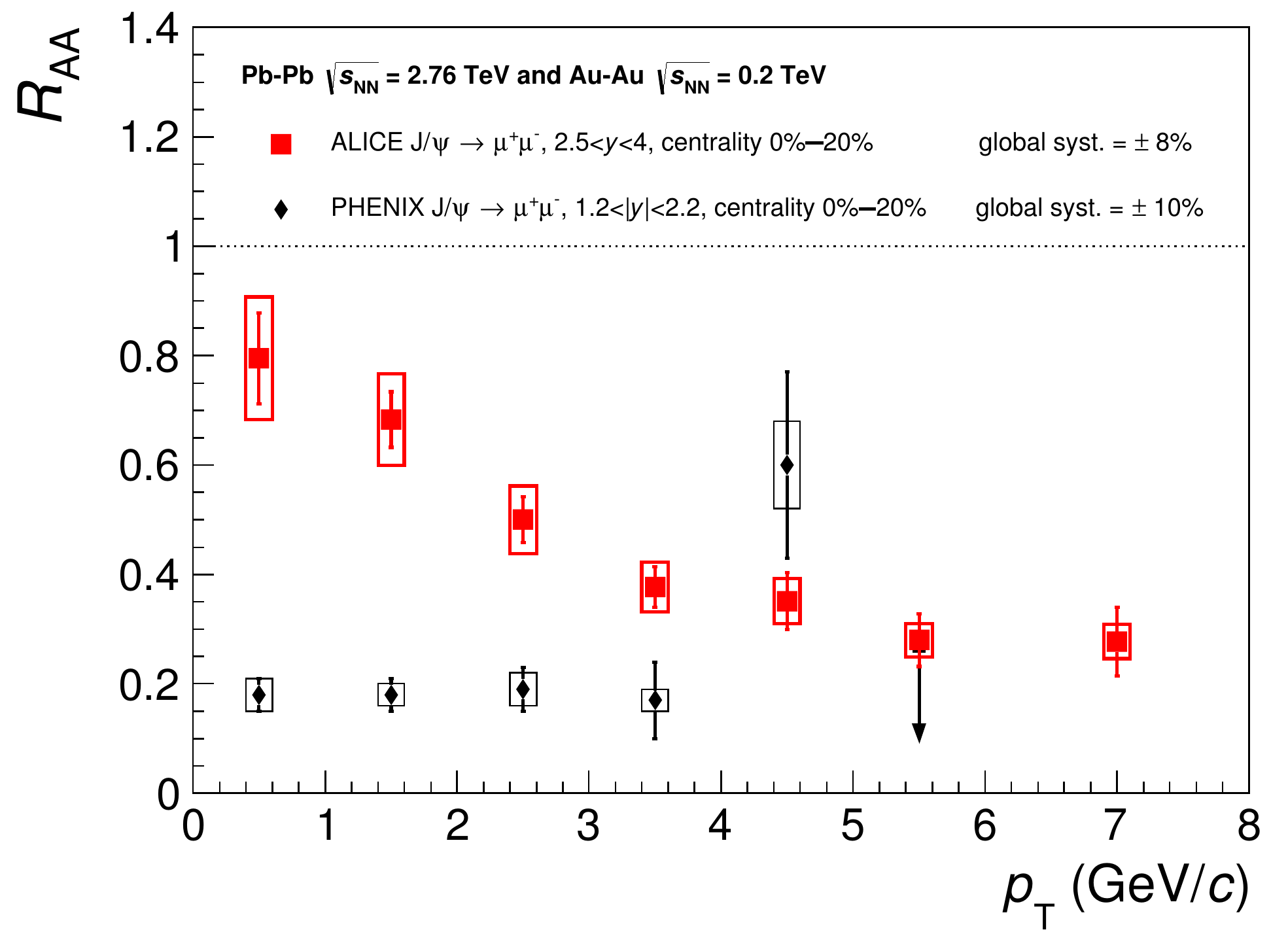}
\caption{Transverse momentum dependence of the nuclear modification factor 
for J/$\psi$ production measured at RHIC \cite{Adare:2006ns} and at the 
LHC \cite{Abelev:2013ila}. Figure taken from \cite{Abelev:2013ila}.} 
\label{fig:raa_jpsi3} 
\end{figure}
%%%%%%%%%%%%%%%%%%%%%%%%%%%%%%%%%%%%%%%%%%%%%%%%%%%%%%%%%%%%%%%%%%%%%%%%%%%%%%

Recent data also provided insight into the transverse momentum
dependence of charmonium production at the LHC as illustrated by the
$p_{T}$ dependence of $R_{\mathrm{AA}}$ measured by ALICE and CMS,
shown in Fig.~\ref{fig:raa_jpsi3}. These data are quite well described
in the above discussed transport models
\cite{Zhao:2011cv,Zhou:2014kka}. There, at small transverse momenta a
large fraction of the J/$\psi$ yield in Pb--Pb collisions comes from
the regeneration mechanism, while this fraction decreases with
increasing $p_{T}$. Within the transport model description,
regeneration is mainly a low-$p_{T}$ phenomenons and this can
account well for the data.

On the other hand, an important assumption underlying the
regeneration process is that the charm quarks reach, during the
evolution of the collision, some degree of equilibration. Inspection
of the centrality dependent shape of transverse momentum spectra
\cite{Abelev:2013ila}, quantified by the mean and variance of the
distributions, lends support to this assumption. Furthermore, the
measurement of a clear signal of J/$\psi$ elliptic flow at the LHC
\cite{ALICE:2013xna} brings an additional argument towards charm quark
thermalization. The J/$\psi$ data at RHIC are compatible with a
vanishing flow signal \cite{Adamczyk:2012pw}, but the errors are large
enough to also allow the magnitude of the flow signal observed at LHC
energy. A non-zero $v_2$ signal was also measured for J/$\psi$
production at the SPS \cite{Prino:2008zz} but was interpreted as a
path-length dependence of Debye screening.

The data by the CMS and ALICE collaborations measured at higher
$p_{T}$ \cite{Chatrchyan:2012np,Abelev:2013ila} exhibit a strong
suppression of J/$\psi$ mesons in Pb--Pb compared to pp
collisions. Importantly, this suppression is similar in magnitude to
that measured for open-charm hadrons. This suppression in the J/$\psi$
signal therefore may also be a result of the energy loss of
high-$p_{T}$ heavy quarks in the hot and dense fireball, leading to
thermalization and, finally, at hadronization, to the formation of
charmonia and open charm mesons.

Based on the new LHC data and the above observations and considerations, 
charmonium production should  be considered a probe of deconfinement of 
heavy quarks rather than a ``thermometer'' of the QGP. Within the 
framework of the thermal approach sketched here, the charmonium states 
become important probes of the phase boundary between the deconfined and 
the hadron phase. The quarkonium data at the LHC have found a natural
explanation in terms of the regeneration model while no other
plausible interpretation was put forward.  

Discriminating between the two pictures of disintegration and
regeneration of charmonia in the QGP versus that of assembly of
charmonia from deconfined charm quarks at the phase boundary
(statistical hadronization) will shed light on fundamental questions
connected with the fate of bound states in a deconfined medium: Can a
J/$\psi$ meson be formed from deconfined charm quarks at temperatures
well above the deconfinement phase transition
\cite{BraunMunzinger:2009ih}? For further discussions about bound
states in a hot medium see also
\cite{Liu:2006nn,Mocsy:2007yj,Laine:2011xr}. Data at the top LHC
energy, including measurements on $\psi(2S)$ production in Pb--Pb
collisions, should clarify these questions. A subject of intense
current research is the study of $\psi(2S)$ production in light
systems such as d--Au collisions at RHIC \cite{Adare:2013ezl} and in
p--Pb at the LHC \cite{Abelev:2014zpa}, where unexpected findings of
suppression indicate possible final-state effects.

\begin{figure}[t]
\begin{tabular}{lr} \begin{minipage}{.49\textwidth}
\hspace{-0.3cm}\includegraphics[width=1.02\textwidth]{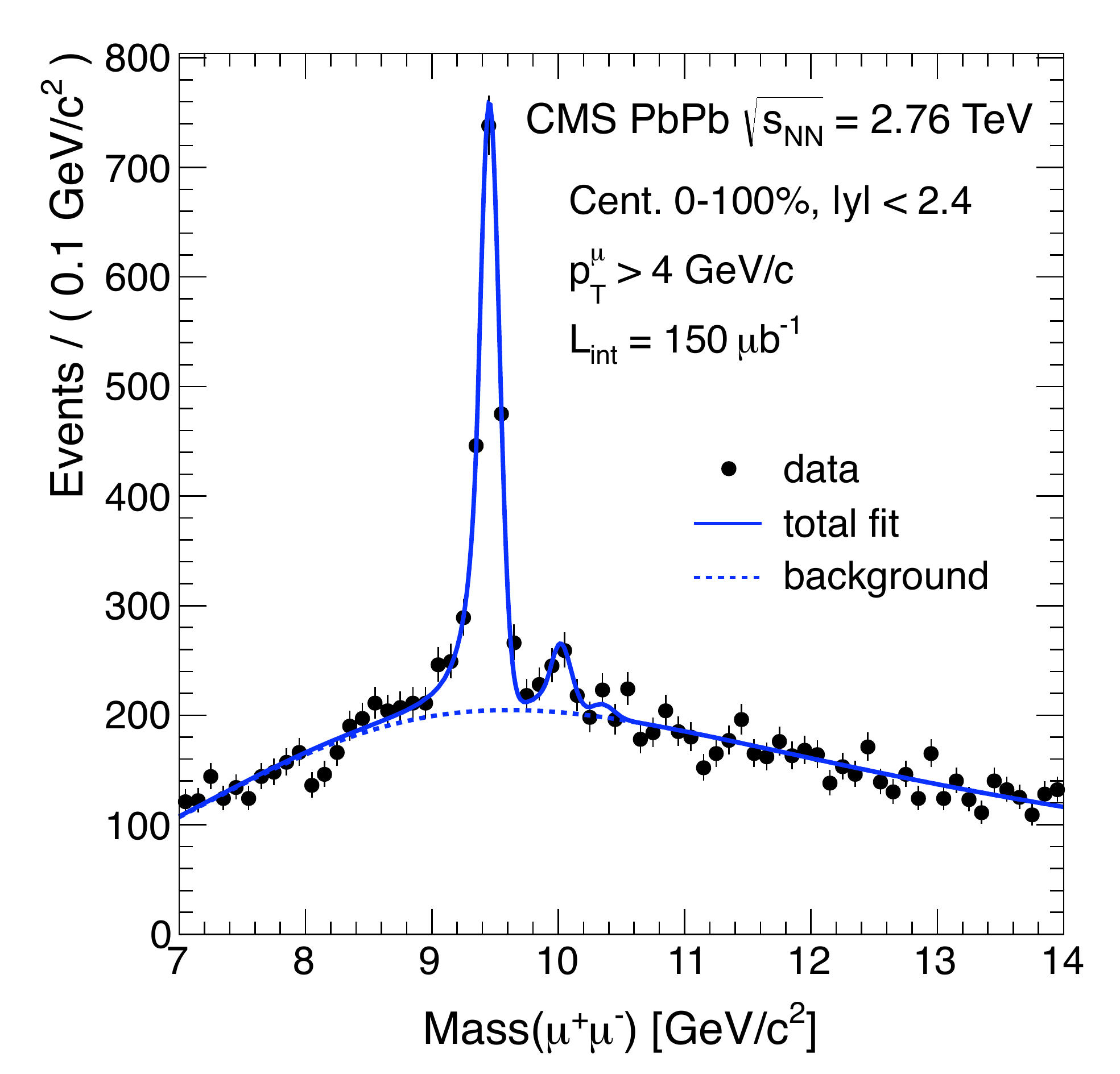}
\end{minipage} & \begin{minipage}{.49\textwidth}
\hspace{-0.3cm}\includegraphics[width=1.02\textwidth]{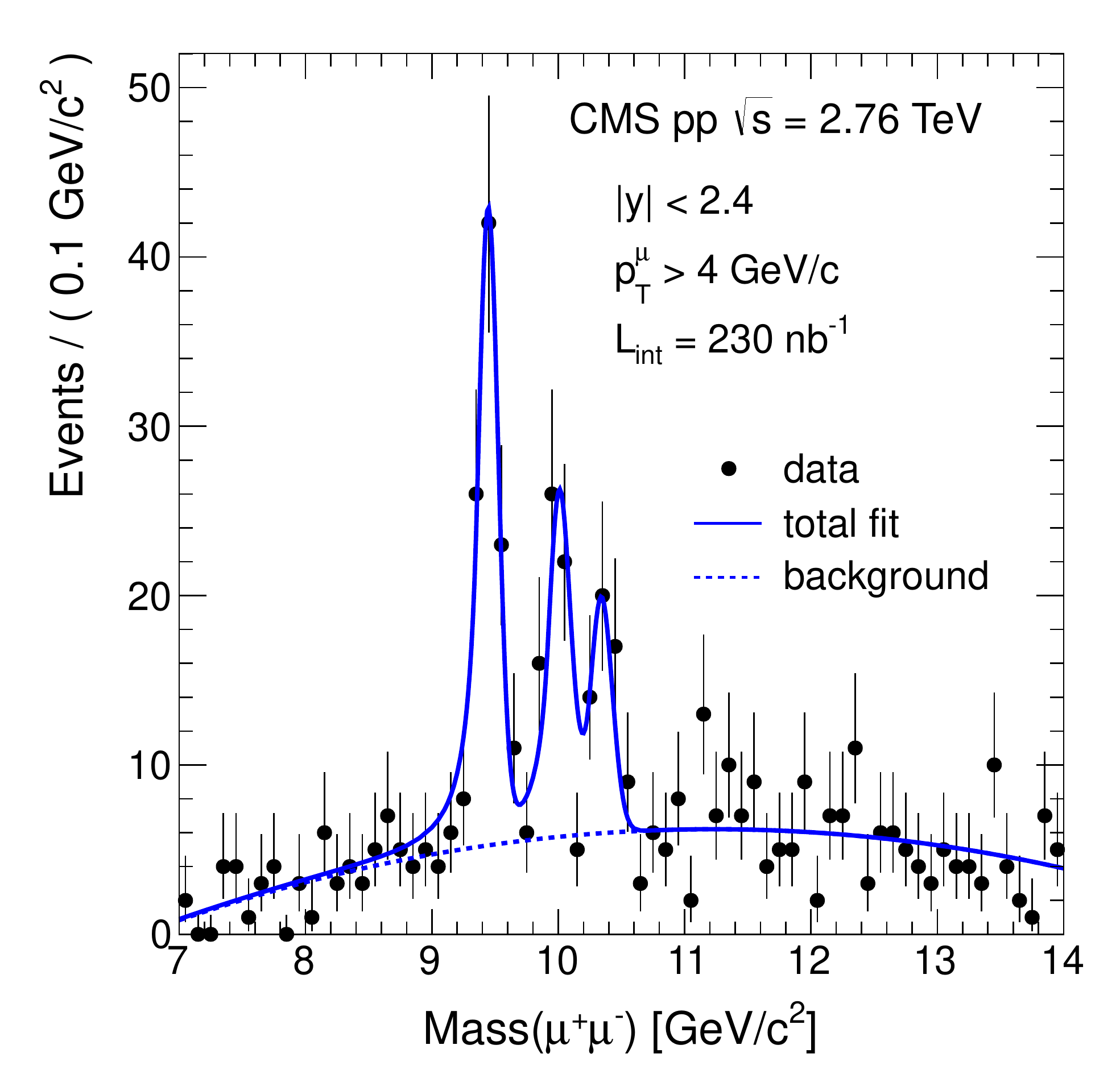}
\end{minipage}\end{tabular}
\caption{Suppression pattern of $\Upsilon$ states measured by the CMS
  collaboration in Pb-Pb collisions
  (left) and pp collisions (right). Figure taken from
  \cite{Chatrchyan:2012lxa}. }
\label{fig:cms_PbPb}
\end{figure}

We close this Section with a short discussion of LHC measurements on
the production of bottomonia (mesons composed of $b\bar{b}$ quarks)
\cite{Chatrchyan:2012lxa,Chatrchyan:2013nza,Abelev:2014nua} and of
similar measurements at RHIC \cite{Adare:2014hje}. The nuclear
modification factor for the $\Upsilon$ states at both RHIC and LHC
clearly shows a suppression pattern
\cite{Chatrchyan:2012lxa,Adare:2014hje} with increasing suppression
from $\Upsilon (1)$ to $\Upsilon (2s)$ and $\Upsilon (3s)$. The CMS
data are displayed in Fig.~\ref{fig:cms_PbPb}. Whether this indicates
sequential suppression due to different binding energies or radii of
the quarkonia is currently the subject of an intense debate. We first
note that the radii and binding energies of J/$\psi$ and $\Upsilon
(2s)$ mesons are similar, but the observed suppression is very
different, casting doubt on a simple Debye screening interpretation.
Further there are the questions of possible thermal equilibration of
beauty quarks as well as on the magnitude of feeding from higher-lying
bottomonia into the 1s state and on the issue of the rapidity
dependence of the nuclear modification factor of bottomonia. Together,
the data from CMS \cite{Chatrchyan:2012lxa} and from ALICE
\cite{Abelev:2014nua} indicate that $R_{\mathrm{AA}}$ peaks at
mid-rapidity although the energy density is largest there. This is a
real challenge to various theoretical models. The situation is
succinctly described in \cite{Andronic:2015wma}.

%%%%%%%%%%%%%%%%%%%%%%%%%%%%%%%%%%%%%%%%%%%%%%%%%%%%%%%%%%%%%%%%%%%%%%%%%%%%%%
%%%%%%%%%%%%%%%%%%%%%%%%%%%%%%%%%%%%%%%%%%%%%%%%%%%%%%%%%%%%%%%%%%%%%%%%%%%%%%

An interesting observation is made in \cite{Andronic:2014sga}: the
ratio $\Upsilon (2S)/\Upsilon (1S)$, is characteristically different
in Pb--Pb compared to pp collisions. Indeed, the data are consistent,
for the most central Pb--Pb collisions, with the value predicted
within the framework of the statistical hadronization model for a
temperature $T=159$ MeV, very close to what was used for the
charmonium case. Is the bottomonium situation also described in the
regeneration picture? More precise data, on the beauty cross section
in Pb--Pb collisions, the question of possible thermalization of
b-quarks, but also in particular for bottomonia on the dependence of
$R_{\mathrm{AA}}$ on transverse momentum and rapidity, are needed to
clarify this interesting possibility.

%%%%%%%%%%%%%%%%%%%%%%%%%%%%%%%%%%%%%%%%%%%%%%%%%%%%%%%%%%%%%%%%%%%%%%%%%%%%%
\section{Outlook}
\label{sec:out}
%%%%%%%%%%%%%%%%%%%%%%%%%%%%%%%%%%%%%%%%%%%%%%%%%%%%%%%%%%%%%%%%%%%%%%%%%%%%%

The relativistic heavy ion programs at RHIC and the LHC have led to
significant advances in our understanding of hot and dense matter.
The experiments have demonstrated that rapid local equilibration takes
place, and this discovery has enabled a systematic program of
determining the properties of strongly interacting matter, such as the
equation of state, the shear and bulk viscosity, the heavy quark
diffusion constant. In addition, it has been established that even very
high energy (several hundreds of GeV) partons lose large fractions of
their energy in the hot fireball, and first signs of deconfinement
have emerged from comparative studies of charmonium production at RHIC
and the LHC.

In the near future, energy and luminosity upgrades at the LHC, and a
suite of detector upgrades at LHC and RHIC, will allow much more
precise measurements of these and related quantities, and explore the
limits of the perfect fluid paradigm in terms of system size,
geometry, and beam energy. In addition to that, experiments at the
ongoing beam energy scan program at RHIC are being performed to search for
critical fluctuations, with the goal of identifying a possible
endpoint of the QCD phase transition. Furthermore, searches will
continue for exotic phenomena such as the existence of exotic bound states
as e.g. dibaryons consisting of a multi-strange baryon with a nucleon, or of
a $\Lambda$ hyperon with two neutrons. Other opportunities for
discovery include the search for quantum anomalies in the quark-gluon
fluid at RHIC and LHC.

In the longer term a number of facilities are expected to come online
that will study the regime of maximum baryon density, which is reached
at energies of a few GeV per baryon in the center of mass. This is the
regime previously explored in the AGS and the SPS fixed
target programs. The new set of experiments planned at the FAIR
facility at GSI, NICA at Dubna, and JPARC, will study collisions in
this energy regime with unprecedented precision, and will make use of
all the knowledge that was gained at the higher energies. The regime
of maximum baryon density presents new challenges: The applicability
of fluid dynamics is not clear, the freeze-out line no longer follows
the phase boundary, and modifications of hadron properties are likely
to be important. Furthermore, the 'nearly instantaneous' collision
time scenario which at collider energies makes the time sequence of
events in a collision transparent will have to be replaced by
collision times comparable to the lifetime of the hot fireball, a
challenge for all theoretical descriptions. Nevertheless, great
opportunities exist and are bounded only by our own ingenuity.
 
Gerry Brown's unique style and insights will continue to guide our
work in future research, and that of many others in the field who were
fortunate to interact with him over the course of his long career.

\section{Acknowledgements}

The authors thank Jiangyong Jia, Klaus Reygers, Hans Specht, Joachim
Stroth, and Zhangbu Xu for helpful communications. Two of us (PBM, JS)
thank thank A. Andronic and K. Redlich for a long time collaboration on
issues related to the thermal/statistical model. V.K. and T.S. are
supported by the Director, Office of Energy Research, Office of High
Energy and Nuclear Physics, Divisions of Nuclear Physics, of the
U.S. Department of Energy under Contract No. DE-AC02-05CH11231 and
DE-FG02-03ER41260, respectively. J.S. is supported in part by the
German BMBF ministry, contract 05P12VHCA1, and by the ExtreMe Matter
Institute EMMI, contract HA216-UHD.

\bibliography{Full_Bib}

\end{document}